\pdfoutput=1
\documentclass[aps,prd,amsmath,floats,floatfix, twocolumn,
superscriptaddress,nofootinbib,showpacs]{revtex4-1}

\usepackage{txfonts}
\usepackage[T1]{fontenc}
\usepackage[utf8]{inputenc}
\usepackage{tensor}
\usepackage{multirow}
\usepackage{textcomp}

\usepackage{verbatim}
\usepackage[utf8]{inputenc}
\usepackage[dvipsnames, usenames]{xcolor}
\definecolor{linkcolor}{rgb}{0.0,0.3,0.5}
\usepackage[hypertexnames=false, unicode, colorlinks=true, linkcolor=linkcolor,
citecolor=linkcolor, filecolor=linkcolor,urlcolor=linkcolor,
pdfusetitle]{hyperref}

\usepackage[all]{hypcap}
\usepackage{graphicx}
\usepackage{xspace}
\usepackage{amsmath,amssymb}
\usepackage[normalem]{ulem} 
\usepackage{bm} 
\usepackage{mathrsfs}
\usepackage[caption=false]{subfig}

\usepackage{microtype}

\graphicspath{%
  {figs/}%
}

\DeclareMathAlphabet{\mathpzc}{OT1}{pzc}{m}{it}

\newcommand{\sytt}{\tensor*[_{-2}]{Y}{_\ell_m}}
\newcommand{\sytmt}{\tensor*[_{-2}]{Y}{_\ell_{,-m}}}

\newcommand{\sy}{\tensor*[_{-2}]{Y}{_\ell_m}(\iota,\beta)}

\newcommand{\h}{h}
\newcommand{\ddh}{\ddot{h}}
\newcommand{\hlm}{h_{\ell m}}
\newcommand{\hlmm}{h_{\ell,-m}}

\newcommand{\etal}{\textit{et al.\ }}

\newcommand{\dphi}{\Delta\Phi_{\rm IS}}

\newcommand{\2}{{(2)}}
\newcommand{\lm}{{\ell m}}
\newcommand{\lp}{I_{22}^m}
\newcommand{\spp}{S_{22}^m}

\DeclareMathOperator{\sech}{sech}
\DeclareMathOperator{\arctanh}{arctanh}
\DeclareMathOperator{\arccoth}{arccoth}
\DeclareMathOperator{\arccot}{arccot}

\begin{document}

\title{Universal features of gravitational waves emitted by  superkick
binary black hole systems}

\newcommand\caltech{\affiliation{TAPIR 350-17, California Institute of
Technology, 1200 E California Boulevard, Pasadena, CA 91125, USA}}
\newcommand{\cornell}{\affiliation{Cornell Center for Astrophysics
    and Planetary Science, Cornell University, Ithaca, New York 14853, USA}} 
\newcommand\cornellPhys{\affiliation{Department of Physics, Cornell
    University, Ithaca, New York 14853, USA}}

\author{Sizheng Ma}
\email{sma@caltech.edu}
\caltech

\author{Matthew Giesler}
\cornell

\author{Vijay Varma}
\thanks{Klarman fellow}
\cornellPhys
\cornell

\author{Mark A. Scheel}
\caltech

\author{Yanbei Chen}
\email{yanbei@caltech.edu}
\caltech

\hypersetup{pdfauthor={Ma et al.}}

\date{\today}

\begin{abstract}
We use numerical relativity to study the merger and ringdown stages of
``superkick'' binary black hole systems (those with equal mass and anti-parallel spins).  We find a universal way to describe the   mass
and current quadrupole gravitational waves emitted by these systems during the merger and ringdown stage: (i) The time evolutions of these waves are
insensitive to the progenitor's parameters (spins) after being normalized by their own
peak values. (ii)  The  peak values, which encode  all the spin information of the progenitor, can be consistently fitted to formulas
inspired by post-Newtonian theory. We find that the universal
evolution of the mass quadrupole wave can be accurately modeled by the
so-called Backwards One-Body (BOB) model. However, the BOB model, in its
present form, leads to a lower waveform match and a significant
parameter-estimation bias for the current quadrupole wave. We also
decompose the ringdown signal into seven overtones, and study  the dependence of mode
amplitudes on the progenitor's parameters. Such dependence is found to be insensitive to the overtone index (up to a scaling factor).
Finally, we use the Fisher matrix technique to investigate how the ringdown waveform can be at least as important for parameter estimation as the inspiral stage. Assuming the Cosmic Explorer, we find the contribution of ringdown portion dominates as the total mass exceeds $\sim 250\,M_\odot$. For massive BBH systems, the accuracy of parameter measurement is improved by incorporating the information of ringdown --- the ringdown sector gives rise to a different parameter correlation from inspiral stage, hence the overall parameter correlation is reduced in full signal.
\end{abstract}

\maketitle

\section{Introduction} 
\label{sec:introduction}
The recently detected gravitational wave (GW) signal, GW190521, is
consistent with the merger of two black holes (BHs) with masses of $85M_\odot$
and $66M_\odot$ \cite{Abbott:2020tfl,Abbott:2020mjq}. The detection of this
event, together with its candidate optical counterpart ZTF19abanrhr
\cite{Graham:2020gwr}, indicates the potential existence of BHs in the mass gap
predicted by (pulsational) pair-instability supernova theory
\cite{Woosley:2016hmi,2019ApJ...878...49W,Abbott:2020mjq}.
A few studies also suggest that this
system could admit a extremely eccentric \cite{Abbott:2020mjq,Gayathri:2020coq}, hyperbolic \cite{Gamba:2021gap}, or a head-on
\cite{Abbott:2020mjq,CalderonBustillo:2020odh} merger interpretation, placing possible constraints on the binary's formation channel \cite{Gayathri:2020coq,CalderonBustillo:2020srq,Gamba:2021gap}. For such
an event, most of the GW detected by the Advanced LIGO
\cite{TheLIGOScientific:2014jea}, VIRGO \cite{TheVirgo:2014hva} and KAGRA
\cite{Akutsu:2018axf,KAGRA:2020tym} network is dominated by the merger and ringdown portions.
This demonstrates the importance of understanding ringdowns for detecting more
GW190521-like cases in the near future \cite{Biscoveanu:2021nvg}. 

The ringdown signal can be treated as a superposition of damped sinusoids,
corresponding to the quasi-normal modes (QNMs) of the final BH
\cite{Kokkotas:1999bd}. Due to the no-hair theorem \cite{PhysRevLett.26.331},
the QNM frequencies and damping time for a spinning BH in general relativity (GR) are fully determined by
its mass and angular momentum. Therefore, measuring a QNM from a GW event can
allow us to determine the properties of the final BH. Alternatively, if
multiple modes are observed at the same time, we can use them to test the
no-hair theorem and general relativity
\cite{Dreyer:2003bv,Berti:2005ys,Berti:2007zu,Gossan:2011ha,Meidam:2014jpa,Berti:2015itd,Berti:2016lat,Baibhav:2017jhs,Baibhav:2018rfk,Berti:2018vdi,Brito:2018rfr,Carullo:2018sfu,Isi:2019aib,Giesler:2019uxc,Capano:2021etf,LIGOScientific:2019fpa,TheLIGOScientific:2016src,Carullo:2019flw,Carullo:2018sfu,Cabero:2017avf,Nagar:2016iwa},
and also constrain modified gravity
\cite{LIGOScientific:2019fpa,TheLIGOScientific:2016src,Cardoso:2019mqo,McManus:2019ulj,Maselli:2019mjd}.


In addition to measuring QNM frequencies, extensive
studies have also been carried out to explore the relationship between
progenitor's parameters and additional ringdown signatures. For instance, the spin
(magnitude and direction) and mass of the remnant BH were fitted to
progenitor's spins $(\bm{\chi}_{1,2})$ and mass ratio $(q=m_{\rm heavy}/m_{\rm light}>1)$
\cite{Rezzolla:2007xa,Rezzolla:2007rz,Rezzolla:2007rd,Buonanno:2007sv,Tichy:2008du,Kesden:2008ga,Barausse:2009uz,Kesden:2010yp,Barausse:2012qz,Healy:2014yta,Healy:2016lce,Jimenez-Forteza:2016oae,Hofmann:2016yih,Healy:2018swt,Varma:2018aht,Varma:2019csw},
as well as  the peak amplitude of GW strain \cite{Ferguson:2019slp}, using
numerical relativity (NR) \cite{Nagar:2016iwa}, the effective-one-body (EOB) approach
\cite{Damour:2007cb}, and also a hybrid way that involves multi-timescale
post-Newtonian integrations and numerical-relativity surrogate models
\cite{Reali:2020vkf}. The gravitational wave frequency at peak amplitude
\cite{Healy:2014eua,Healy:2018swt} and the peak GW luminosity
\cite{Healy:2016lce,Healy:2018swt,Taylor:2020bmj} were both found to have a clean dependence on the progenitor's parameters. The above facts clearly
imply that the initial conditions (e.g., at merger) for the progenitor are encoded in the ringdown portion of GW, including QNM frequencies and amplitudes. Therefore it's not surprising that the ringdown can be used to learn about the component properties.


\begin{figure}[ht]
    \centering
    \subfloat[SKu]{\includegraphics[width=\columnwidth,height=6.9cm]{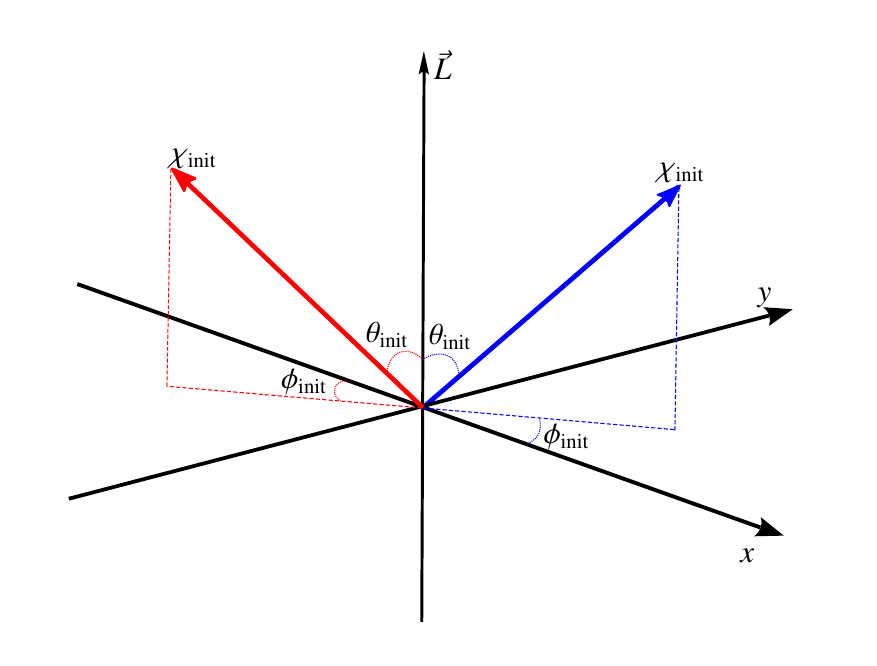}}\\
    \subfloat[SKd]{\includegraphics[width=\columnwidth,height=7cm]{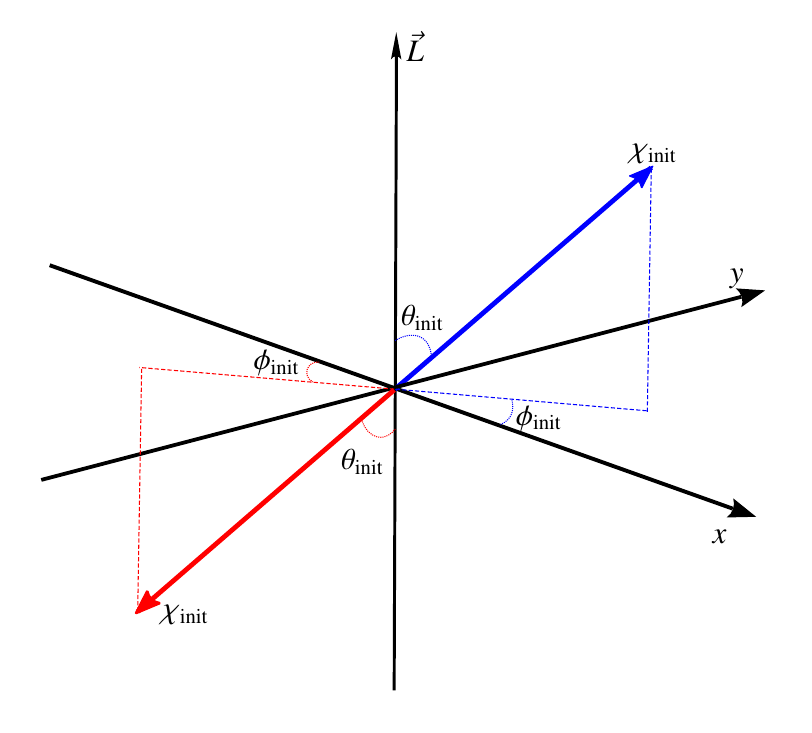}}
    \caption{Sketches for a SKu (a) and a SKd (b) system. Two arrows (in different colors) represent two individual spins. The letter ``u'' and ``d'' refer to the up- and down-state for the red arrow. Both SKu and SKd systems have equal mass BHs
  with the same dimensionless spin magnitude $\chi_{\rm init}$. For SKd, two
  individual spins are anti-parallel, whereas for SKu, only the orbital-plane
  components are opposite. SKd and SKu are fully characterized by three
  parameters: $(\chi_{\rm init},\theta_{\rm init},\phi_{\rm init})$, where
  $\theta_{\rm init}$ stands for the polar angle of one of the holes (relative
  to the orbital angular momentum), and $\phi_{\rm init}$ the azimuthal angle
  of the in-plane spin measured from the line of two BHs. Three parameters are
  specified at a reference time in the inspiral regime (labeled by the
  subscript `init').}
    \label{fig:configuration}
\end{figure}

Apart from conveying the importance of ringdown studies, the detection of the
candidate optical counterpart of GW190521 has also provided us with a new
scheme to measure the gravitational recoil \cite{Graham:2020gwr}. General
relativity predicts  that a system is kicked after merger due to the linear
momentum carried away by GW
\cite{1973ApJ...183..657B,Peres:1962zz,1961RSPSA.265..109B}. By applying
various methods, including NR, post-Newtonian (PN) theory
\cite{PhysRevD.46.1517,Kidder:1995zr,Blanchet:2005rj}, EOB
\cite{Schnittman:2007sn}, and the close-limit approximation
\cite{Sopuerta:2006wj}, several studies showed that the kick velocity is a
result of the asymmetry between different GW modes \cite{Brugmann:2007zj}, or
alternatively, the beating between the mass and current quadrupole waves
\cite{Schnittman:2007ij,RevModPhys.52.299}, caused by the unequal mass
\cite{Gonzalez:2006md,Herrmann:2006ks,Baker:2006vn,Lousto:2007db,Baker:2007gi}
and spins
\cite{Koppitz:2007ev,Gonzalez:2007hi,Campanelli:2007ew,Campanelli:2007cga,Lousto:2007db,Herrmann:2007ac,Baker:2007gi,Tichy:2007hk,Herrmann:2007ex}.
In particular, the superkick (SK) \cite{Campanelli:2007cga,Gonzalez:2007hi,Campanelli:2006uy,Lousto:2011kp,Lousto:2019lyf}
configurations lead to relatively large kick velocities. In our paper, we adopt two types of SK configurations: SKu and SKd, whose sketches are shown in Fig.~\ref{fig:configuration}. Both systems have
equal mass, spin magnitude, and tilt angles. As for SKd, two individual spins are
anti-parallel,
whereas for SKu, only the spin components in the orbital plane are opposite. SKu and SKd are fully characterized by $(\chi_{\rm init},\phi_{\rm init},\theta_{\rm
init})$, where the subscript refers to a reference time in the inspiral regime:
$\chi_{\rm init}$ is the magnitude of the dimensionless spin; $\theta_{\rm
init}$ is the polar angle of one of the holes (relative to the orbital angular
momentum $\vec{L}$); $\phi_{\rm init}$ is the azimuthal angle
between the in-(orbital)plane spin and the separation vector pointing from the lighter to the heavier BH.  

During the evolution, the effect of frame-dragging from two anti-parallel
in-plane spins moves the center of mass up and down in the inertial frame
\cite{Pretorius:2007nq}. This process is halted as the common horizon forms
\cite{Keppel:2009tc,Lovelace:2009dg,Gralla:2013ela}, and the
kick is imparted. In addition, the SKu system usually emits more energy and
linear angular momentum than SKd because of the orbital hang-up effect, which arises due to the need to radiate way additional angular momentum before the binary can merge~\cite{Campanelli:2006uy}, and leads to a larger kick. Recently simulations
showed that the kick for the SKu system could be as large as 5000 km/s (if extrapolated to the maximal spin) \cite{Lousto:2011kp,Lousto:2019lyf}. Such a large kick will
lead to important astrophysical consequences
\cite{Merritt:2004xa,Bonning:2007vt,Volonteri:2007et,Komossa:2012cy}, as well
as  Doppler shifts in GWs \cite{Chamberlain:2018snj}, which could be detected
by current and future detectors \cite{Gerosa:2016vip,Varma:2020nbm}. Numerous
studies have been implemented to fit kick velocities to progenitor's spins and
mass ratio
\cite{Campanelli:2007cga,Campanelli:2007ew,Healy:2016lce,Healy:2018swt,Rezzolla:2007xa,Kesden:2010ji,Healy:2014yta}.
In particular, the development of numerical relativity surrogate model
\cite{Blackman:2015pia,Blackman:2017dfb,Blackman:2017pcm,Varma:2018aht,Varma:2018mmi,Varma:2019csw}
has allowed a systematic study to explore much larger parameter space
\cite{Gerosa:2018qay}.

\begin{table}
    \centering
    \caption{A summary of 12 of our NR simulations with \texttt{SpEC}. All systems are in the SKd configuration, with the individual dimensionless spin $\chi_{\rm init}=0.4$, $\theta_{\rm init}=\pi/2$, and $\phi_{\rm init}\in[-\pi,\pi]$. The reference (initial) orbital frequency is chosen to be 0.0175 (in the unit of total mass). The first and second columns are the name of runs used in this paper, while the third column corresponds to the name in the Simulating eXtreme Spacetimes Collaboration catalog. The fourth column gives $\phi_{\rm init}$. The last three columns correspond to the mass, kick velocity, and spin of the final BH. A summary of SKu configurations is in Table \ref{table:NR-pars-all}.}
    \begin{tabular}{c c c c c c c} \hline\hline
\multicolumn{3}{c}{Run label} & \multirow{2}{*}{$\phi_{\rm init}$ (rad)} & \multirow{2}{*}{$m_f/M$} & $v_f^{z}$ & \multirow{2}{*}{$\chi_f$}   \\
 & This paper & SXS:BBH & & & $(\times10^{-3})$ \\ \hline
\multirow{15}{*}{SKd4} & `01' & 2451 & 2.25 &0.952  &2.36   &0.686  \\ \cline{2-7}
& `02' & 2452 & -3.04  &0.951  &$-4.73$  & 0.684 \\ \cline{2-7}
& `03' & 2453 & -1.70  &0.951  &1.71 &0.685 \\ \cline{2-7}
& `04' & 2454 & 0.66   &0.951  &$-4.45$ &0.683  \\ \cline{2-7}
& `05' & 2455 & 1.30  &0.951  &$-1.68$ & 0.685 \\ \cline{2-7}
& `06' & 2456 & 2.88  &0.951 & 4.75& 0.684 \\ \cline{2-7}
& `07' & 2457 &-2.58  & 0.951 & 4.47&0.683  \\ \cline{2-7}
& `08' & 2458 &-1.07  & 0.952 & $-2.11$&0.686  \\ \cline{2-7}
& `09' & 2459 &-2.93  & 0.951 & 4.94&0.683  \\ \cline{2-7}
& `10' & 2460 &-1.78  & 0.951 & 1.24&0.686  \\ \cline{2-7}
& `11' & 2461 &-1.36  & 0.952 & $-1.41$&0.686  \\ \cline{2-7}
& `12' & 2462 &0.21 & 0.951 & $-4.93$&0.683  \\ 
 \hline\hline
     \end{tabular}
     \label{table:NR-pars}
\end{table}

Interestingly, GW190521 was found to be consistent with a large in-plane spin
configuration. Its kick posterior is much broader and is consistent with $0-3500$ km/s  \cite{Abbott:2020mjq}. 
Meanwhile, its potential optical counterpart was predicted to have a kick velocity of $\sim 200~{\rm km~s^{-1}}$ \cite{Graham:2020gwr}. In the future, it is still likely to detect GW events with non-negligible gravitational recoils, and even SK-like binaries \cite{Yu:2020iqj}. Accordingly, in this paper, we aim to explore the features
of ringdown for SKd binaries carefully and relate them to the phenomenon of
gravitational recoil.  Specifically, we shall focus on the amplitudes of QNMs
\cite{Hughes:2019zmt,Apte:2019txp,Lim:2019xrb}, as well as mass and current
quadrupole waves \cite{Schnittman:2007ij}, and study how those features depend
on the progenitor's parameters. Comparing to a generic BBH
system, a SKd system has several advantages that can ease the difficulty of
analysis. (i) The parameter space for a SKd binary is 3D, i.e., $(\chi_{\rm
init},\phi_{\rm init},\theta_{\rm init})$, instead of generally 7D. (ii)  SKd configurations have a
high level of symmetry. Subsequently, the orbital angular momentum is
non-precessing, and the spin direction of the remnant BH is fixed
during the merger. This allows us to conveniently choose coordinates in which
only the $(2,2)$ and $(2,-2)$ modes dominate. (iii) The
mass and spin of the remnant BH are not impacted by varying $(\chi_{\rm
init},\phi_{\rm init},\theta_{\rm init})$, nor are the QNM
frequencies. Hence we can study the mode excitation (complex) amplitudes exclusively while avoiding
changes in the mode frequencies. 

In our study, we use waveforms generated by the Spectral Einstein Code
(\texttt{SpEC}) \cite{spec}, and two NR surrogate models, also based on
\texttt{SpEC}: NRSur7dq4, NRSur7dq4Remnant \cite{Varma:2018aht,Varma:2019csw}.
In particular, NRSur7dq4 is a waveform model  valid for mass ratio $<4$ and
dimensionless spin magnitudes $<0.8$, while NRSur7dq4Remnant is a model that
predicts the  mass, spin and kick velocity of the remnant BH from the parameter
of individual merging BHs. 
Meanwhile, we have in total 35 NR simulations where systems are either in the
SKd (Table \ref{table:NR-pars}) or the SKu (Table \ref{table:NR-pars-all})
configuration. The dimensionless spin of BH ranges from 0.4 to 0.95. Those
runs will be
available in the Simulating eXtreme Spacetimes (SXS) Collaboration
catalog
\cite{Boyle:2019kee,Mroue:2013xna}. We have checked that our NR runs agree with the predictions of NRSur7dq4, with mismatches $\sim 10^{-5}-10^{-4}$. For each simulation, we evolve with
three numerical resolutions. Among those cases, the largest kick is $\sim
4050~{\rm km~s^{-1}}$ (Table \ref{table:NR-pars-all}).

\begin{figure*}[htb]
    \centering
    \subfloat[mass quadrupole wave]{\includegraphics[width=\columnwidth,height=10cm]{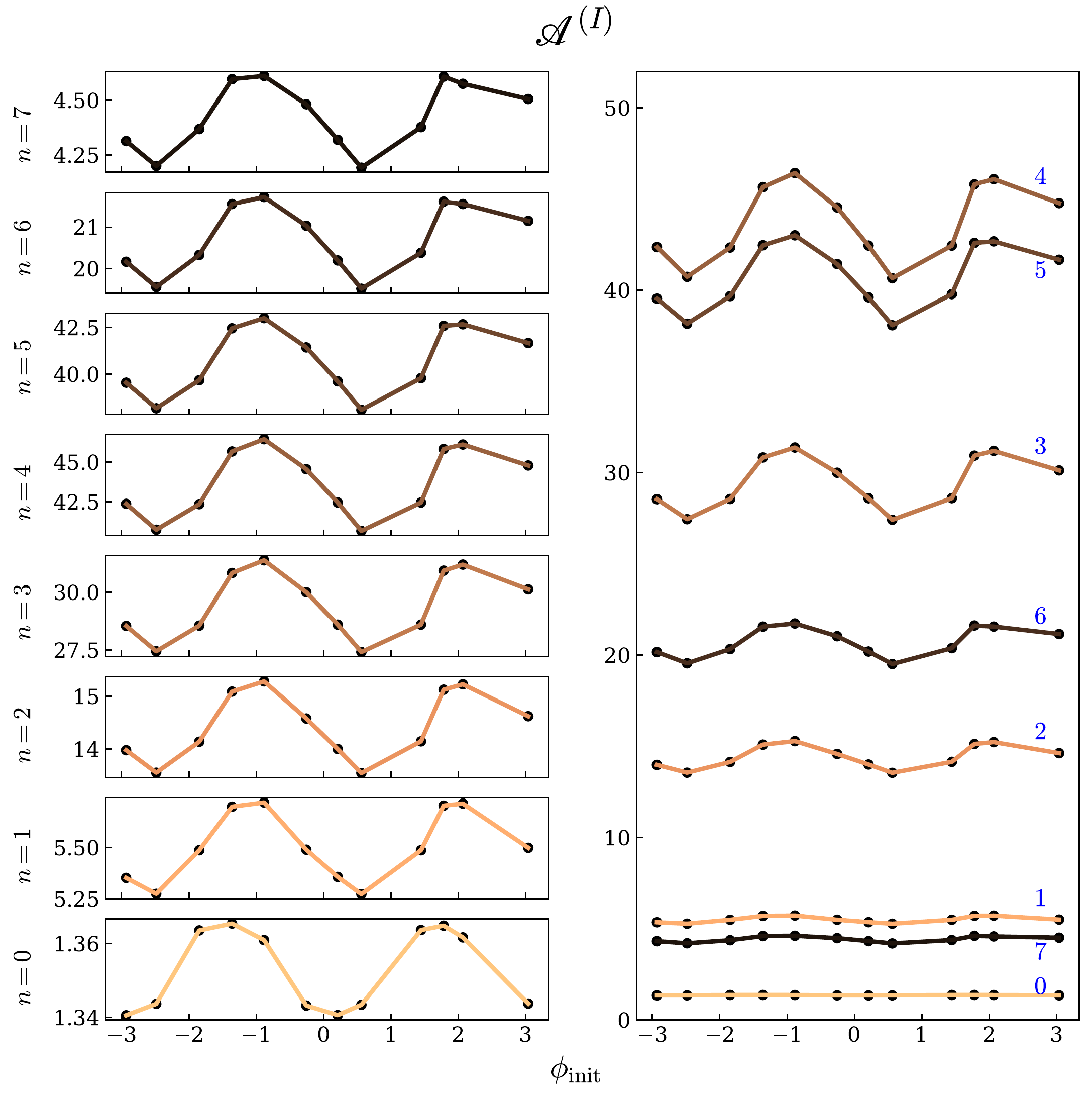}}
    \subfloat[current quadrupole wave]{\includegraphics[width=\columnwidth,height=10cm]{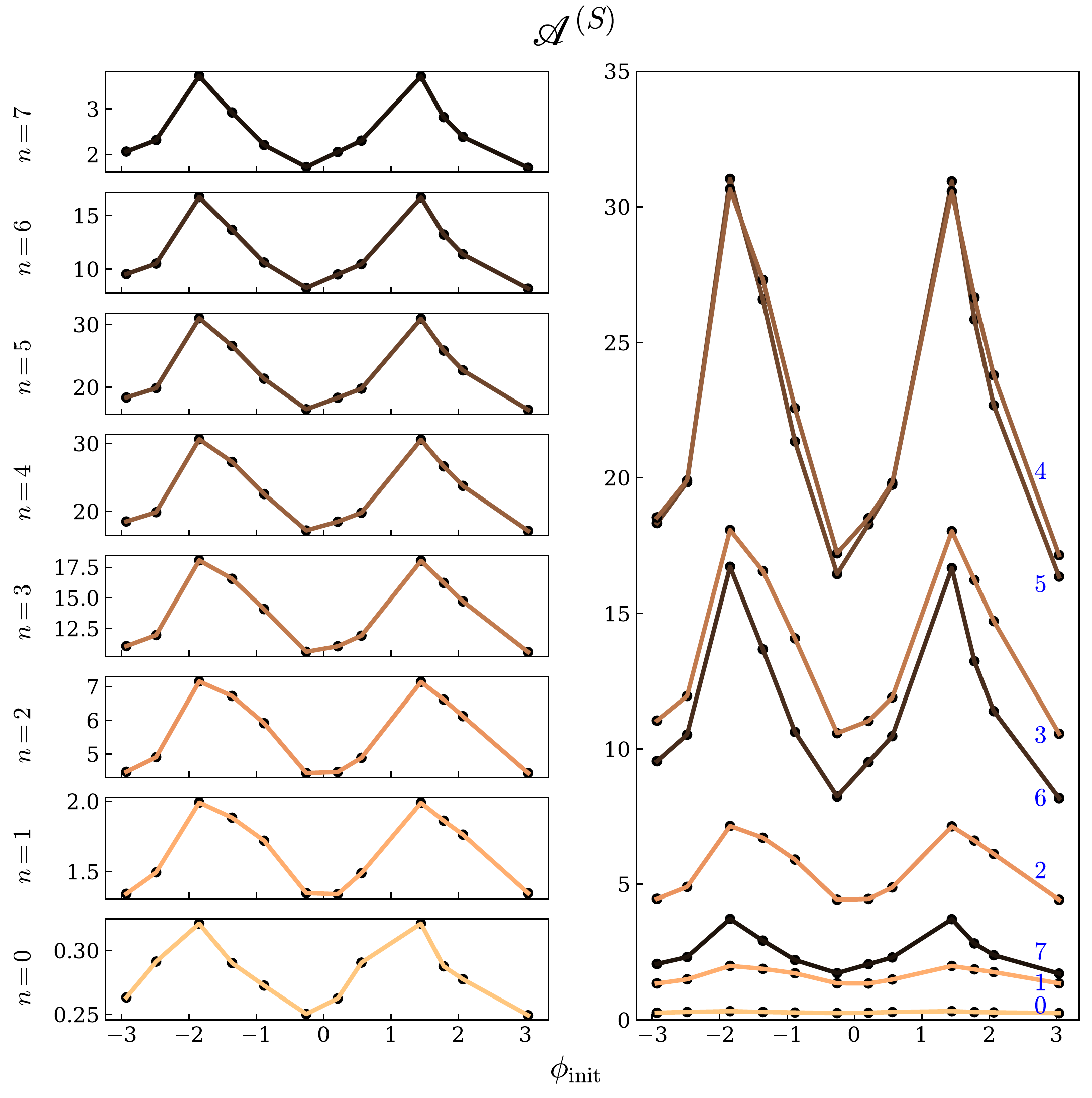}}
  \caption{QNM magnitudes versus $\phi_{\rm init}$ for mass
  ($\mathscr{A}^{(I)}$) and current ($\mathscr{A}^{(S)}$) quadrupole waves.
  Data are from 12 of our NR simulations listed in Table \ref{table:NR-pars}.
  All BBH systems are in the SKd configuration. Fig.~\ref{fig:overtone-phiinit-nr} (a) corresponds
  to $\mathscr{A}^{(I)}$, where the left eight panels are the zoom-in plot for
  each overtone. The overtone index $n$ is in descending order. Similarly, Fig.~\ref{fig:overtone-phiinit-nr} (b) corresponds to $\mathscr{A}^{(S)}$. The spectra peak at
  $n=4$ (because the n=4 amplitude is largest), and patterns are roughly periodic with a period $2\pi$. Examining the zoomed in plots, it can be seen that approximately,
  the patterns are the same for all $n$ (up to a scaling factor).
  }
 \label{fig:overtone-phiinit-nr}
\end{figure*}

This paper is organized as follows. In Sec.~\ref{sec:overtone}, we decompose
ringdown into QNMs (7 overtones) and explore the dependence of mode amplitudes
on the progenitor's parameters. In Sec.~\ref{sec:multipolar}, we study the
phenomenon of radiative mass and current quadrupole waves and relate them to
kick velocity. Then in Sec.~\ref{sec:BOB}, we apply the backward-one-body (BOB)
model, conceived recently by McWilliams \cite{McWilliams:2018ztb}, to SK binaries.
Sec.~\ref{sec:fisher} focuses on parameter estimation, where we use the Fisher
information matrix formalism to discuss the parameter correlations in the
ringdown signal. Finally, in Sec.~\ref{sec:conclusion} we summarize our
results.

Throughout this paper we use the geometric units with $G = c = 1$. We use $M$
to refer to the initial total mass of the binary system. All GW waveforms are
aligned in the time domain such that $t=0M$ corresponds to the time of the peak
of the total amplitude $\sqrt{\sum_{l,m}|h_{lm}|^2}$.


\begin{figure*}[htb]
        \includegraphics[height=6cm,clip=true]{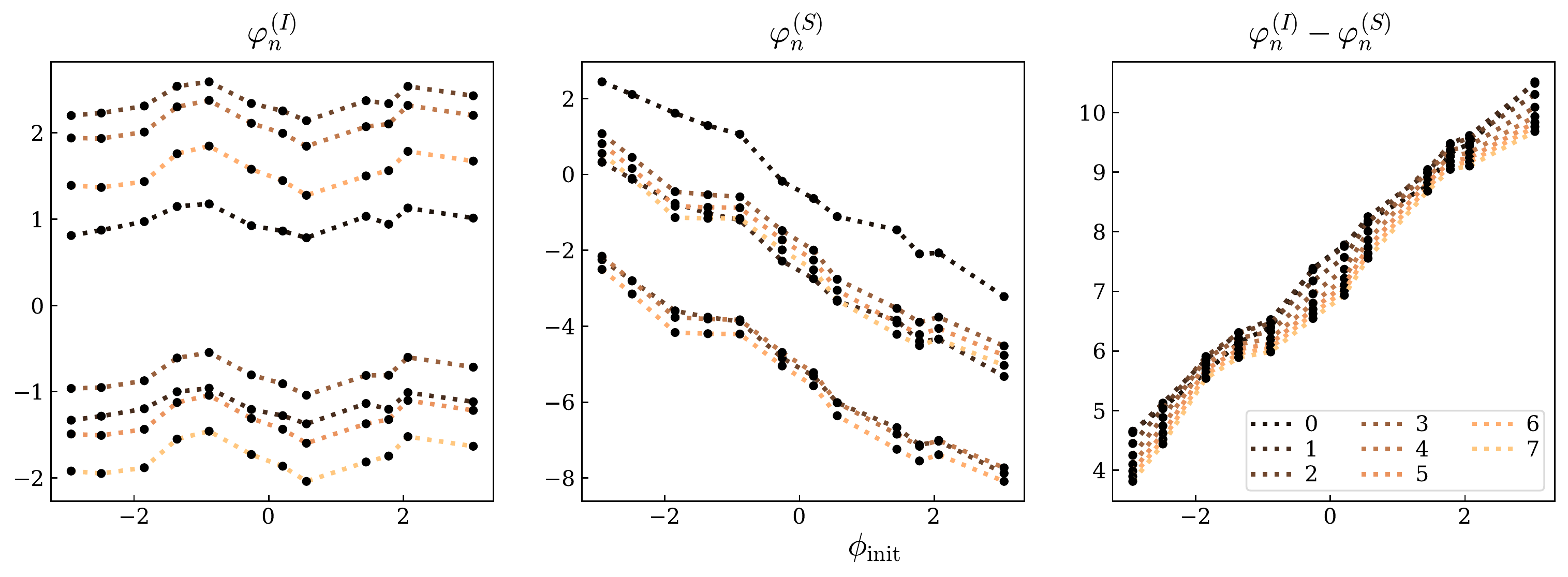} 
        \caption{The dependence of $\varphi_n^{(I)}$, $\varphi_n^{(S)}$, as well as their difference, on $\phi_{\rm init}$. It turns out that $\varphi_n^{(I)}$ is roughly 
        insensitive to $\phi_{\rm init}$, whereas $\varphi_n^{(S)}$ is approximately linear in $\phi_{\rm init}$.}
 \label{fig:overtone-phiinit-nr-phase}
\end{figure*}

\section{Multipole decomposition of the waveform and quasi-normal mode excitations}
\label{sec:overtone}

In this section, we decompose the ringdown signal into QNMs and study how each mode is excited. 

\subsection{Multipole decomposition of the waveform}
\label{sec:overtone_phi_deps}

In a spherical polar coordinate system, with an observer located at the 
$(\iota,\beta)$ direction, following the widely used convention for defining the $+$ and $\times$ polarizations of the gravitational wave \cite{misner1973gravitation}, one can define a complex strain 
\begin{align}
    \h(t,\iota,\beta)&=\h_+(t,\iota,\beta)-i\h_\times(t,\iota,\beta)
\end{align}
and further decompose it into a sum over a set of spin-weighted spherical harmonics $\sy$:
\begin{align}
\h(t,\iota,\beta)&=\h_+(t,\iota,\beta)-i\h_\times(t,\iota,\beta) \notag \\
&=\sum_{\ell=2}^\infty \sum_{m=-\ell}^{\ell}\frac{1}{D}\hlm(t) \sy, \label{spherical-harm}
\end{align}
where $D$ is the distance between the
source and the observer.
Meanwhile, it is also natural to group $h_{\ell,m}$ and $h_{\ell,-m}$ into mass and current quadrupole waves~\cite{RevModPhys.52.299}, writing
\begin{subequations}
\begin{align}
&I_\lm=\frac{1}{\sqrt{2}}[\hlm+(-1)^m\h^*_{\ell,-m}], \\
&S_\lm=\frac{i}{\sqrt{2}}[\hlm-(-1)^m\h^*_{\ell,-m}].
\end{align}
\label{rad-moment}%
\end{subequations}
Here $I_\lm$ ($S_\lm$) is the mass (current) quadrupole wave,
proportional to the $\ell$-th order time derivative of the mass (current) $\ell$-pole moment. 
For the SKd configuration, $h_{2,\pm2}$ always dominate over other modes, hence we shall primarily focus on these two modes.

\subsection{QNM excitation in multipolar modes}
As discussed in Ref.~\cite{Giesler:2019uxc}, the ringdown portion of $h_{2,\pm2}$ of a non-precessing system can be modeled as a sum of QNMs, as early as $t=0M$, which is defined as the moment of time at which $\sqrt{\sum_{l,m}|h_{lm}|^2}$ peaks.  The expansion reads: 
\begin{equation}
\begin{aligned}
&\h_{22}=\sum_{n=0}^N\mathcal{A}_{22n}e^{i\psi_{22n}} e^{-i\omega_{22 n}t} \\
&\h_{2,-2}=\sum_{n=0}^N\mathcal{A}_{2,-2n}e^{i\psi_{2,-2n}} e^{i\omega^*_{22 n}t}
\end{aligned}
\quad\quad  t\geq 0M, \label{qnm-overtone}
\end{equation}
where  $\mathcal{A}_{22n}e^{i\psi_{22n}}$ and
$\mathcal{A}_{2,-2n}e^{i\psi_{2,-2n}} $ are the complex  amplitudes of the
$n$-th overtone, while $\omega_{22n}$ and $-\omega_{22n}^*$  are the mode
frequencies.  Note that $\omega_{22n}$ and $-\omega_{22n}^*$ have opposite real
parts and equal imaginary parts; both correspond to the prograde $\ell =2$
quasi-normal mode. In Eq.~(\ref{qnm-overtone}) we have adopted the
approximation that the angular wavefunction of the $(2,2)$ mode is given by the
spin-weighted spherical harmonics instead of the spin-weighted spheroidal harmonics --- the spherodial-spherical mixing
\cite{1973ApJ...185..649P,Berti:2014fga} can be ignored because of the moderate
spin of final BHs ($\sim0.68$) studied in this paper. In this way, both the
prograde, $\omega_{22}$, and the retrograde, $\omega_{2,-2}$, modes share the
same angular wavefunction. Meanwhile, the retrograde modes $\omega_{2,-2n}$ and
$-\omega_{2,-2n}^*$ [see Eq.(3.6) of Ref.~\cite{Lim:2019xrb}], are negligible
in our case.

Inserting Eqs.~\eqref{qnm-overtone} to Eqs.\ (\ref{rad-moment}) we have
\begin{equation}
\begin{aligned}
&I_{22}=\sum_{n=0}^N\mathscr{A}_{n}^{(I)}e^{i\varphi_n^{(I)}} e^{-i\omega_{22 n}t}, \\
&S_{22}=\sum_{n=0}^N\mathscr{A}_{n}^{(S)} e^{i\varphi_n^{(S)}} e^{-i\omega_{22 n}t},
\end{aligned}
\quad\quad  t\geq 0M, \label{IS-overtone}
\end{equation}
with 
\begin{subequations}
\begin{align}
&\mathscr{A}_{n}^{(I)}e^{i\varphi_n^{(I)}}=\frac{1}{\sqrt{2}}(\mathcal{A}_{22n}e^{i\psi_{22n}} +\mathcal{A}_{2,-2n}e^{-i\psi_{2,-2n}}), \\
&\mathscr{A}_{n}^{(S)} e^{i\varphi_n^{(S)}}=\frac{i}{\sqrt{2}}(\mathcal{A}_{22n}e^{i\psi_{22n}} -\mathcal{A}_{2,-2n}e^{-i\psi_{2,-2n}}). 
\end{align}
\label{IS-22-conversion}%
\end{subequations}
To give an example, we fit the ringdown portion of SKd4 set of NR simulations (Table
\ref{table:NR-pars}) with 7 overtones, following the procedure of
Ref.~\cite{Giesler:2019uxc}. We use unweighted linear least squares to fit the
mode amplitudes and use nonlinear least squares to fit the final spin and mass.
The mode frequency $\omega_{22n}$ is obtained from a Python package
$\textsf{qnm}$ \cite{Stein:2019mop}.

First focusing on $I_{22}$ and $S_{22}$, we plot $\mathscr{A}_n^{(I)}$ (mass)
and $\mathscr{A}^{(S)}_n$ (current) as functions of $\phi_{\rm init}$ in
Fig.~\ref{fig:overtone-phiinit-nr}. We can see
$\mathscr{A}_n^{(I)}>\mathscr{A}^{(S)}_n$ for any $n$, and both of them peak at
$n=4$. Patterns have a rough period $\pi$. An interesting feature is that the
dependence on $\phi_{\rm init}$ is similar for all overtones (up to a scaling
factor). The analogous universal feature for EMRI was explored by Lim
\etal\cite{Lim:2019xrb}. After a proper normalization [see their Eq.~(5.1)],
the angular dependence of mode amplitudes is insensitive to the mode indices
[see their Fig.~12]. Similarly, for the phase of mode amplitude $\varphi_n^{(I)}$ and $\varphi_n^{(S)}$, as shown in Fig.~\ref{fig:overtone-phiinit-nr-phase}, their dependence on $\phi_{\rm init}$ is also insensitive to the overtone index $n$. 

The features of $\mathscr{A}_n$ and $\varphi_n$ allow us to conclude that the dependence of QNM amplitudes for $I_{22}$ and $S_{22}$ on $\phi_{\rm init}$ can be factored out from the temporal sector, i.e.,
\begin{subequations}
\begin{align}
   & I_{22}(\phi_{\rm init}, t) \sim I_{22}^m(\phi_{\rm init}) T_I(t), \\
   & S_{22}(\phi_{\rm init}, t) \sim \spp(\phi_{\rm init})e^{-i \phi_{\rm init}} T_S (t),
\end{align}
\label{IS-universal-overtone}%
\end{subequations}
where $T_I(t)$ and $T_S(t)$ are two complex functions, corresponding to the temporal evolution of the mass and current quadrupole waves, respectively. Since $T_I(t)$ and $T_S(t)$ do not depend on $\phi_{\rm init}$, they represent the common features of all SKd binaries. We will explore the features of $T_I(t)$ and $T_S(t)$ in Sec.~\ref{sec:BOB}.

On the other hand, the progenitor configuration, at least $\phi_{\rm init}$, is encoded mainly in two functions $I_{22}^m(\phi_{\rm init})$ and $\spp(\phi_{\rm init})$. Figure~\ref{fig:overtone-phiinit-nr-phase} exhibits that to the leading order, $\varphi_n^{(I)}$ is insensitive to $\phi_{\rm init}$,
while $\varphi_n^{(S)}\propto-\phi_{\rm init}$. As a result, $I_{22}^m(\phi_{\rm init})$ and $S_{22}^m(\phi_{\rm init})$ can be regarded approximately as two real functions. Thus the phase
difference between $I_{22}$ and $S_{22}$, $\dphi$, is roughly linear in $\phi_{\rm
init}$.  We will explore Eq.~(\ref{IS-universal-overtone}) more carefully later
in Sec.~\ref{sec:multipolar}, as well as extending to the full $(\chi_{\rm
init}, \theta_{\rm init}, \phi_{\rm init})$ parameter space.


\begin{figure}[htb]
         \includegraphics[width=0.95\columnwidth,height=6.4cm,clip=true]{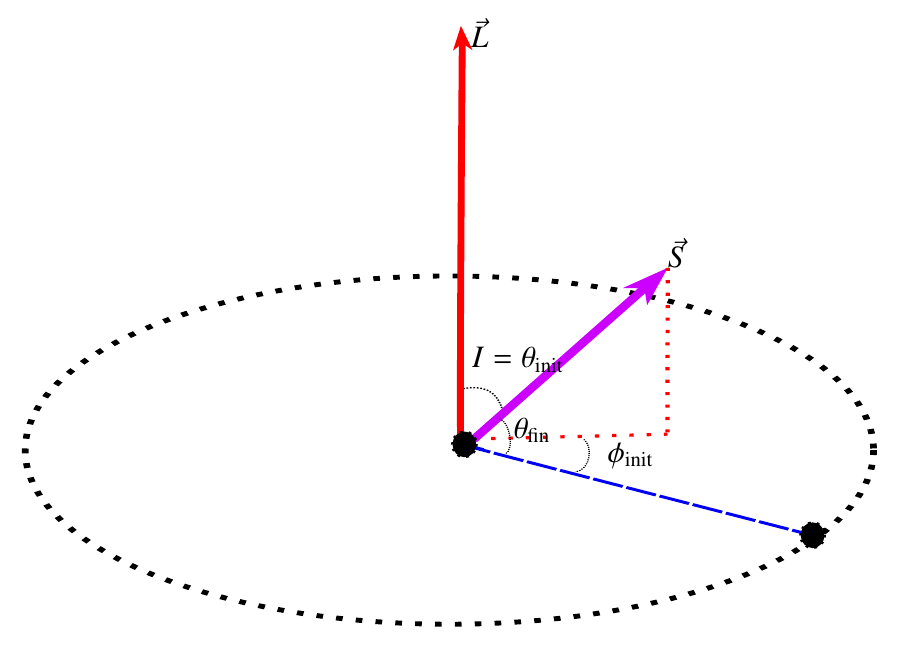}
  \caption{An illustration for the EMRI-parameterization $(I,\theta_{\rm fin})$ and the SKd-parameterization $(\theta_{\rm init},\phi_{\rm init})$. The origin is chosen to be one of the BHs. Following the discussion of Hughes \etal\cite{Hughes:2019zmt,Lim:2019xrb,Apte:2019txp}, $I$ is defined to be the angle between $\bm{L}$ (the red arrow) and $\bm{S}$ (the purple arrow), while $\theta_{\rm fin}$ is the angle between the $\bm{S}$ (the purple arrow) and the orbital separation vector (the blue dashed line). For the SKd-parameterization, $\phi_{\rm init}$ is the angle between the in-plane spin (the red dashed horizontal line) and the orbital separation vector (the blue dashed line), whereas $\theta_{\rm init}$ is the angle between $\bm{L}$ (the red arrow) and $\bm{S}$ (the purple arrow). The connection between two parameterizations is given by Eq.~(\ref{I-theta-fin}).}
 \label{fig:Hughes_EMRI}
\end{figure} 

\begin{figure*}[htb]
        \includegraphics[width=\columnwidth,height=7.8cm,clip=true]{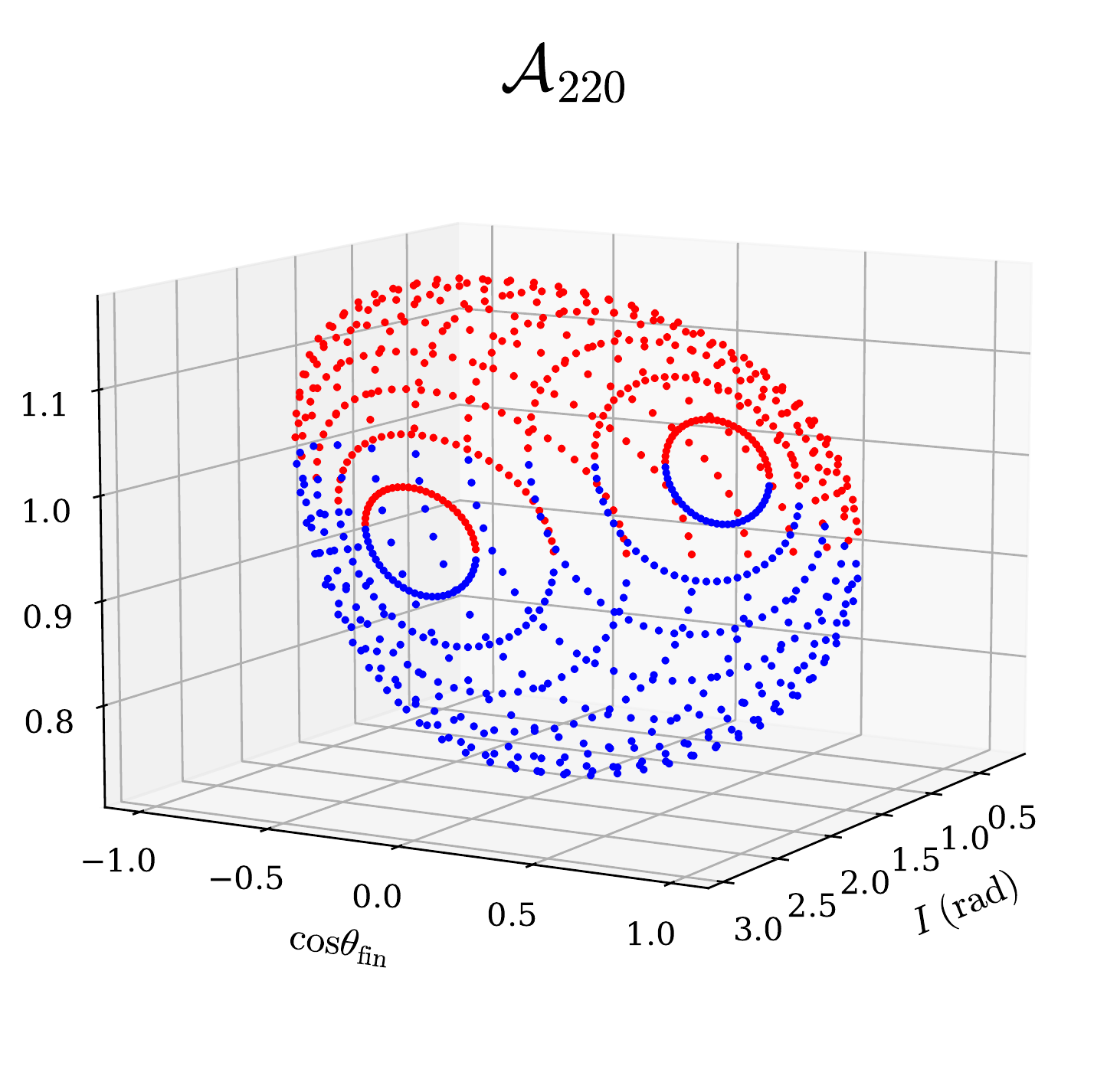} 
        \includegraphics[width=\columnwidth,height=7.8cm,clip=true]{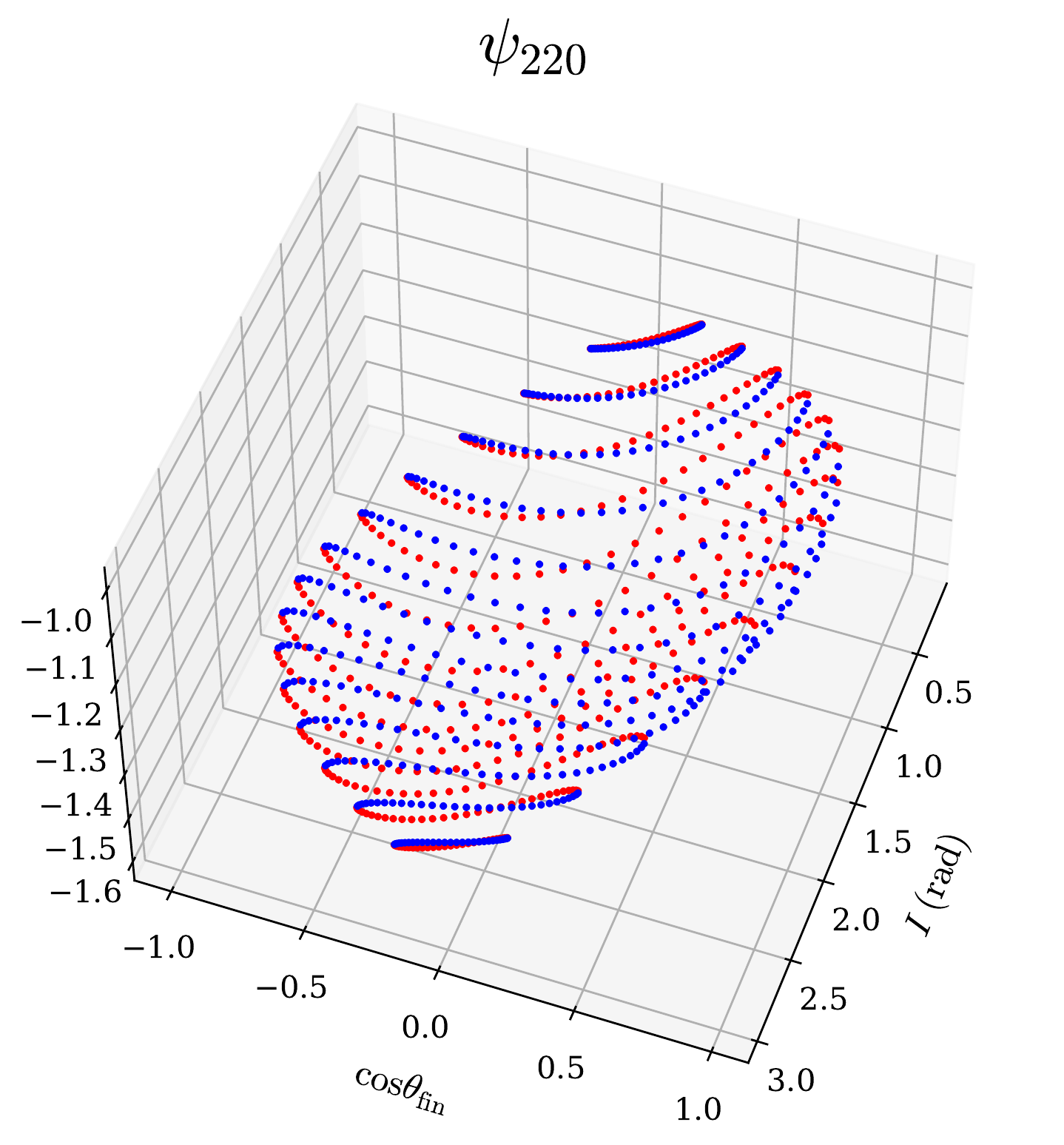}  \\    
        \includegraphics[height=7.4cm,clip=true]{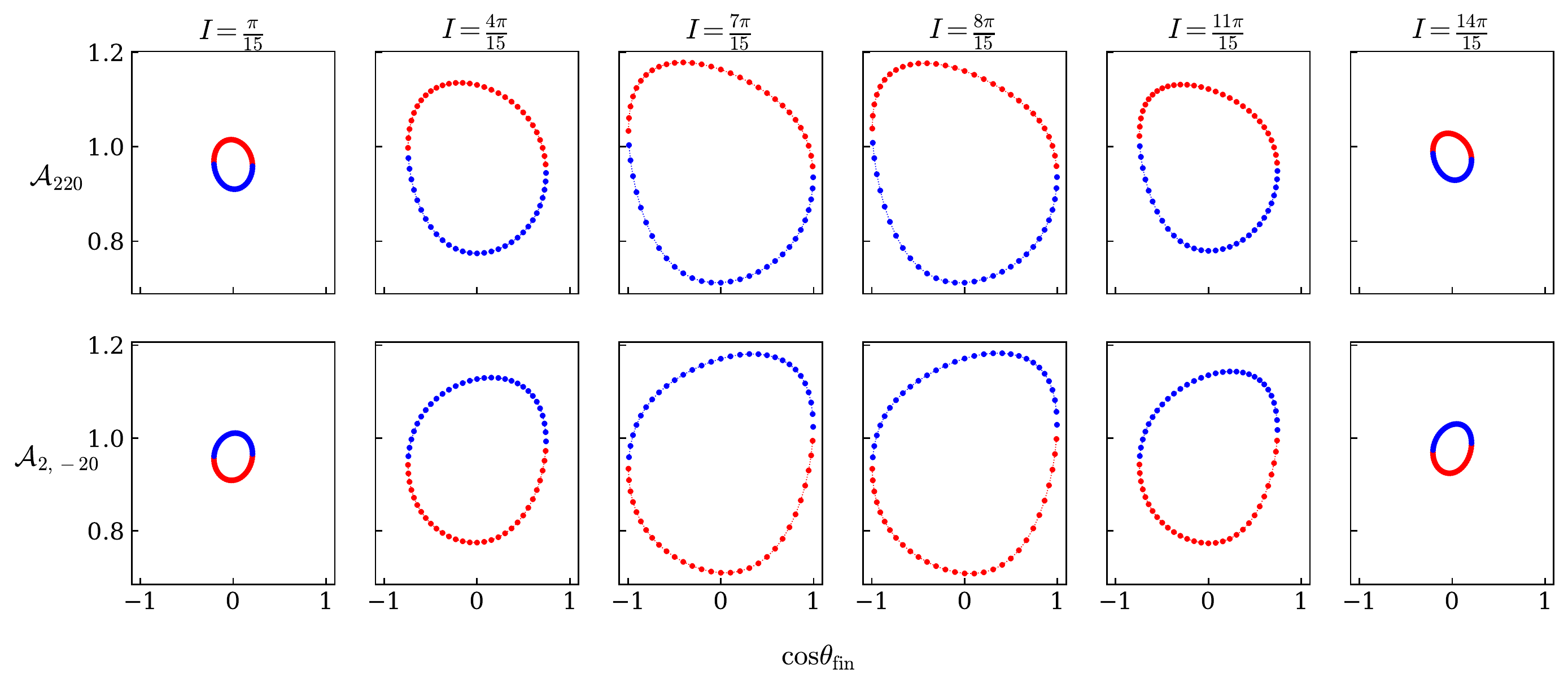} \\
        \includegraphics[height=7.4cm,clip=true]{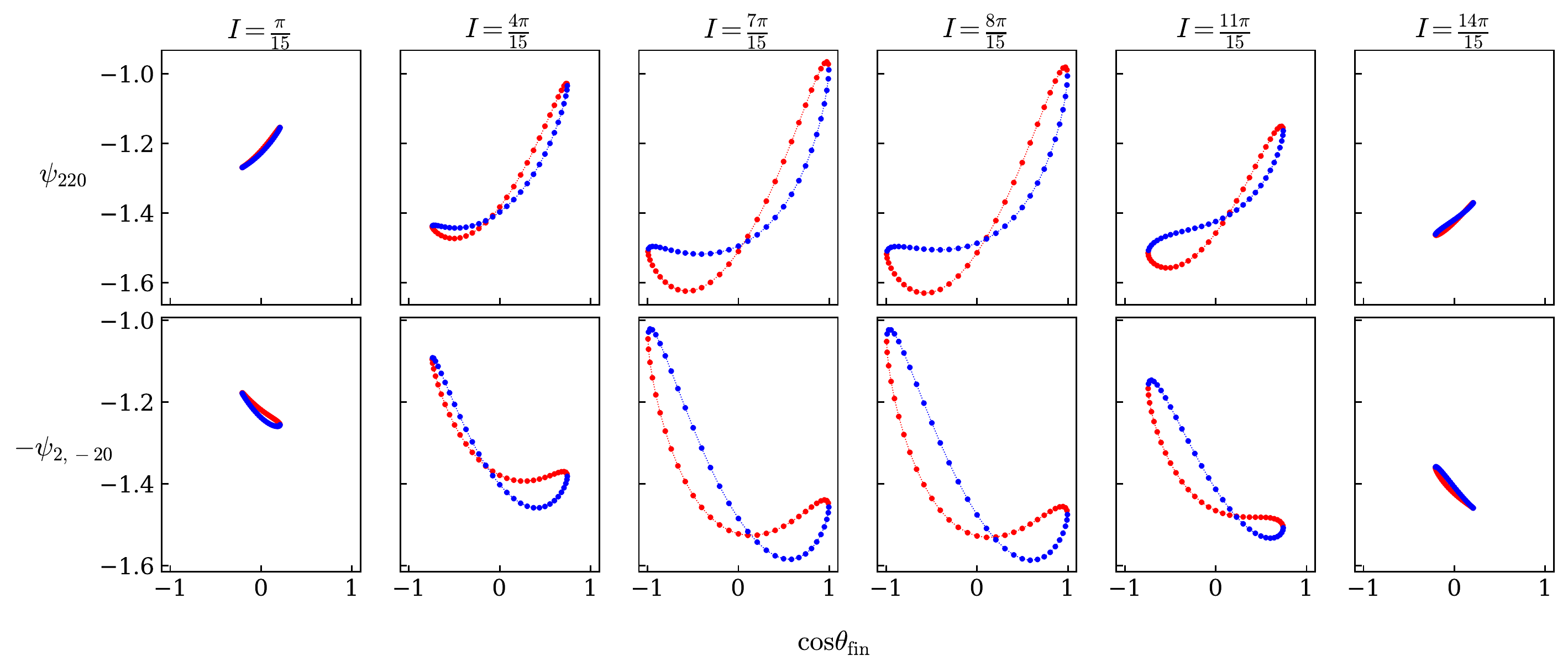} 
  \caption{The fundamental mode amplitude and phase versus initial spin configuration $(I,\cos\theta_{\rm fin})$. Those two independent variables are chosen since they coincide with the variables used in Ref.~\cite{Hughes:2019zmt} [see Eq.~(\ref{I-theta-fin})]. Data are obtained from NRSur7dq4. All BBH systems are in the SKd configuration with $\chi_{\rm init}=0.4$. Points are drawn with two colors, where blue stands for $\sin\phi_{\rm init}<0$ while red for $\sin\phi_{\rm init}>0$. The second and fourth rows are results of $h_{2,2}$ for some $I-$slices, while the third and fifth rows correspond to $h_{2,-2}$.}
 \label{fig:overtone-h-scott}
\end{figure*}

\subsection{Full $(\theta_{\rm init},\phi_{\rm init})$ dependence and correspondence with the extreme mass-ratio case}
\label{sec:theta_phi_scott}
In the case of EMRI, Hughes \etal\cite{Hughes:2019zmt,Lim:2019xrb,Apte:2019txp} investigated the ringdown spectra of $h_{2,\pm2}$ modes rather than $I_{22}$ and $S_{22}$. In order to make a connection to their studies, we now turn our attention to $h_{2,\pm2}$. 

\subsubsection{Mapping between SKd and EMRI system parameters}
\label{sec:Mapping between SK and EMRI system parameters}

\begin{figure*}[htb]
    \includegraphics[width=0.3\textwidth,height=4.6cm,clip=true]{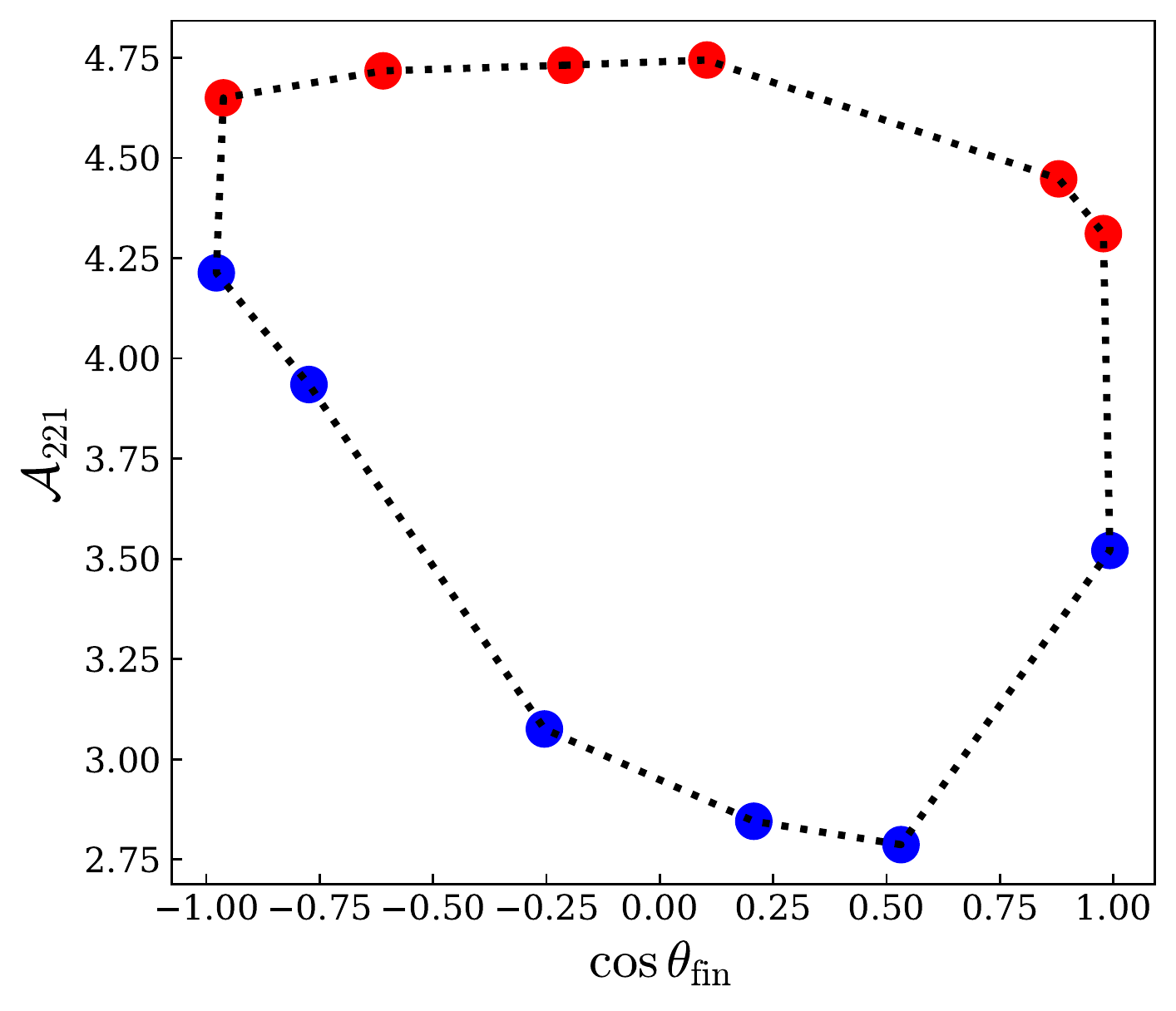}
    \includegraphics[width=0.3\textwidth,height=4.6cm,clip=true]{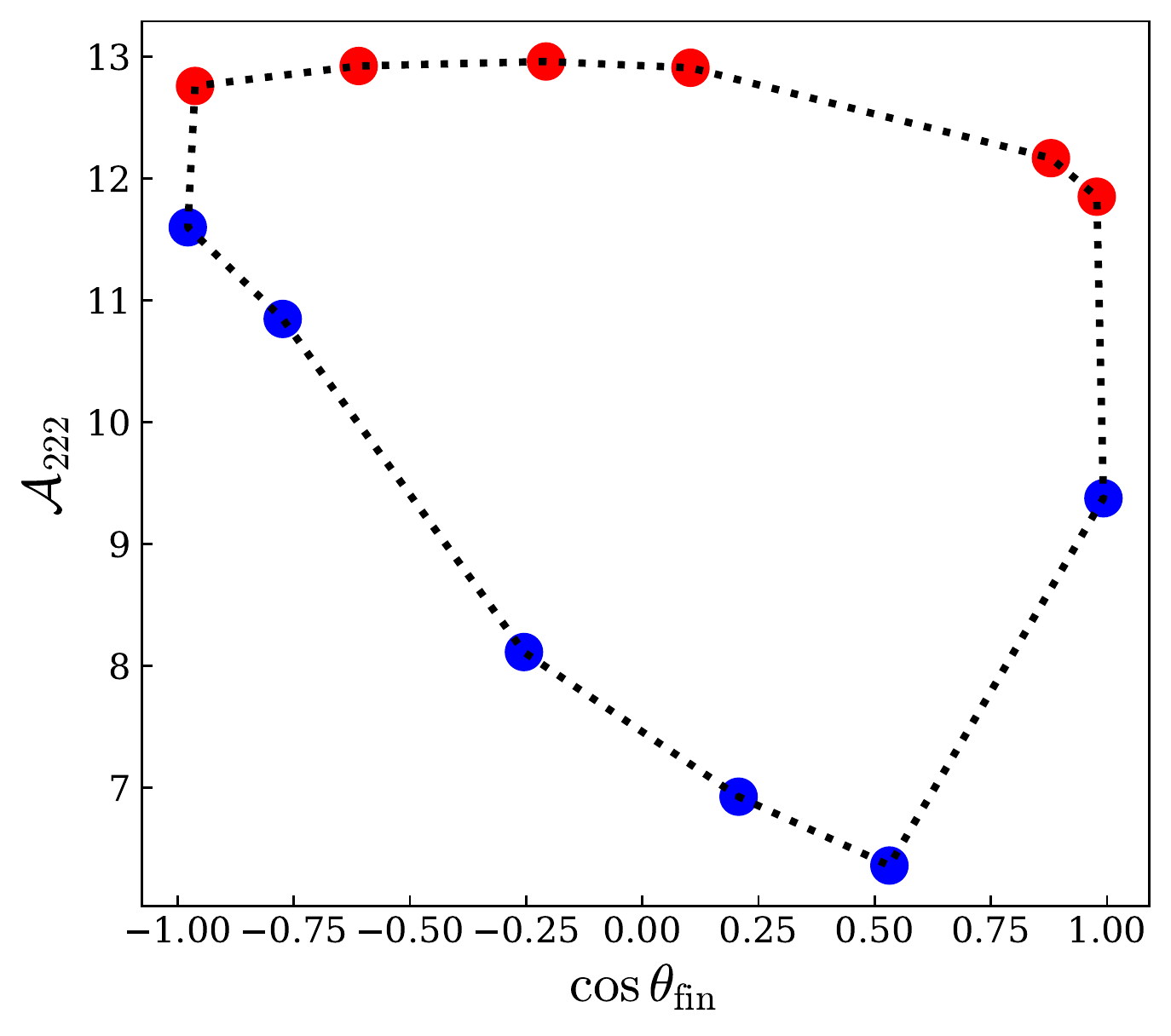}
    \includegraphics[width=0.3\textwidth,height=4.6cm,clip=true]{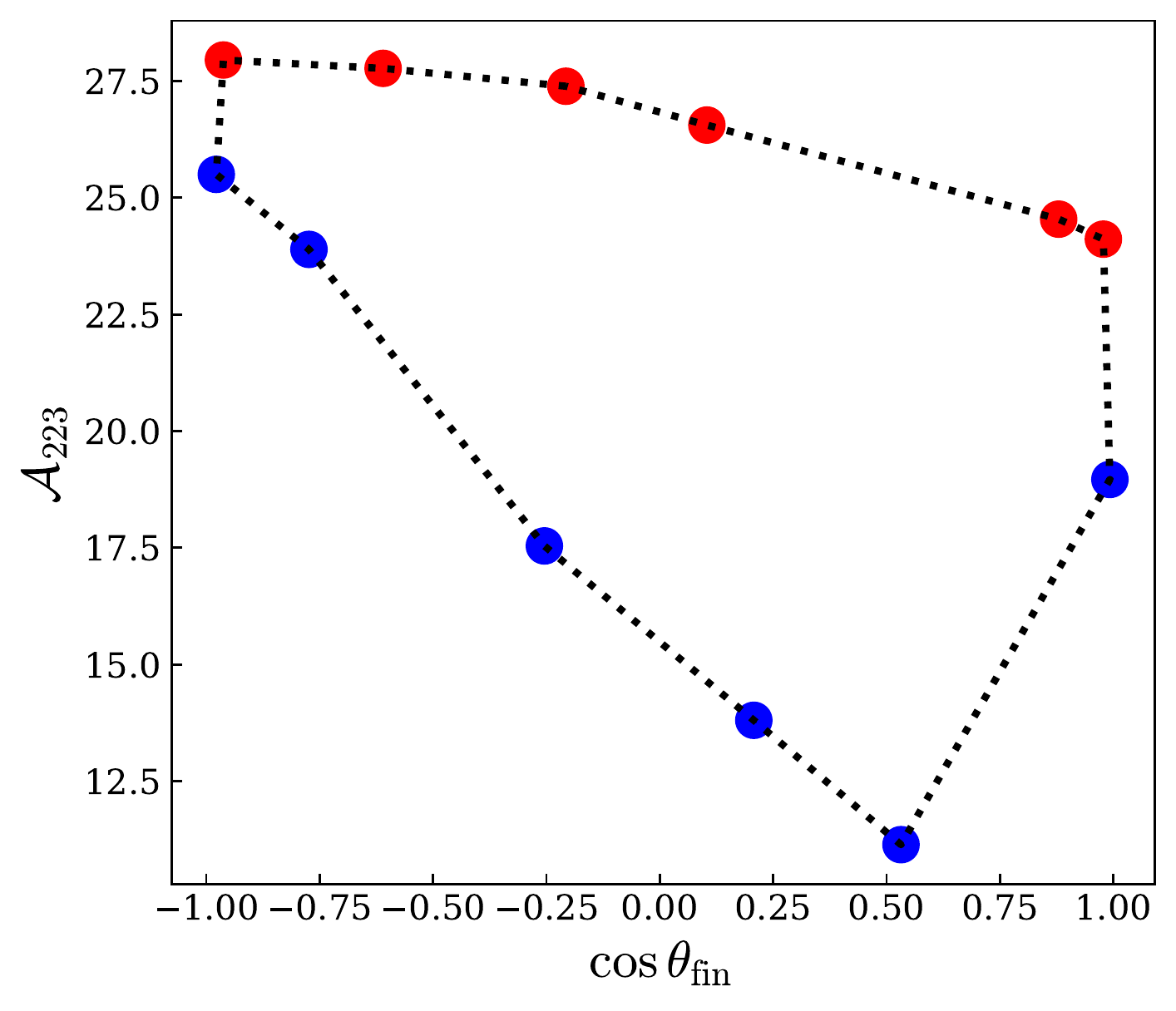} \\
    \includegraphics[width=0.3\textwidth,height=4.6cm,clip=true]{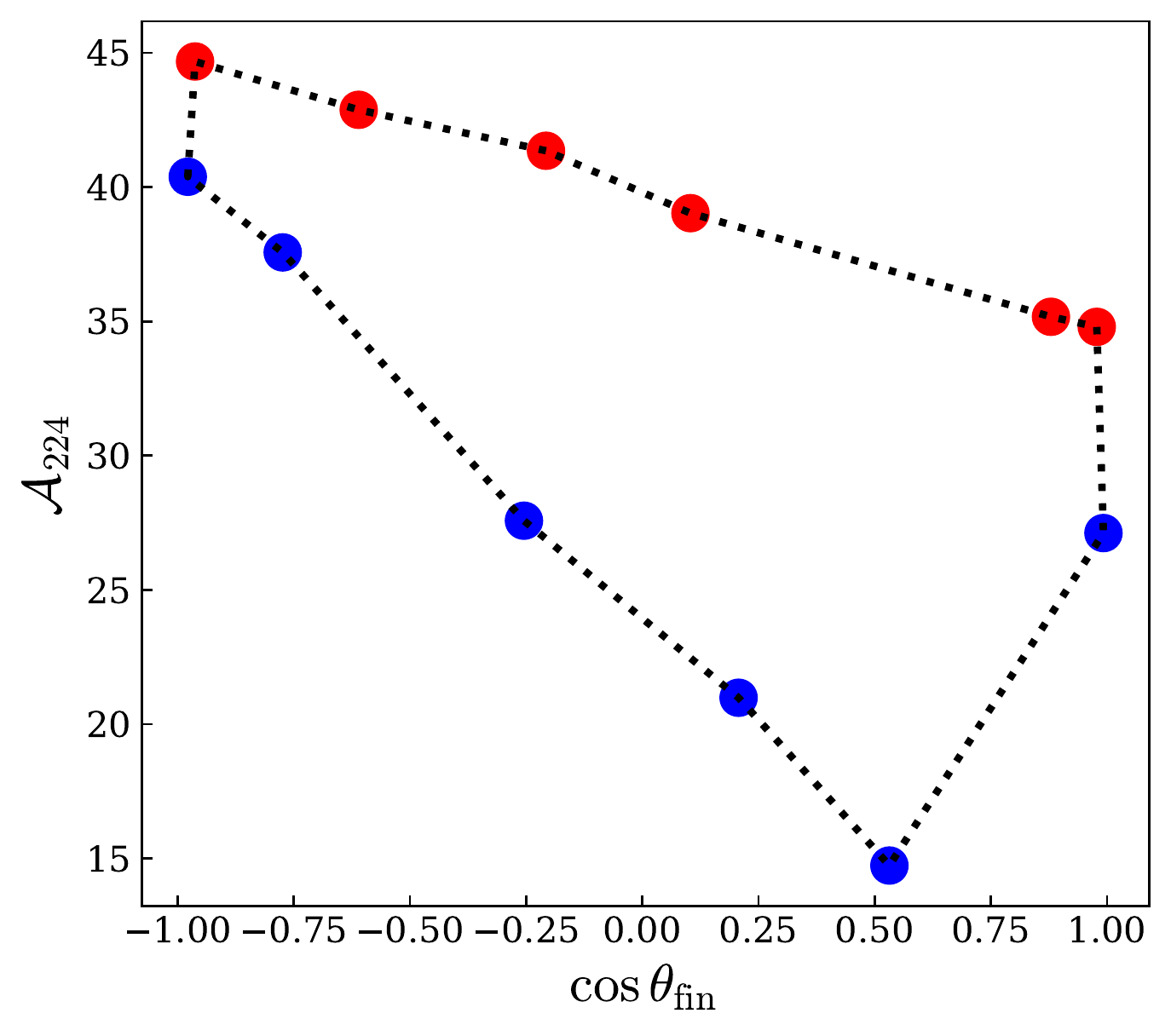}
    \includegraphics[width=0.3\textwidth,height=4.6cm,clip=true]{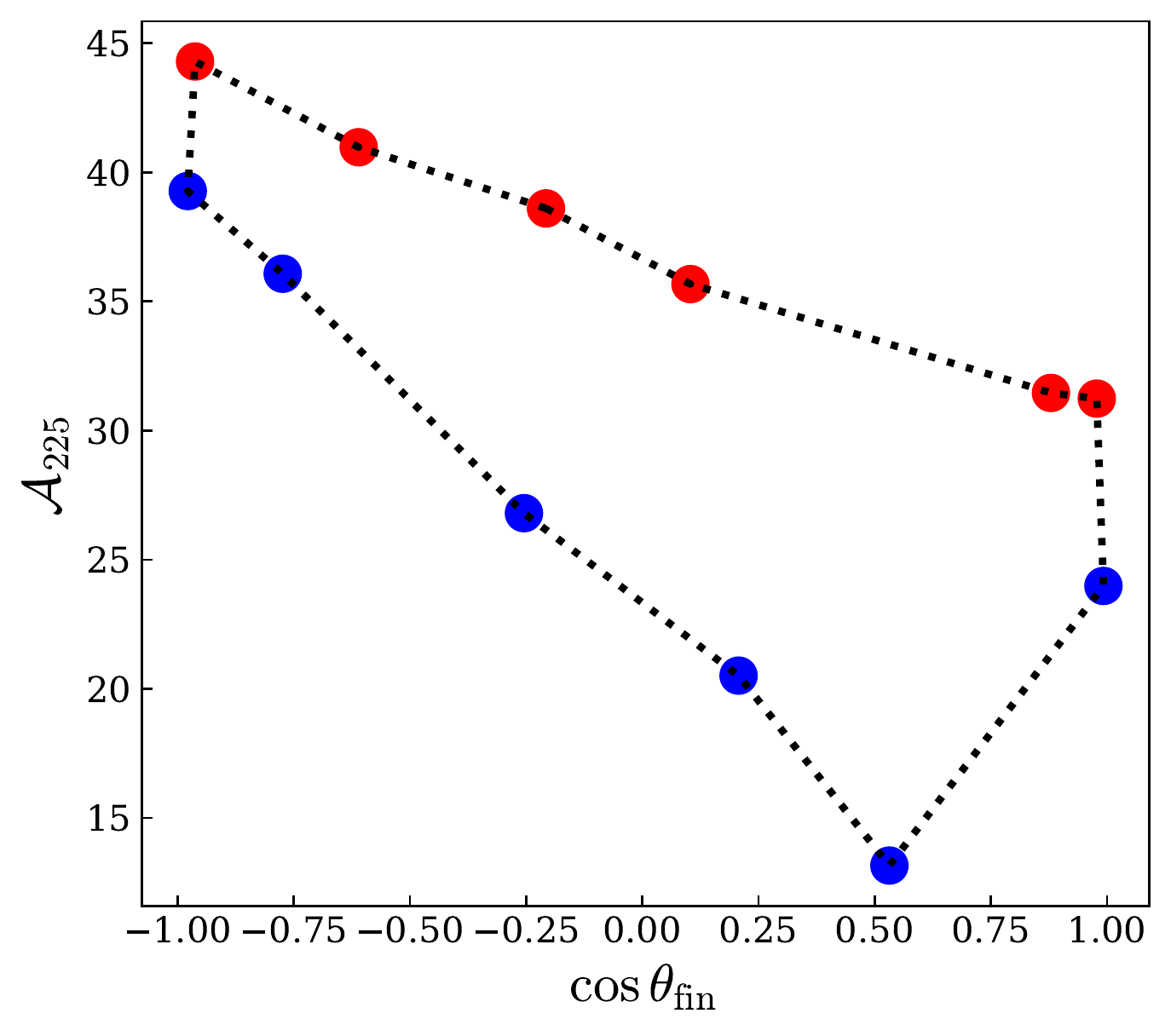}
    \includegraphics[width=0.3\textwidth,height=4.6cm,clip=true]{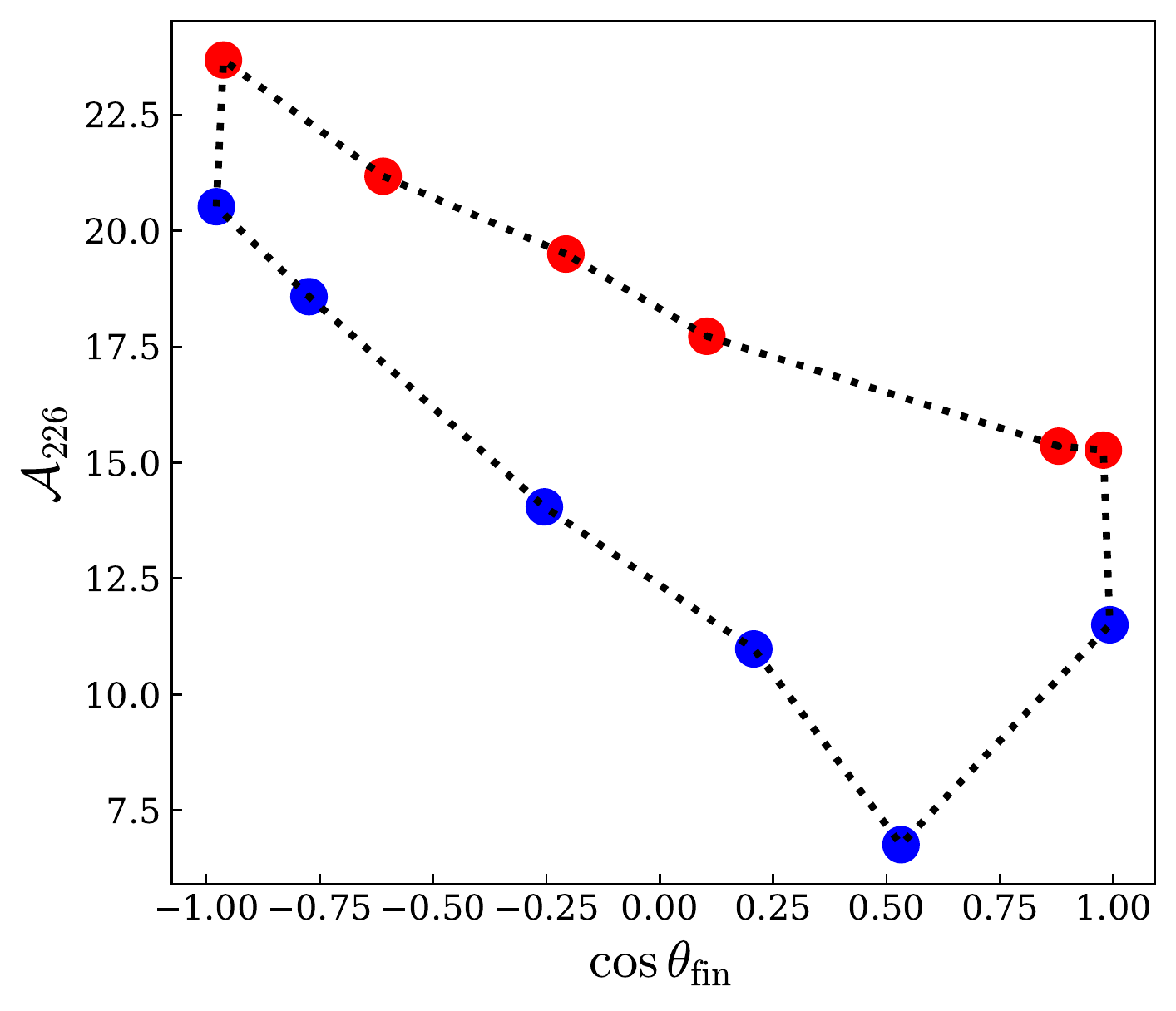}
  \caption{ The overtone mode amplitudes $\mathcal{A}_{2,+2n}~(n=1-6)$ versus $\cos\theta_{\rm fin}$, with the same convention as Fig.~\ref{fig:overtone-h-scott}. The data are from our SKd4 runs listed in Table~\ref{table:NR-pars}, which corresponds to $I=\pi/2$.
  }
 \label{fig:A_nr_hugues}
\end{figure*}

Hughes \etal\cite{Hughes:2019zmt,Lim:2019xrb,Apte:2019txp} parameterized EMRIs
with two geometric quantities $\theta_{\rm fin}$ and $I$ [see Fig.~1 of
Ref.~\cite{Hughes:2019zmt}], where $I\in[0,\pi]$ is the angle between the spin
of the primary BH and the orbital angular momentum, while $\theta_{\rm fin}$ is
the angle between the spin of primary BH and the orbital separation vector (at
the moment of plunge), satisfying $|\cos\theta_{\rm fin}|\leq\sin I$. For SKd systems, we can find the counterparts of $(I,\theta_{\rm fin})$ if we treat one of the BHs as the ``primary'' object. Below we still use the same notation, namely $(I,\theta_{\rm fin})$, to refer to these two angles. As shown in Fig.~\ref{fig:Hughes_EMRI}, we pick the primary BH to be the center of the coordinates. $I$ is still defined to be the angle between $\bm{L}$ (the red arrow) and $\bm{S}$ (the purple arrow), while $\theta_{\rm fin}$ remains to be the angle between the $\bm{S}$ (the purple arrow) and the orbital separation vector (the blue dashed line). The relations between $(I,\theta_{\rm fin})$ and our parameterization (namely $\theta_{\rm init}$ and $\phi_{\rm init}$, see Fig.~\ref{fig:configuration}) read
\begin{align}
& I=\theta_{\rm init}, &  \cos\theta_{\rm fin}=\sin\theta_{\rm init}\cos\phi_{\rm init}.
\label{I-theta-fin}
\end{align}
We want to emphasize there are two major difference in the parameterization of EMRIs and SKds. First, the parameters for EMRIs are defined at the moment of plunge, whereas in our
case, it becomes difficult to find well-defined quantities at the merger, thus
we use the initial geometry instead (at a reference time during the inspiral
stage). Second, for EMRIs, $\pi>I>\pi/2$ represents the retrograde motion of the small body, and hence the retrograde QNMs dominate in the ringdown signal. By contrast, only the prograde QNMs are excited for SKd systems [see Eq.~(\ref{qnm-overtone})].

With the purpose of exploring full parameter space of $I$ and $\theta_{\rm
fin}$, we now use the surrogate model NRSur7dq4. Comparing against NR ringdowns, even though NRSur7dq4 has
mismatches of order $\sim 3\times10^{-4}$, we find that it is not accurate
enough to reproduce the correct final mass and spin, in agreement with
Ref.~\cite{Finch:2021iip}. Mismatches of order $10^{-6}$ in the ringdown may be
necessary to achieve this. Therefore we fix the values of the final mass and spin
to the NR values (coming from NRSur7dq4Remnant) while fitting the mode amplitudes to NRSur7dq4. In addition,
we consider only the fundamental mode ($n=0$). 

The results for $\mathcal{A}_{220}$ and $\psi_{220}$ are shown in the first row
of Fig.~\ref{fig:overtone-h-scott}. Similar to
Refs.~\cite{Hughes:2019zmt,Lim:2019xrb}, we use two colors to stand for the
sign of $\dot{\theta}_{\rm fin}$, which was used in the EMRI case to represent
the moving direction at the plunge ($\dot{\theta}_{\rm fin}>0$ means that the
small particle moves toward the south pole of the Kerr BH, and vice versa). In
our case, $\dot{\theta}_{\rm fin}$ is determined by the sign of $\sin\phi_{\rm
init}$. Comparing to Fig.~3 of Ref.~\cite{Hughes:2019zmt}, we can see the
dependence is similar, although the absolute value of $\mathcal{A}_{220}$
differs.

In the second and third rows of Fig.~\ref{fig:overtone-h-scott}, we present how mode amplitudes depend on $\cos\theta_{\rm fin}$ for several $I$ slices [Eq.~(\ref{I-theta-fin})]. Those are direct analogs Fig.~4 of Ref.~\cite{Lim:2019xrb}. It is interesting to note that $\mathcal{A}_{2,+20}$ and $\mathcal{A}_{2,-20}$ are symmetric about the axis of $\cos\theta_{\rm fin}=0$, so are the patterns for $\psi_{2,+20}$ and $-\psi_{2,-20}$. The other intriguing feature is that the patterns for $I$ and $\pi-I$ are similar. 

For overtones $\mathcal{A}_{2,+2n}~(n>0)$, NRSur7dq4 is not accurate enough to provide any prediction, so we use our SKd4 runs instead (see Table \ref{table:NR-pars}), which corresponds to the $I=\pi/2$ slice. We translate our previous results in Fig.~\ref{fig:overtone-phiinit-nr} and \ref{fig:overtone-phiinit-nr-phase} to the cases of $h_{2,\pm2}$ based on Eq.~(\ref{IS-22-conversion}). Results are shown in Fig.~\ref{fig:A_nr_hugues}. We can see the patterns for high-$n$ are more distorted.

\subsubsection{Understanding the QNM excitation of $h_{2,\pm2}$ in terms of $(I_{22},S_{22})$}

It turns out that the features in the amplitudes $\mathcal{A}_{2,\pm 2,0}$ that we discussed in Sec.~\ref{sec:Mapping between SK and EMRI system parameters} can be understood based on what we have learned about $(I_{22},S_{22})$. In order to translate our previous results about $(I_{22},S_{22})$ to $h_{2,\pm2}$, we use the inverse of Eq.~(\ref{IS-22-conversion})
\begin{align}
    &\mathcal{A}_{2,\pm2n}e^{\pm i\psi_{2,\pm2n}}=\frac{1}{\sqrt{2}} \left[\mathscr{A}_{n}^{(I)}e^{i\varphi_n^{(I)}}\mp i\mathscr{A}_{n}^{(S)}e^{i\varphi_n^{(S)}}\right], 
\end{align}
and hence
\begin{align}
    \mathcal{A}_{2,\pm2n}^2=\frac{1}{2}\left[ \mathscr{A}_{n}^{(I)2}+\mathscr{A}_{n}^{(S)2}\mp2\mathscr{A}_{n}^{(I)}\mathscr{A}_{n}^{(S)}\sin(\varphi_n^{(I)}-\varphi_n^{(S)})\right]. \label{scott-explain}
\end{align}
As we shall explore later in Secs.~\ref{sec:ip-surrogate} and \ref{sec:spp-surrogate} [see Eqs.~(\ref{I22-fit-leading}) and (\ref{s-pn})], we have two dependencies 
\begin{align}
    &\mathscr{A}_{n}^{(I)}(I,\theta_{\rm fin})\sim{\rm const.}
    +\mathcal{O}(v^4), & \mathscr{A}_{n}^{(S)}(I,\theta_{\rm fin})\sim v^2\sin I+\mathcal{O}(v^4). \label{overtone_fit}
\end{align}%
where we have omitted specific numerical coefficients that are independent from $I$ and $\theta_{\rm fin}$, and $v^2$ is a parameter to keep track of the order of approximation (In fact, as we shall show in Sec.~\ref{sec:ip-surrogate}, $v$ is the orbital velocity that is widely used in the post-Newtonian theory). Furthermore, we have
\begin{equation}
    \varphi_n^{(I)} -\varphi_n^{(S)} \sim \phi_{\rm init} +\mathrm{const.}
\end{equation}
Using the above simple dependences of $[\mathscr{A}_n^{(I)},\mathscr{A}_n^{(S)},\varphi_n^{(I)} -\varphi_n^{(S)}]$ on $I$ and $\phi_{\rm init}$, we obtain:
\begin{align}
    &\mathcal{A}_{2,\pm2n}\sim {\rm const.}\pm v^2\sin I\sin(\phi_{\rm init}+{\rm const.})+\mathcal{O}(v^4).
    \label{a2m2-pn-preview}
\end{align}
As a result, for each $I$-slice (i.e., $\theta_{\rm init}$-slice), the $\mathcal{A}_{2,\pm20}-\cos\theta_{\rm fin}$ pattern is an approximate Lissajous-like curve (with identical frequencies), distorted by the higher order term containing $v^4$. The variation depends on $I$, which vanishes when $I=0,\pi$, and is maximal when $I=\pi/2$. Physically speaking, $\mathcal{A}_{2,\pm2n}$ depends sensitively on $\phi_{\rm init}$ when the spins of two BHs lie entirely in the orbital plane [see Fig.~\ref{fig:configuration}], but does not change with $\phi_{\rm init}$ as the spins are (anti-)parallel with the orbital angular momentum.

In addition, Eq.~(\ref{a2m2-pn-preview}) implies that $\mathcal{A}_{220}$ and $\mathcal{A}_{2,-20}$ are related by a transformation $\phi_{\rm init}\to \phi_{\rm init}+\pi$, i.e., $\cos\theta_{\rm fin}\to-\cos\theta_{\rm fin}$ [see Eq.~(\ref{I-theta-fin})]. This transformation represents the interchange of the in-plane spins for two BHs [see Fig.~\ref{fig:configuration}]. In fact, as we shall study in Sec.~\ref{sec:parity}, this conclusion can be generalized to the entire evolution regime (not only the ringdown phase). The symmetry of the SKd system results in [see Eq.~(\ref{hlm-thetaphi})]
\begin{align}
&\hlm (\pi-I,\phi_{\rm init})=(-1)^m\hlm (I,\phi_{\rm init}),  \notag \\
&\hlm (I,\phi_{\rm init}+\pi)=(-1)^\ell\hlmm^* (I,\phi_{\rm init}), \notag 
\end{align}
i.e., $\mathcal{A}_{2,\pm2n}$ remains unchanged when $I\to \pi-I$ (two BHs interchange their $z$-component spins), and $\mathcal{A}_{2,+2n}\to\mathcal{A}_{2,-2n}$ and $\psi_{2,+20}\to-\psi_{2,-20}$ when $\phi_{\rm init}\to\phi_{\rm init}+\pi$  (two BHs interchange their in-plane spins)\footnote{Equivalently, $\cos\theta_{\rm fin}\to-\cos\theta_{\rm fin}$}. In Fig.~\ref{fig:overtone-h-scott}, we can clearly see the patterns for $\mathcal{A}_{220}$ and $\mathcal{A}_{2,-20}$, as well as the patterns for $\psi_{2,+20}$ and $-\psi_{2,-20}$, are symmetric about the $\cos\theta_{\rm fin}=0$ axis. Meanwhile, the patterns for $\mathcal{A}_{2\pm20}$ are symmetric about the $I=\pi/2$ axis.

\section{The feature of mass and current quadrupole waves}
\label{sec:multipolar}

In the last section, we explored how QNMs are excited with different initial parameters $(\theta_{\rm init},\phi_{\rm init})$. We now aim to study  features of ringdown more quantitatively. In particular, we focus on the mass $(I_{22})$ and current $(S_{22})$ quadrupole waves of SKd systems, and relate their features to $(\chi_{\rm init},\theta_{\rm init},\phi_{\rm init})$. Moreover, since kick velocity is one of the important quantities that reflects SKd systems\textquotesingle \ properties, we also include it to our study. 


\subsection{A brief review}
\label{sec:multipolar-review}

This subsection briefly reviews some facts about the gravitational recoil. In particular, we relate the kick velocity to the radiative mass and current quadrupole waves.
\begin{figure}[htb]
         \includegraphics[width=0.95\columnwidth,height=5.6cm,clip=true]{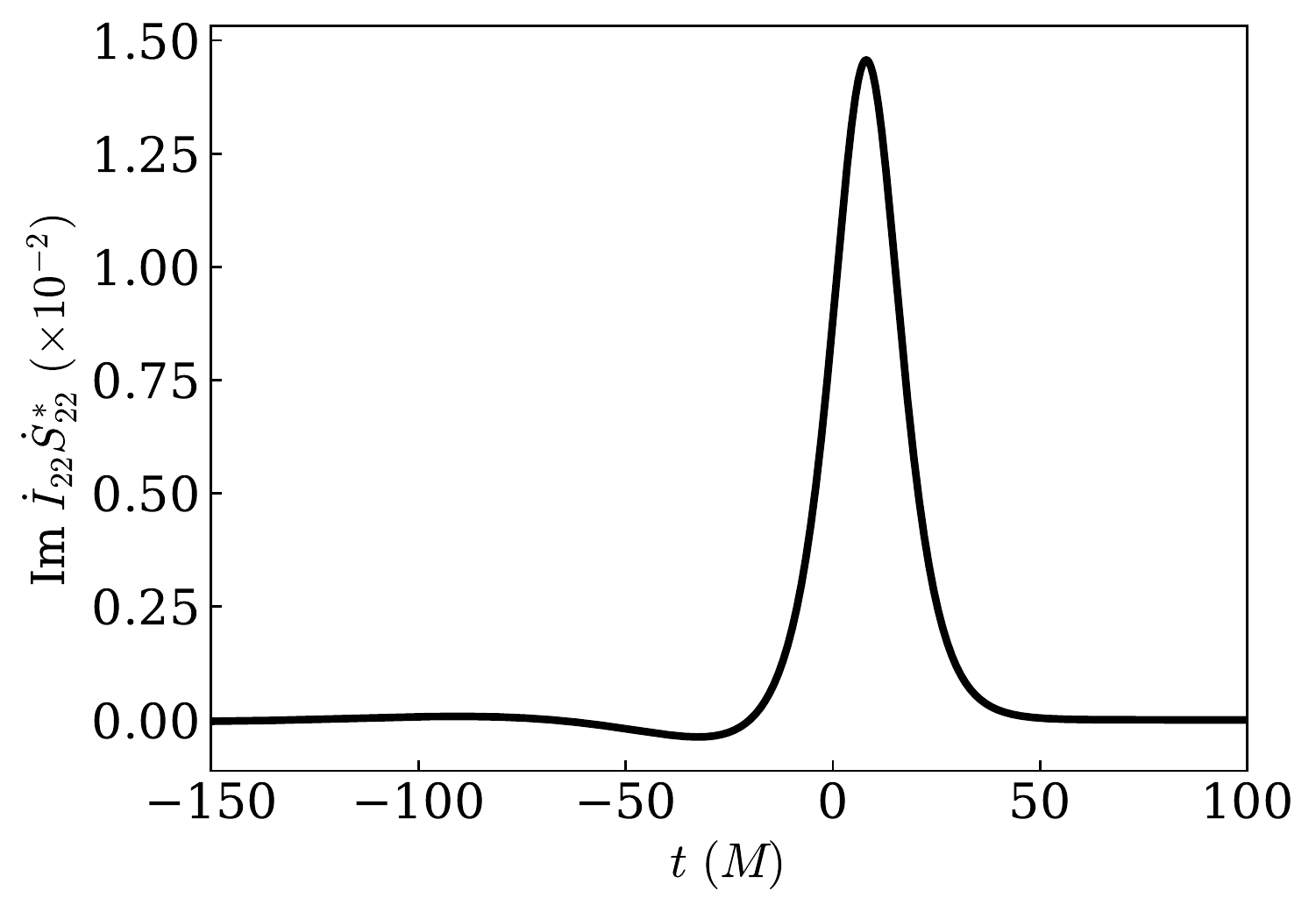}
  \caption{The integrand of Eq.~(\ref{vfinal}) for the SKd4--`06' system. The majority kick velocity is accumulated around $t\sim 0M$, and the final kick is $4.75\times10^{-3}$.}
 \label{fig:vt}
\end{figure} 

\begin{figure*}[htb]
        \includegraphics[width=\columnwidth,height=6.6cm,clip=true]{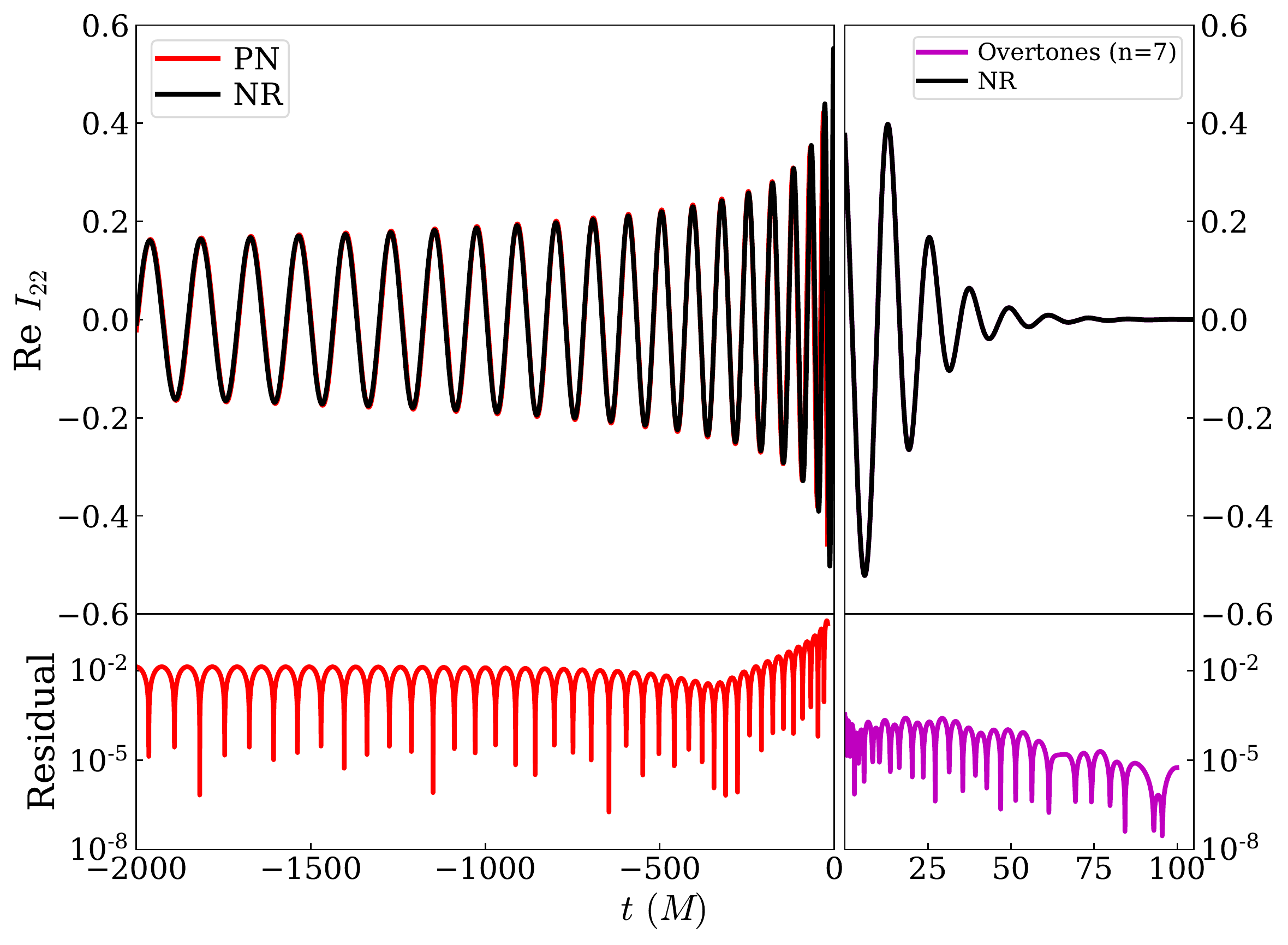}
        \includegraphics[width=\columnwidth,height=6.6cm,clip=true]{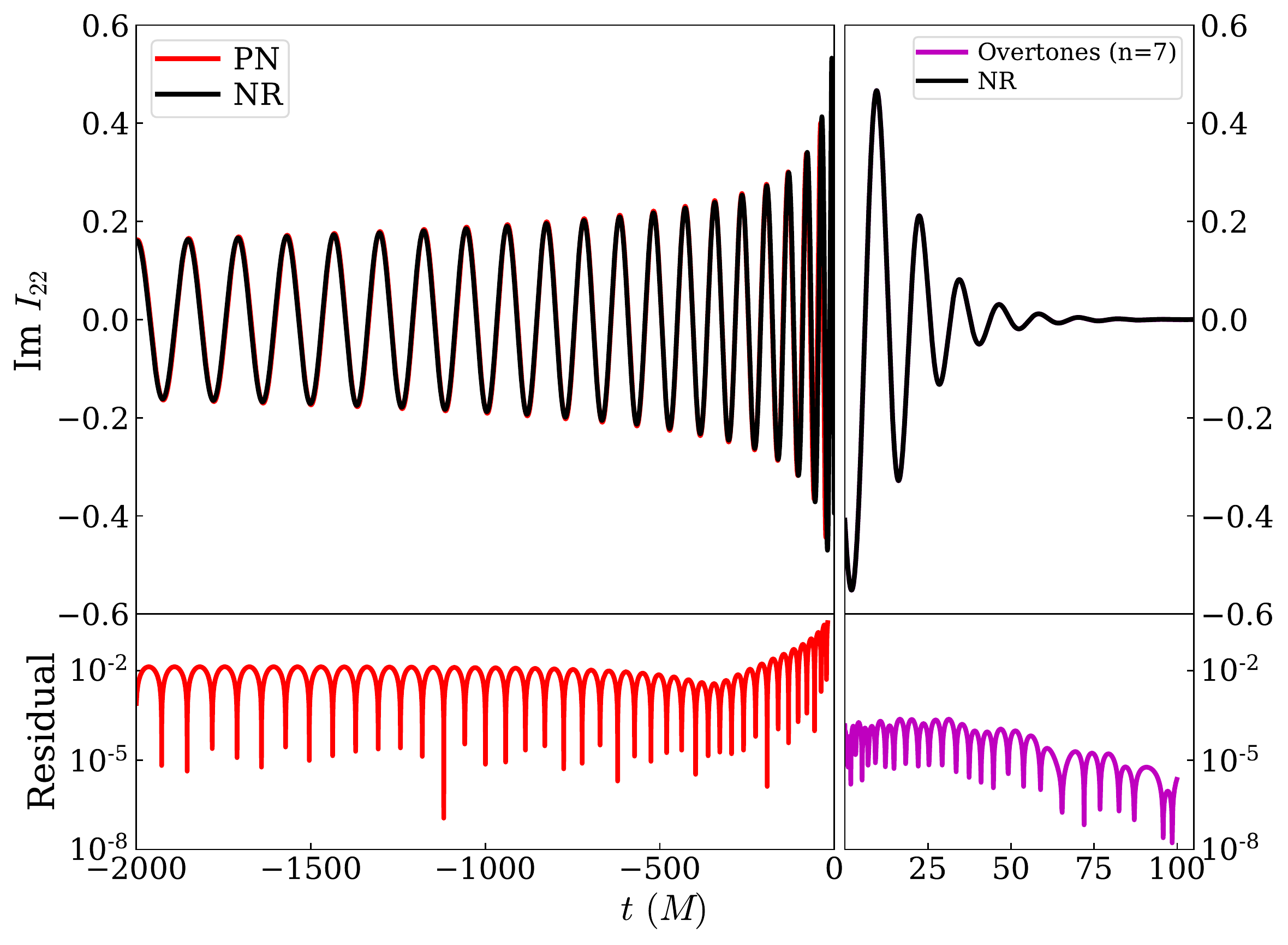} \\
        \includegraphics[width=\columnwidth,height=6.6cm,clip=true]{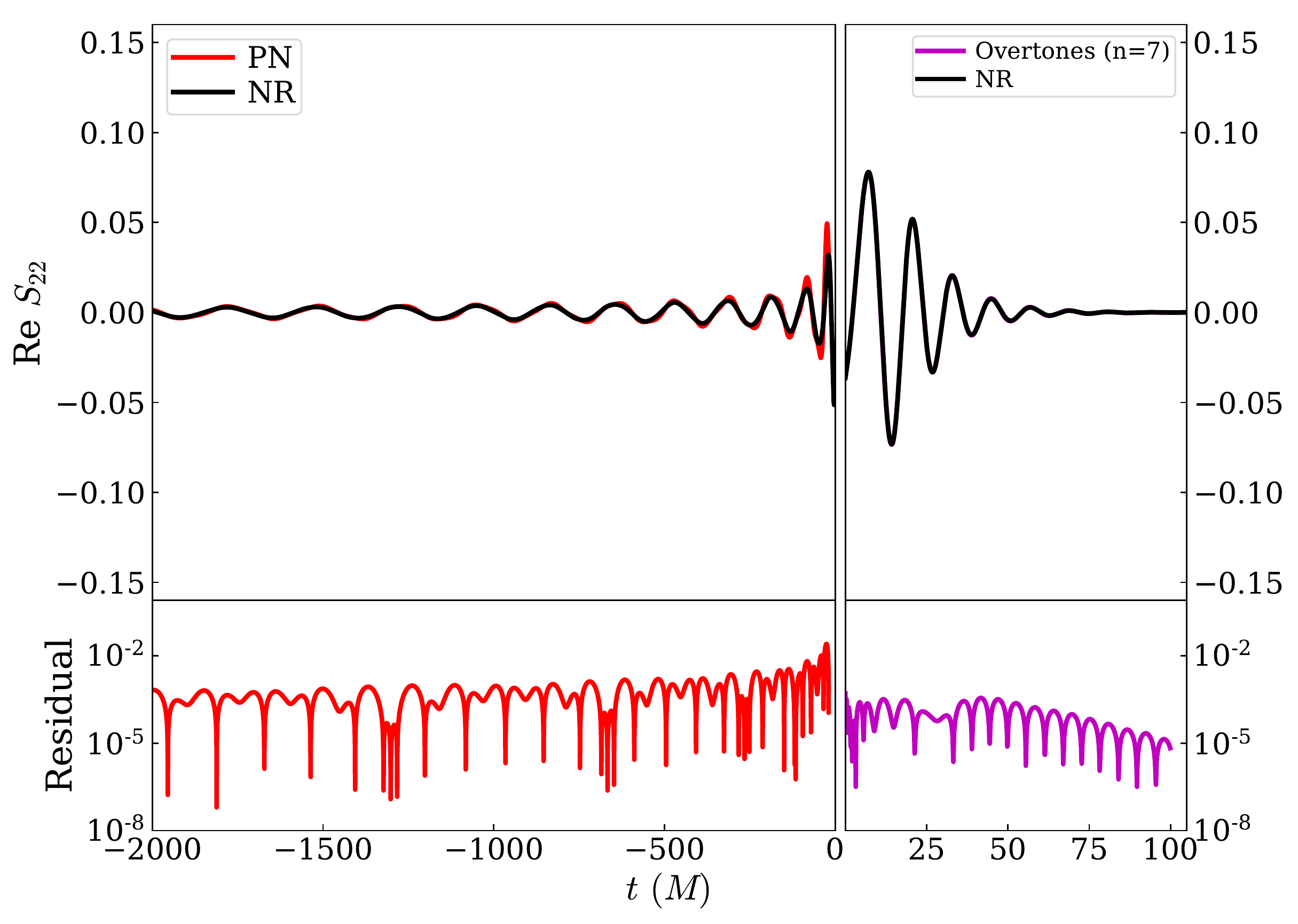}
        \includegraphics[width=\columnwidth,height=6.6cm,clip=true]{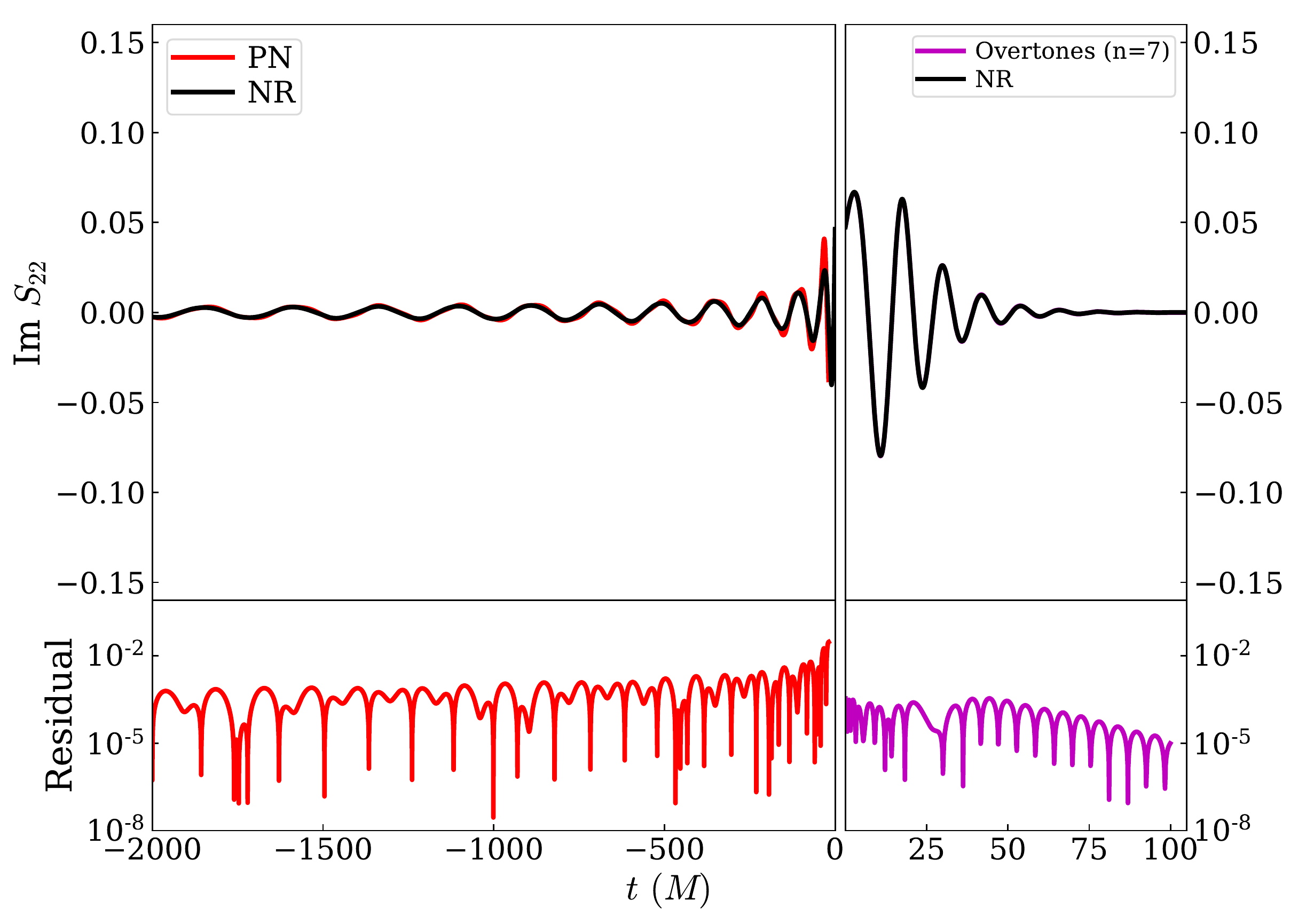}
  \caption{The time evolution of $I_{22}$ (upper row) and $S_{22}$ (bottom row) for the SKd4--`03' system, where $t=0M$ stands for the peak of strain amplitude. The mass ($I_{22}$) and current ($S_{22}$) quadrupole waves are compared to PN formulas [Eq.~(\ref{pn-SI})] during the inspiral stage, and to QNMs (7 overtones) in the ringdown regime.
}
 \label{fig:chi-0.6-compare}
\end{figure*}

It has been shown that for a SKd system, the kick magnitude can be estimated with a simple formula \cite{Gonzalez:2007hi,Campanelli:2007cga,Campanelli:2007ew}
\begin{align}
v_f\sim \chi_{\rm init}\sin(\phi_{\rm init}-\phi_{\rm init}^{(0)})\sin\theta_{\rm init}, \label{map-kick}
\end{align}
where $\phi_{\rm init}^{(0)}$ is a constant. 
Eq.~(\ref{map-kick}) is based on the computation of linear momentum carried away by GW \cite{Ruiz:2007yx}
\begin{align}
\dot{P}_z=\lim_{D\to\infty}\frac{1}{24\pi}(\dot{h}_{22}\dot{h}_{22}^*-\dot{h}_{2,-2}\dot{h}_{2,-2}^*), \label{Pdot}
\end{align}
where $*$ stands for complex conjugate, $D$ is the distance between the source and
the observer, and the $z$-axis is in the direction of
orbital angular momentum. Here we have ignored the effects of other modes since
they are negligible. 

In terms of $I_{22}$ and $S_{22}$ [Eq.~(\ref{rad-moment})], Eq.~(\ref{Pdot}) can also be written as
\begin{align}
\dot{P}_z=-\frac{1}{12\pi}{\rm Im}~\dot{I}_{22}\dot{S}_{22}^*,\label{pdot}
\end{align}
and the final kick velocity is given by
\begin{align}
m_fv_f&=\frac{1}{12\pi}{\rm Im}~\int \dot{I}_{22}\dot{S}_{22}^*dt \notag \\
&=\frac{1}{12\pi}{\rm Im}~\int |\dot{I}_{22}||\dot{S}_{22}|e^{i\Phi_{\dot{I}\dot{S}}}dt, \label{vfinal}
\end{align}
with $\Phi_{\dot{I}\dot{S}}$ the phase difference between $\dot{I}_{22}$ and $\dot{S}_{22}$.
Note that the change of sign from Eq.~(\ref{pdot}) to (\ref{vfinal}) is a result of linear momentum conservation. In Fig.~\ref{fig:vt}, we show the time evolution of the above-mentioned integrand for SKd4--`06' (cf.~Table \ref{table:NR-pars}). We can see that most of the kick velocity is accumulated around $t\sim 0M$.

\begin{figure*}[htb]
        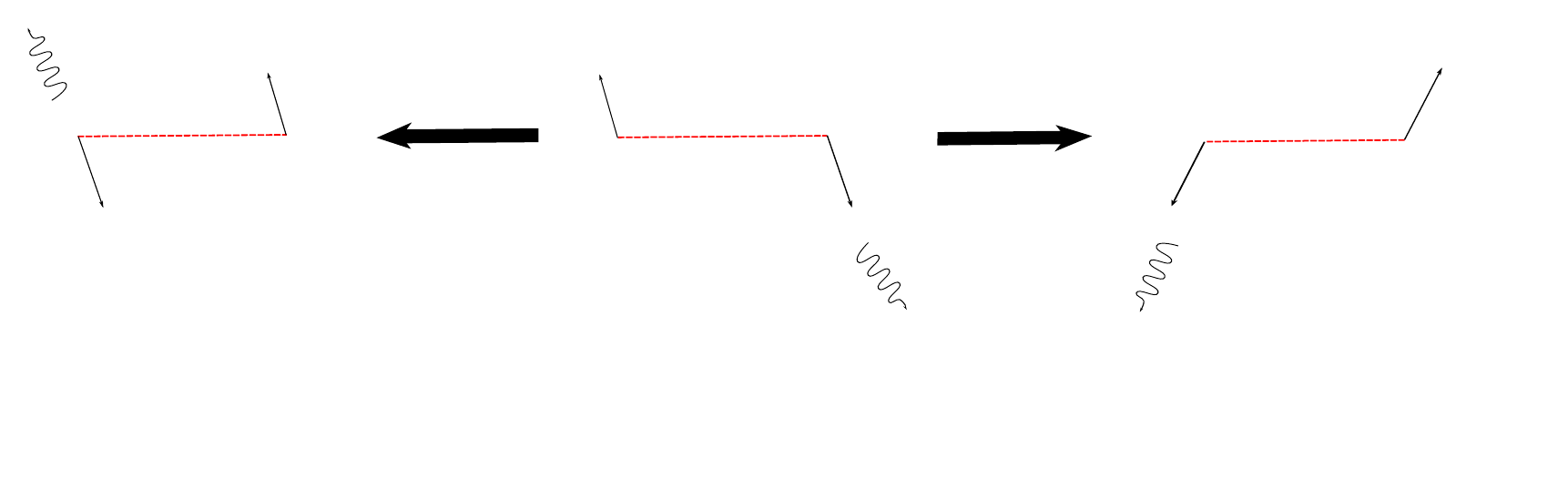
  \caption{Parity inversion of a SKd BBH system. We use arrows to represent the spin direction of BHs, and wavy lines to stand for the GW propagating direction. The complex strain of a SKd system is totally determined by two extrinsic parameters $(\iota,\beta)$, and three intrinsic parameters $(\chi_{\rm init},\theta_{\rm init},\phi_{\rm init})$. Here the intrinsic parameters are the spin of the left BH. The properties of the right BH are determined based on the SKd condition. Figs.~\ref{fig:parity} (a) and (b) are related by a parity inversion: two BHs exchange their locations while having their own spins fixed. As a result,
   the GW propagation direction and both spins change sign, i.e., $(\iota,\beta)\longleftrightarrow(\pi-\iota,\pi+\beta)$ and $(\theta_{\rm init},\phi_{\rm init})\longleftrightarrow(\pi-\theta_{\rm init},\pi+\phi_{\rm init})$. Figs.~\ref{fig:parity} (b) and (c) are related by a $\pi$-rotation about the orbital angular momentum. Thus we have $(\iota,\beta)\longleftrightarrow(\iota,\pi+\beta)$ and $(\theta_{\rm init},\phi_{\rm init})\longleftrightarrow(\pi-\theta_{\rm init},\phi_{\rm init})$.}
 \label{fig:parity}
\end{figure*}

During the inspiral stage, it was shown that $I_{22}$ and $S_{22}$ are related to the source quadrupole moments\footnote{Hereafter we shall not distinguish the source quadrupole moment and the (radiative) quadrupole wave since it will not cause any confusion.}. At the leading order, from Refs.~\cite{Schnittman:2007ij,Porto:2010zg}, we write 
\begin{subequations}
\begin{align}
&I_{22}(t)=-\frac{M}{2}\sqrt{\frac{2\pi}{5}}\frac{d^2}{dt^2}r(t)^2 e^{-2i\phi(t)}, \label{I22-pn} \\
&S_{22}(t)=\sqrt{\frac{2\pi}{5}}\chi \sin\theta(t)\frac{d^2}{dt^2}r(t)e^{-i\phi(t)-i\phi_{\rm pre}(t)},\label{S22-pn}
\end{align}
\label{pn-SI}%
\end{subequations}
where $M$ is the total mass of the BBH system; $\chi$ is the dimensionless spin
of an individual BH; $\phi(t)$ and $r(t)$ are the orbital phase and separation,
respectively; $\theta(t)$ is the polar angle of the spin; and $\phi_{\rm
pre}(t)$ is the precession angle (the azimuthal angle of the in-plane spin
component). Note that at the initial time $t_{\rm init}$
\begin{align}
    &\theta_{\rm init}\coloneqq\theta(t_{\rm init}), & \phi_{\rm init}\coloneqq\phi(t_{\rm init}). \notag 
\end{align}
For instance, we choose SKd4--`03' (see Table \ref{table:NR-pars}) and compare
its radiative multipolar waves $I_{22}$ and $S_{22}$ to PN formulas in
Eq.~(\ref{pn-SI}). We read off the values of $r(t)$, $\theta(t)$, and $\phi(t)$ directly from the outputs of NR simulation. The results are shown in Fig.~\ref{fig:chi-0.6-compare}. For
comparison, we also fit the ringdown signal with QNMs (7 overtones), starting
from $t=0M$. We can see the Newtonian formulas can accurately model the phase
evolution up to $t\sim-250M$. Meanwhile, both $I_{22}$ and $S_{22}$ are
described by 7 overtones accurately from $t=0M$.

In the rest of this section, we shall discuss how $I_{22}$ and $S_{22}$ depend
on $(\chi_{\rm init},\theta_{\rm init},\phi_{\rm init})$, and apply our
understanding to the gravitational recoil.

\subsection{Symmetry properties of $I_{22}$ and $S_{22}$}
\label{sec:parity}

Before exploring the detailed relations between $(I_{22},S_{22})$ and
$(\chi_{\rm init},\theta_{\rm init},\phi_{\rm init})$, we first take advantage
of the symmetry of SKd systems, and study its implication on $(I_{22},S_{22})$.
As shown in Fig.~\ref{fig:parity}, there are three SKd binaries, where (a) and
(b) are related by a parity transformation, i.e., two BHs interchange their
locations while having their spin directions fixed, recalling that spin is an
axial vector, which is not changed by the parity transformation. On the other
hand, we rotate the whole system in (b) about the orbital angular momentum by
$\pi$, and obtain (c). We use wavy lines to stand for the GW propagating
direction, and $(\iota,\beta)$ are the coordinates of the observer in (b), as
defined in Eq.~(\ref{spherical-harm}). The coordinates of observers in (a) and
(c) are transformed accordingly. As discussed in Eq.~(\ref{spherical-harm}),
$\h(t,\iota,\beta)$ can be decomposed into the extrinsic part $\sy$ and the
intrinsic part $\hlm(\theta_{\rm init},\phi_{\rm init})$\footnote{We use
$(\theta_{\rm init},\phi_{\rm init})$ to stand for the spin of BH on the left.
The other spin is determined uniquely by the SKd condition.}. Here we omit
$\chi_{\rm init}$ in the argument of $\hlm$ since it has no impact on the
transformation in question. 

\begin{figure*}[htb]
        \includegraphics[width=\textwidth,clip=true]{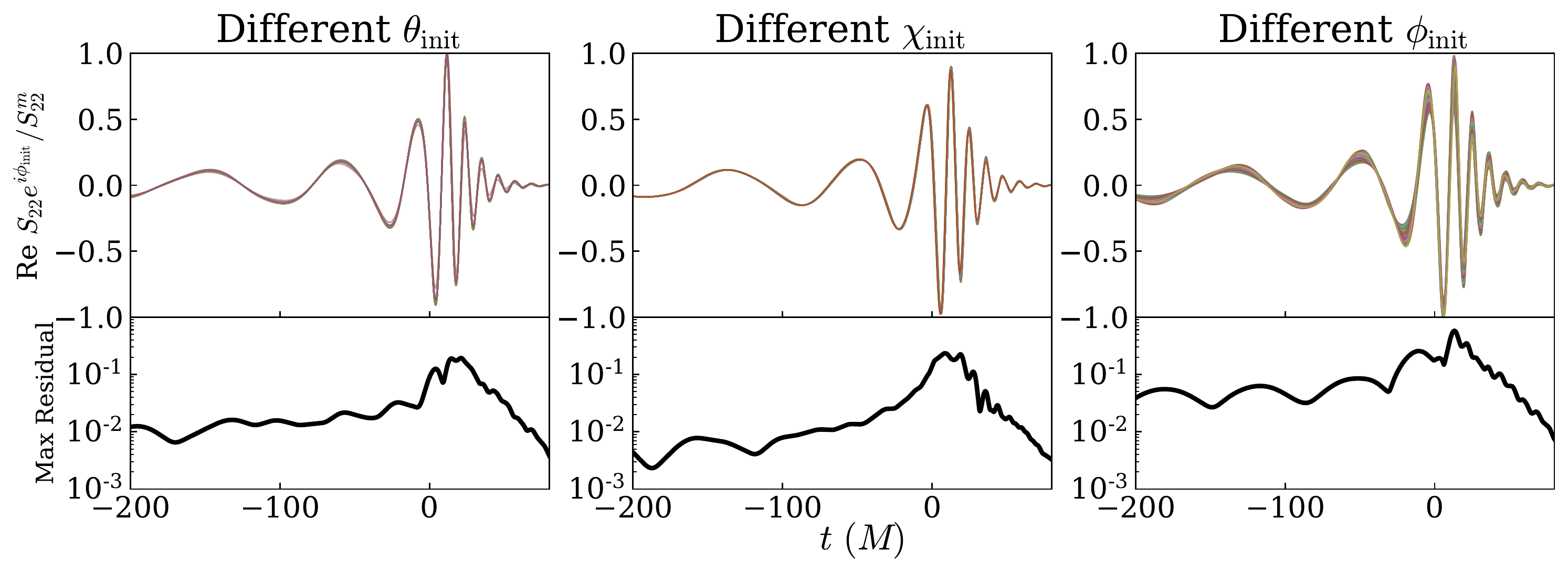}
        \includegraphics[width=\textwidth,height=5.7cm,clip=true]{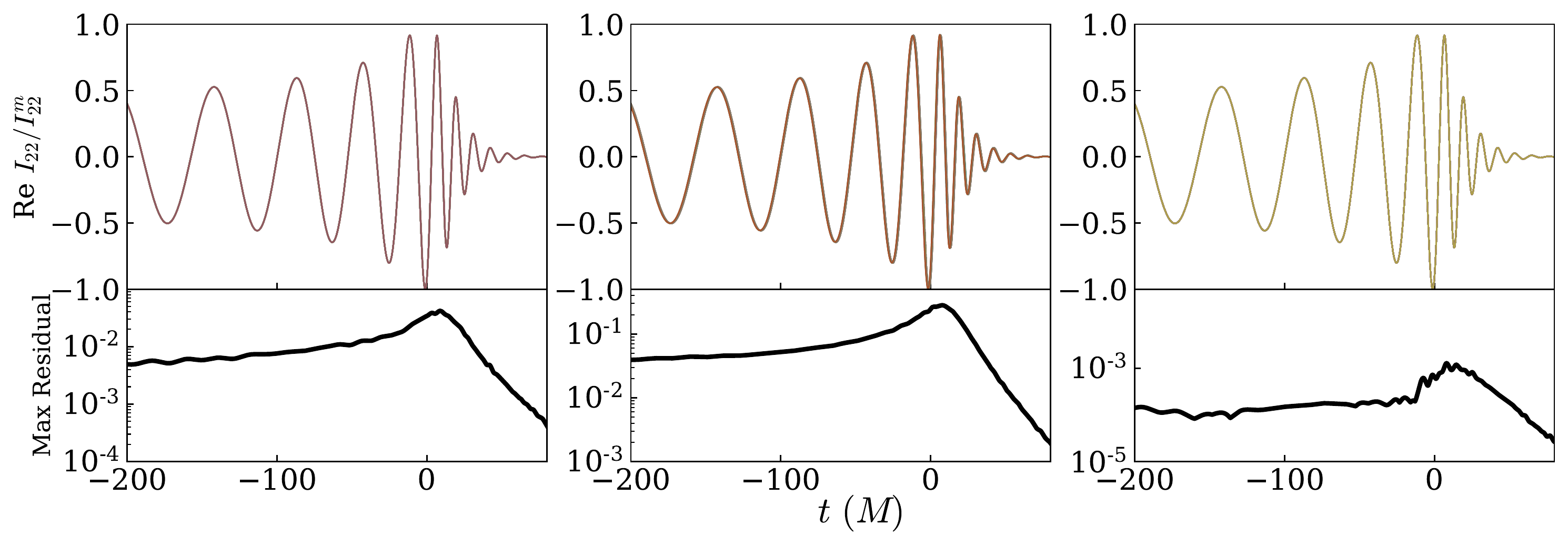}
        \includegraphics[width=\textwidth,height=5.5cm,clip=true]{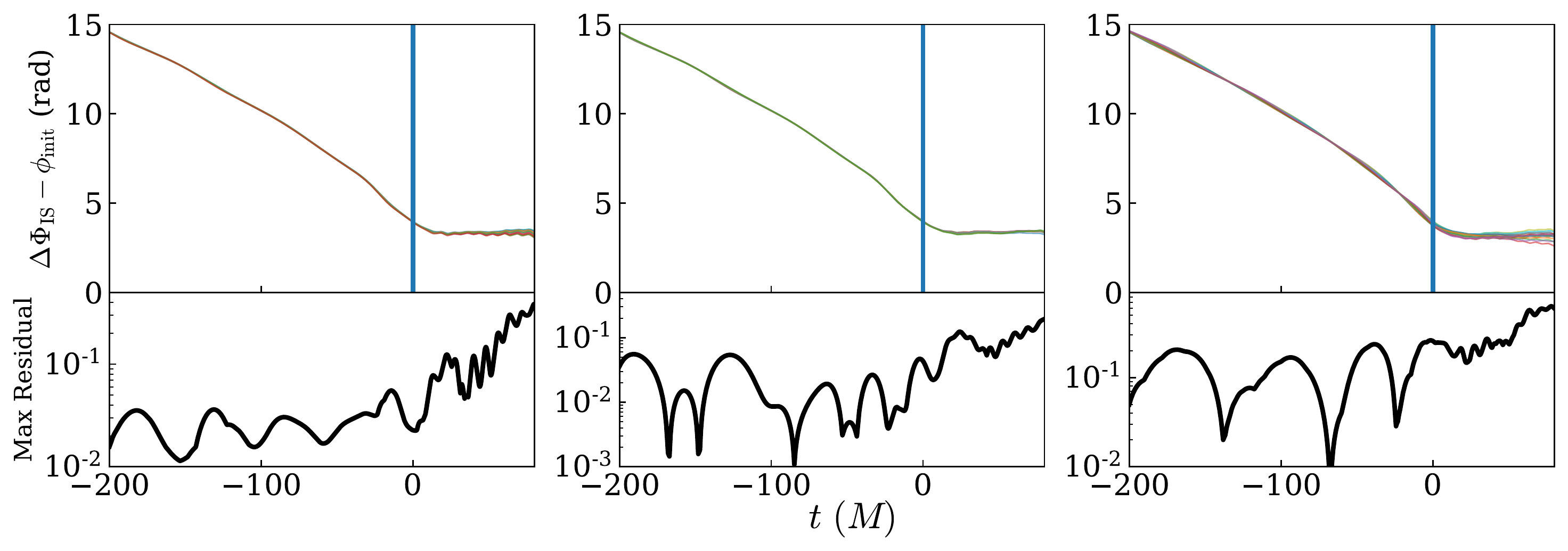}
\caption{The time evolution of the real part of the normalized $S_{22}$ (upper
row), the real part of the normalized $I_{22}$ (middle row), as well as
$\dphi-\phi_{\rm init}$ (bottom row), using the SKd BBH configuration from
NRSur7dq4. The imaginary part is similar. We sample in total 180 cases with
different $\theta_{\rm init}\in[0,\pi]$ (left column), $\chi_{\rm init}\in[0,0.8]$ (middle column),
and $\phi_{\rm init}\in[0,2\pi]$ (right column), and plot them on top of each other. `Max Residual' is defined to be the
maximum difference of all cases at each time step. The normalized $I_{22}$ and
$S_{22}$ are insensitive to $(\chi_{\rm init},\theta_{\rm init},\phi_{\rm
init})$, to the level of $\sim0.1\%-30\%$.}
 \label{fig:phasediff-surr}
\end{figure*}

Fig.~\ref{fig:parity} (a) and (b) are related by a parity transformation, hence we have (see Appendix \ref{app:parity} for more details)
\begin{align}
\h^{(a)}=\h^{(b)*},
\end{align}
i.e.,
\begin{align}
&\hlm (\theta_{\rm init},\phi_{\rm init})\sy \notag \\
&=\hlm^* (\pi-\theta_{\rm init},\phi_{\rm init}+\pi)\sytt^*(\pi-\iota,\pi+\beta).
\end{align}
Using the fact that
\begin{align}
\sytt^*(\pi-\iota,\pi+\beta)=(-1)^{\ell+m}\sytmt(\iota,\beta),
\end{align}
we obtain
\begin{align}
&\hlm (\theta_{\rm init},\phi_{\rm init}) =(-1)^{\ell+m}\hlmm^* (\pi-\theta_{\rm init},\phi_{\rm init}+\pi). \label{parity-de-1}
\end{align}
On the other hand, Fig.~\ref{fig:parity} (b) and (c) are related by a global rotation. Therefore, the observable $\h(t,\iota,\beta)$ should not be affected
\begin{align}
\h^{(b)}=\h^{(c)},
\end{align}
i.e.,
\begin{align}
&\hlm (\theta_{\rm init},\phi_{\rm init})\sy \notag \\
&=\hlm (\pi-\theta_{\rm init},\phi_{\rm init})\sytt(\iota,\pi+\beta).
\label{transform_of_theta}
\end{align}
Recalling that
\begin{align}
\sytt(\iota,\pi+\beta)=(-1)^m\sytt(\iota,\beta),
\label{spin_weighted_theta}
\end{align}
we then have
\begin{subequations}
\begin{align}
&\hlm (\pi-\theta_{\rm init},\phi_{\rm init})=(-1)^m\hlm (\theta_{\rm init},\phi_{\rm init}), \label{parity-de-2} \\
&\hlm (\theta_{\rm init},\phi_{\rm init}+\pi)=(-1)^\ell\hlmm^* (\theta_{\rm init},\phi_{\rm init}),
\end{align}
\label{hlm-thetaphi}%
\end{subequations}
where the first line is the result of Eqs.~(\ref{transform_of_theta}) and (\ref{spin_weighted_theta}), and the second line comes from the combination of Eq.~(\ref{parity-de-1}) and (\ref{parity-de-2}). Eqs.~(\ref{hlm-thetaphi}) give the transformation of $\hlm$ under $\theta_{\rm init}\to\pi-\theta_{\rm init}$ (two BHs interchange their $z$-component spins) and $\phi_{\rm init}\to\phi_{\rm init}+\pi$ (two BHs interchange their in-plane spins). As we discussed in Sec.~\ref{sec:theta_phi_scott}, Eq.~(\ref{hlm-thetaphi}) directly leads to several features revealed in Fig.~\ref{fig:overtone-h-scott}: the patterns for $\mathcal{A}_{2\pm20}$, as well as $\psi_{2,+20}$ and $-\psi_{2,-20}$, have a  reflective symmetry about the $\cos\theta_{\rm fin}=0$ axis; and the patterns for $\mathcal{A}_{2\pm20}$ are symmetric about the $I=\pi/2$ axis.

We then apply Eqs.~(\ref{hlm-thetaphi}) to the case of $I_{\ell m}$ and $S_{\ell m}$, [see Eqs.~(\ref{rad-moment})]
\begin{subequations}
\begin{align}
&I_{\ell m}(\pi-\theta_{\rm init},\phi_{\rm init})=(-1)^mI_{\ell m} (\theta_{\rm init},\phi_{\rm init}),\\
&S_{\ell m}(\pi-\theta_{\rm init},\phi_{\rm init})=(-1)^mS_{\ell m} (\theta_{\rm init},\phi_{\rm init}),\\
&I_{\ell m}(\theta_{\rm init},\phi_{\rm init}+\pi)=(-1)^{\ell+m} I_{\ell m} (\theta_{\rm init},\phi_{\rm init}),\\
&S_{\ell m}(\theta_{\rm init},\phi_{\rm init}+\pi)=(-1)^{\ell+m+1} S_{\ell m} (\theta_{\rm init},\phi_{\rm init}),
\end{align}
\label{IS-map}%
\end{subequations}
One can find the counterpart of Eqs.~(\ref{IS-map}) for EMRIs in Eq.~(4.6) of Ref.~\cite{Lim:2019xrb}.
Those relations imply that the dependence of $I_{22}$ and $S_{22}$ on $\theta_{\rm init}$ is symmetric about $\theta_{\rm init}=\pi/2$ axis, whereas the dependence of $|I_{22}|$ and $|S_{22}|$ on $\phi_{\rm init}$ have a period $\pi$\footnote{Here we use the absolute value for future convenience.}. We shall see these features shortly from numercal results.

\subsection{Time dependence of $I_{22}$ and $S_{22}$}
\label{sec:universality}
After the study of $(I_{22},S_{22})-(\chi_{\rm init},\theta_{\rm init},\phi_{\rm init})$ dependence enforced by the symmetry, we are in a position to carry out more detailed analyses. Based on the discussion around Eq.~(\ref{IS-universal-overtone}), for the post-merger evolution of $I_{22}$ and $S_{22}$, their $\phi_{\rm init}$ dependence can be factored out. In particular, the spin sector of $I_{22}$ is described by a function $I_{22}^m(\phi_{\rm init})$, and that of $S_{22}$ is given by $\spp(\phi_{\rm init})e^{-i\phi_{\rm init}}$. In fact, those features are also consistent with PN predictions, as shown in Eqs.\ (\ref{pn-SI}): To the leading PN order, $I_{22}$ is independent of $(\chi_{\rm init},\theta_{\rm init},\phi_{\rm init})$, whereas $S_{22}\sim\chi_{\rm init}\sin\theta_{\rm init}e^{-i\phi_{\rm init}}$. In light of the facts, it is reasonable to conjecture that the separability between the spin sector (including $\chi_{\rm init},\theta_{\rm init},\phi_{\rm init}$) and the temporal sector is preserved throughout the entire process, i.e., 
\begin{subequations}
\begin{align}
&I_{22}(t,\chi_{\rm init},\theta_{\rm init},\phi_{\rm init})=I_{22}^m(\chi_{\rm init},\theta_{\rm init},\phi_{\rm init}) T_I(t), \\
&S_{22}(t,\chi_{\rm init},\theta_{\rm init},\phi_{\rm init})=S_{22}^m(\chi_{\rm init},\theta_{\rm init}, \phi_{\rm init})e^{-i\phi_{\rm init}} T_S(t),
\end{align}
\label{universality}%
\end{subequations}
where $T_I(t)$ and $T_S(t)$ are two complex functions of time, which are normalized such that they each is equal to 1 at the moment when its magnitude is at maximum. As a result, $I_{22}^m$ and $S_{22}^m$ are in fact the peak values of $I_{22}$ and $S_{22}$, respectively, i.e.,
\begin{subequations}
\begin{align}
&I_{22}^m(\chi_{\rm init},\theta_{\rm init},\phi_{\rm init}) =\max_t |I_{22}(t,\chi_{\rm init},\theta_{\rm init},\phi_{\rm init})|, \\
&S_{22}^m(\chi_{\rm init},\theta_{\rm init},\phi_{\rm init})=\max_t |S_{22}(t,\chi_{\rm init},\theta_{\rm init},\phi_{\rm init})|.
\end{align}
\label{IpSp}%
\end{subequations}
We want to emphasize that Eq.~(\ref{IpSp}) is an approximation based on the observation we made in Fig.~\ref{fig:overtone-phiinit-nr-phase}, namely to the leading order $\varphi_n^{(I)}$ is insensitive to $\phi_{\rm init}$,
while $\varphi_n^{(S)}\propto-\phi_{\rm init}$. This fact allows us to treat $I_{22}^m$ and $S_{22}^m$ as two real functions [see the context below Eq.~(\ref{IS-universal-overtone})]. The higher order corrections will lead to additional phase factors for both $I_{22}^m$ and $S_{22}^m$. This is beyond the scope of this work.

To test the accuracy of Eqs.~(\ref{universality}) and (\ref{IpSp}),  we use NRSur7dq4 to obtain $I_{22}$ and $S_{22}$ with different initial spin configurations. They are normalized by $\lp$ and $\spp e^{-i\phi_{\rm init}}$, respectively. The results are shown in the first two rows of Fig.~\ref{fig:phasediff-surr}. To avoid redundancy, we present only the real part since the imaginary part is similar. As we can see, the normalized $I_{22}$ with different $(\chi_{\rm init},\theta_{\rm init},\phi_{\rm init})$  evolves in a similar way, so does the normalized $S_{22}$. The residuals imply that Eqs.~(\ref{universality}) are accurate to $\sim0.1\%-30\%$ throughout the entire evolution. We remark that the accuracy is limited by the approximation adopted in Eq.~(\ref{IpSp}), where $I_{22}^m$ and $S_{22}^m$ are treated as two real functions and their phases (higher order effects) are not included. If we omit these additional phase terms that are functions of $(\chi_{\rm init},\theta_{\rm init}, \phi_{\rm init})$, there will be a non-negligible increase in the residual. In fact, if we consider only the absolute value of the normalized $I_{22}$ and $S_{22}$, the residual can be decreased by a factor of $1.6\sim100$.

Nevertheless, the progenitor's information is primarily described by the peak value of mass and current quadrupole waves, $I_{22}^m$ and $S_{22}^m$. On the other hand, the temporal evolution, $T_I(t)$ and $T_S(t)$, encode the common feature of SKd systems.
In particular, as we discussed in Sec.~\ref{sec:multipolar-review}, the phase difference between $T_I(t)$ and $T_S(t)e^{-i\phi_{\rm init}}$, denoted by $\dphi$
\begin{align}
\dphi\equiv\arg (T_I)-\arg (T_Se^{-i\phi_{\rm init}})\equiv \arg (I_{22})-\arg (S_{22}), \label{delta_phi_IS_universal}
\end{align}
is closely related to the gravitational recoil.

We have introduced three quantities, $\lp$, $\spp$ and $\dphi$, which are important characteristics of SKd systems. In the rest of this section, we aim to study $\lp$, $\spp$ and $\dphi$ more carefully and more quantitatively. In particular, we will show that $\lp$ and $\spp$ are subject to the periodic condition in both $\theta_{\rm init}-$ and $\phi_{\rm init}-$ directions, as enforced by the symmetry in Eqs.~(\ref{IS-map}). 

\begin{figure}[htb]
        \includegraphics[width=\columnwidth,height=6.1cm,clip=true]{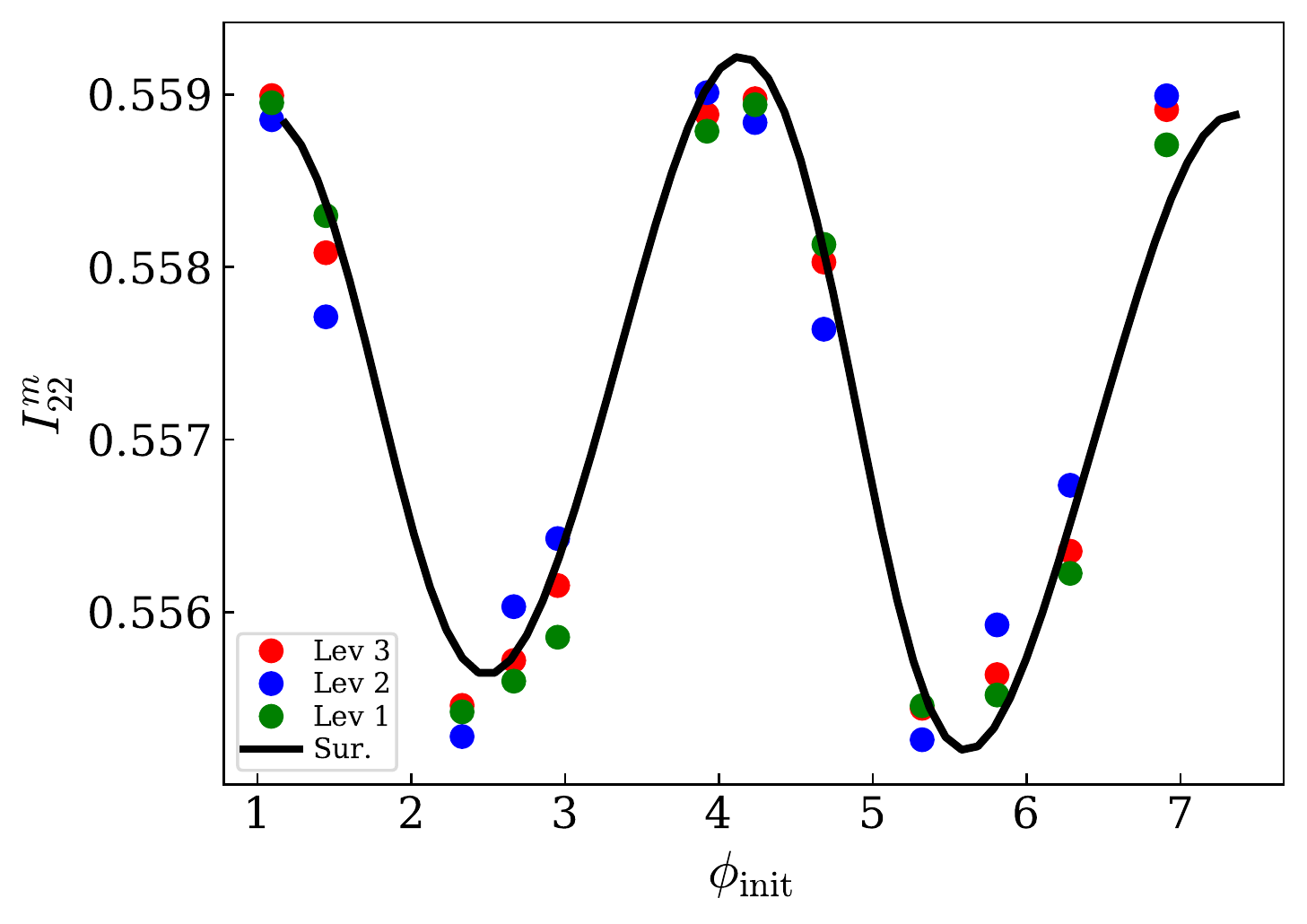}
  \caption{The peak value of mass quadrupole wave $\lp$ as a function of $\phi_{\rm init}$. We use SKd4 systems listed in Table \ref{table:NR-pars}. The black curve is from NRSur7dq4, whereas points are from NR simulations. Colors (labeled by Lev) correspond to numerical resolutions, where "Lev 1" stands for the lowest resolution. Predictions of NRSur7dq4 are consistent with NR results: $\lp$ oscillates with $\phi_{\rm init}$ on the level of $\sim 0.36\%$, around a base value $\sim 0.557$. }
 \label{fig:I22-hangup-NR}
\end{figure}

\subsection{The peak of mass quadrupole wave $\lp$}
\label{sec:ip-surrogate}
We saw that $\lp$ is an important characteristic quantity for SKd systems. In fact, it was shown that the remnant BH spin is already encoded in the peak amplitude of the gravitational wave strain \cite{Ferguson:2019slp}. Therefore, it is instructive to study how $\lp$ depends on $(\chi_{\rm init},\theta_{\rm init},\phi_{\rm init})$.

 We first look at our SKd4 NR runs listed in Table \ref{table:NR-pars}. Fig.~\ref{fig:I22-hangup-NR} shows $\lp$ as a function of $\phi_{\rm init}$. We can see that $\lp$ does depend weakly on $\phi_{\rm init}$ for all three numerical resolutions, which verifies that the dependence is not a numerical artifact. For comparison purposes, we also show the prediction of NRSur7dq4 with the same BBH system but varying $\phi_{\rm ini}$. Two results are close. With different $\phi_{\rm init}$, $\lp$ varies on the level of $\sim0.36\%$, around a base value $\sim0.557$. As discussed earlier, $\lp-\phi_{\rm init}$ relation is expected to have a period of $\pi$ [Eq.~(\ref{IS-map})]. However, the black curve is slightly asymmetric. We attribute this to the numerical error of NRSur7dq4. Furthermore, the change of $\lp$ is much smaller than the base value, which is qualitatively consistent with PN predictions, because the variation caused by spin is 2PN \cite{Porto:2010zg} smaller than the leading contribution from the orbital mass quadrupole moment [Eq.~(\ref{I22-pn})].

%
%

\begin{figure}[htb]
        \includegraphics[height=7.1cm,clip=true]{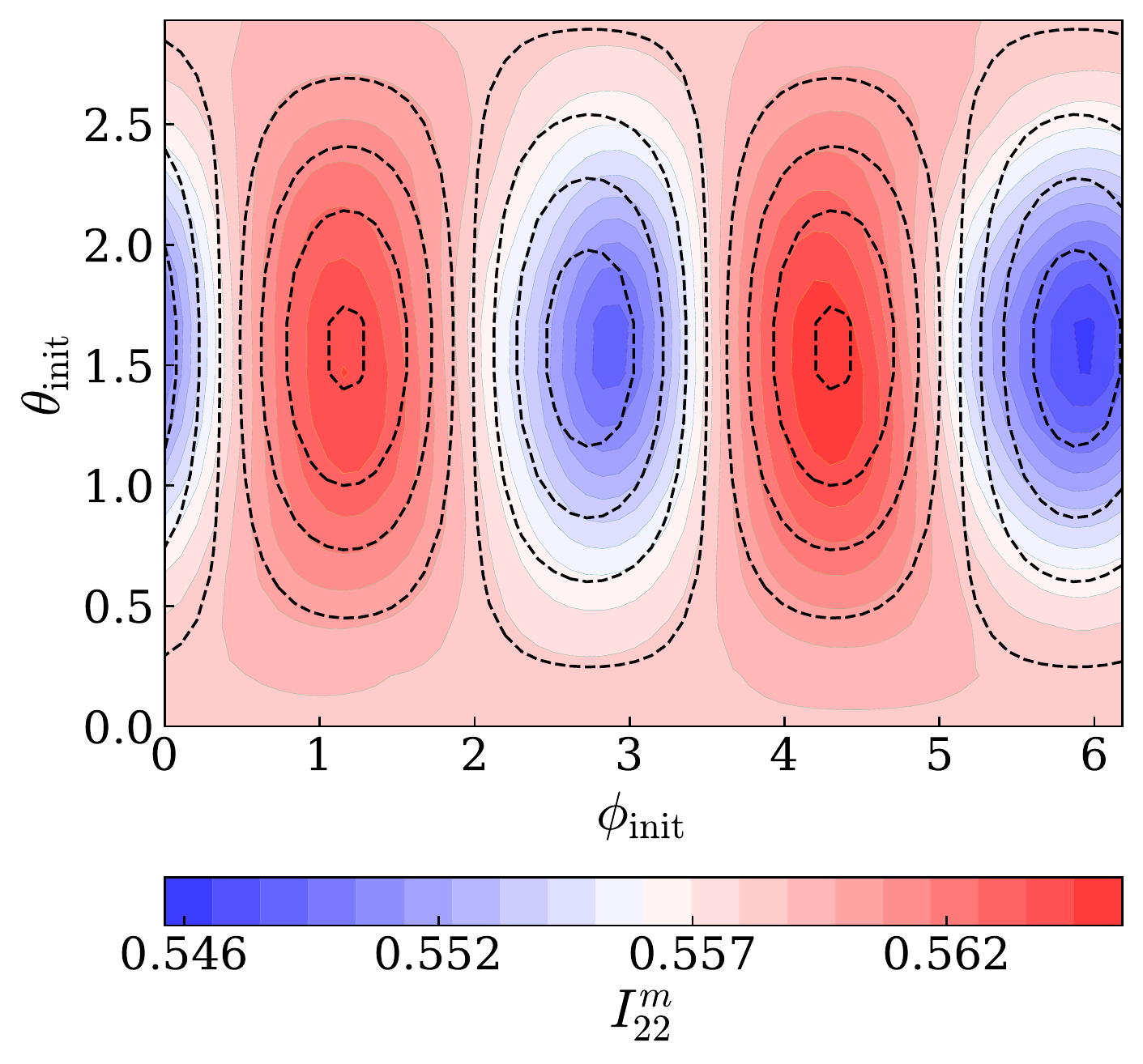} 
  \caption{The peak value of mass quadrupole wave $\lp$ as a function of $(\theta_{\rm init},\phi_{\rm init})$, with $\chi_{\rm init}=0.8$ (SKd configuration). Results are from NRSur7dq4. The pattern is symmetric about $\theta_{\rm init}=\pi/2$, and has a period $\pi$ in the $\phi_{\rm init}$-direction, consistent with Eq.~(\ref{IS-map}). The contours with dashed lines are  the prediction of PN-inspired counterpart in Eq.~(\ref{I22-fit-leading}).}
 \label{fig:mass-2d-distr-no-spin}
\end{figure}


To explore a larger parameter space, we use NRSur7dq4 and plot $\lp$ as a function of $(\theta_{\rm init},\phi_{\rm init})$ in Fig.~\ref{fig:mass-2d-distr-no-spin}, with $\chi_{\rm init}=0.8$. The pattern exhibits quadrupolar structure, i.e., symmetric about $\theta_{\rm init}=\pi/2$, and has a period $\pi$ in the $\phi_{\rm init}$-direction. This is consistent with what we obtained in Eq.~(\ref{IS-map}). 

To have a better understanding of $I_{22}^m$, we use PN prediction of mass quadrupole wave during the inspiral stage \cite{Porto:2010zg}
\begin{align}
\mathcal{I}_{22}=\mathcal{I}_{\rm orb}+\mathcal{I}_{S1}+\mathcal{I}_{S2}, \label{I22-total}
\end{align}
where
\begin{subequations}
\begin{align}
&\mathcal{I}_{\rm orb}=-\frac{M}{2}\sqrt{\frac{2\pi}{5}}r^2e^{i\phi}, \\
&\mathcal{I}_{S1}= \sqrt{\frac{2\pi}{5}}\frac{M^3}{16}\chi^2_{\rm init}\sin^2\theta_{\rm init}e^{-2i\phi_{\rm init}}, \label{Is1}\\
&\mathcal{I}_{S2}= -\sqrt{\frac{2\pi}{5}}\frac{M^3}{16}\chi_{\rm init}^2v^2, \label{Is2}
\end{align}%
\label{PN-mass-peak}%
\end{subequations}
with $v\sim\sqrt{M/r}$ the velocity of an individual BH. In Eq.~(\ref{I22-total}), the leading contribution from the orbital sector $\mathcal{I}_{\rm orb}$ is modified by the spin sector $\mathcal{I}_{S1,2}$. As discussed in Ref.~\cite{Kamaretsos:2011um,Kamaretsos:2012bs}, the amplitudes of ringdown waveforms in different $(\ell,m)$ modes are related to those of the corresponding modes during the inspiral stage. Therefore, we can write down a fitting formula for $I_{22}^m-(\chi_{\rm init},\theta_{\rm init},\phi_{\rm init})$ relation, inspired by Eq.~(\ref{I22-total}) and the definition of $I_{22}$ in Eq.~(\ref{rad-moment}),
\begin{align}
&\lp=Q_{\rm orb}+Q_{S1}\chi_{\rm init}^2\sin^2\theta_{\rm init}\sin2(\phi_{\rm init}+\phi_0)\notag \\
&+Q_{S2}\chi_{\rm init}^2 , \label{I22-fit-leading}
\end{align}
where $Q_{\rm orb},Q_{S1},Q_{S2}$ are constants. Their fitted values are listed in Table \ref{table:mass-fit-value}, and the contours of Eq.~(\ref{I22-fit-leading}) are plotted as dashed lines in Fig.~\ref{fig:mass-2d-distr-no-spin}. We note that Eq.~(\ref{I22-fit-leading}) was applied to understand the features of $\mathscr{A}_{n}^{(I)}$ in Sec.~\ref{sec:theta_phi_scott} [see Eq.~(\ref{overtone_fit})], where we used the fact that $\mathscr{A}_{n}^{(I)}$ is insensitive to the overtone index $n$ and we ignored the mixing between overtones.

Three terms in Eq.~(\ref{I22-fit-leading}) correspond to $\mathcal{I}_{\rm orb}$, $\mathcal{I}_{S1}$ and $\mathcal{I}_{S2}$, respectively. They imply that
\begin{subequations}
\begin{align}
&Q_{\rm orb}/Q_{S1}\sim|\mathcal{I}_{\rm orb}|\chi_{\rm init}^{2}/|\mathcal{I}_{S1}|\sim8\frac{r^2}{M^2}\sim72,  \\
&Q_{S2}/Q_{S1}\sim|\mathcal{I}_{S2}^{ij}|\sin^2\theta_{\rm init}/|\mathcal{I}_{S1}^{ij}|\sim v^2\sim0.3,
\end{align} 
\label{mass-I-ratio}%
\end{subequations}
where the formula is evaluated at $r=3M$, i.e., the radius of the light ring. In fact, values in Eq.~(\ref{mass-I-ratio}) are close to the fitted result listed in Table \ref{table:mass-fit-value}. Therefore, the peak of mass quadrupole momentum $\lp$, as an important characteristic of the ringdown phase, is still qualitatively consistent with the prediction of PN theory.

Although Eq.~(\ref{I22-fit-leading}) can predict the major pattern of $\lp-(\theta_{\rm init},\phi_{\rm init})$ relation, a correction term
\begin{align}
\sim \chi_{\rm init}^4\sin^4\theta_{\rm init} f(\sin\phi_{\rm init},\cos\phi_{\rm init}),
\end{align}
is still needed if one wants to further recover sub-leading features. Here $f(\sin\phi_{\rm init},\cos\phi_{\rm init})$ is a function of $\phi_{\rm init}$, corresponding to higher PN correction. 


\begin{table}
    \centering
    \caption{The coefficients in Eq.~(\ref{I22-fit-leading}) by fitting  to the NRSur7dq4 data. The values of $Q_{\rm orb}/Q_{S1}$ and $Q_{S2}/Q_{S1}$ are close to the PN predictions in Eqs.~(\ref{mass-I-ratio}).}
    \begin{tabular}{c c c c c c c c} \hline\hline
$Q_{\rm orb}$ & $Q_{S2}$  & $Q_{S1}$ & $\tan2\phi_0$ & $Q_{\rm orb}/Q_{S1}$ & $Q_{S2}/Q_{S1}$ \\ \hline
0.557 & $2.72\times10^{-3}$ & $12.2\times10^{-3}$ & $-0.98$ & 45.7 & 0.22 \\ \hline\hline
     \end{tabular}
     \label{table:mass-fit-value}
\end{table}

\subsection{The peak of current quadrupole wave $\spp$}
\label{sec:spp-surrogate}
\begin{figure}[htb]
        \includegraphics[width=\columnwidth,height=7.1cm,clip=true]{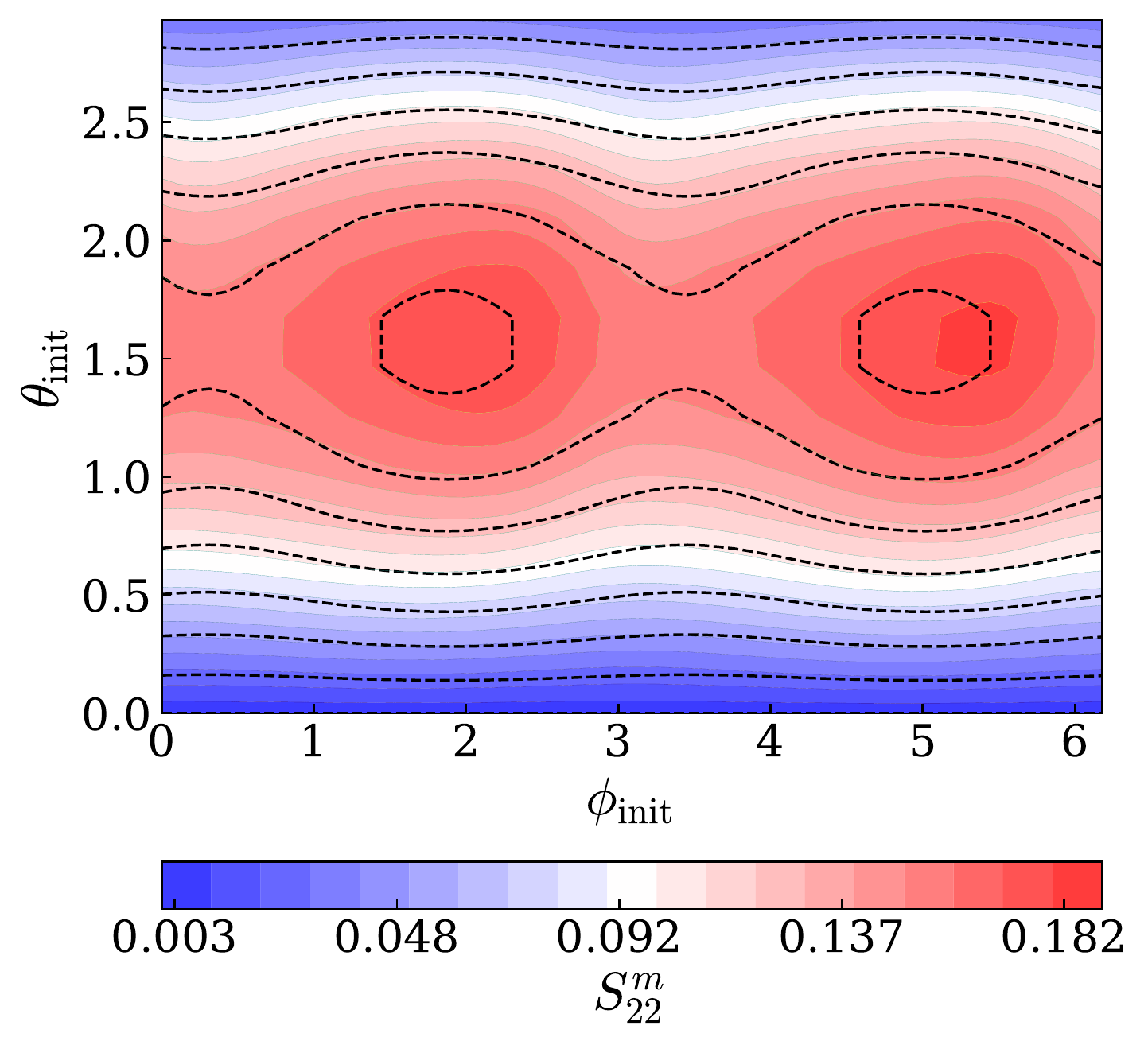}
  \caption{The peak value of current quadrupole wave $\spp$ as a function of $(\theta_{\rm init},\phi_{\rm init})$, with $\chi_{\rm init}=0.8$ (SKd configuration). The data are from NRSur7dq4, while dashed lines are the prediction of the PN-inspired counterpart in Eq.~(\ref{s-pn}).}
 \label{fig:curr-2d-distr-hangup}
\end{figure}

We now turn our attention to $\spp$. In Fig.~\ref{fig:curr-2d-distr-hangup}, we use NRSur7dq4 and plot $\spp-(\theta_{\rm init},\phi_{\rm init})$ with $\chi_{\rm init}=0.8$. The pattern is still symmetric about $\theta_{\rm init}=\pi/2$ and has a period $\pi$ in the $\phi_{\rm init}$-direction, consistent with Eq.~(\ref{IS-map}). We repeat our previous process and use PN predictions to understand the pattern. With PN theory, we have \cite{Porto:2010zg}
\begin{align}
\mathcal{S}_{22}=\mathcal{S}_{22}^{(1)}+\mathcal{S}_{22}^{(2)},
\end{align}
where
\begin{subequations}
\begin{align}
&\mathcal{S}_{22}^{(1)}\sim \chi_{\rm init}\sin\theta_{\rm init} r e^{-i\phi_{\rm init}}, \\
&\mathcal{S}_{22}^{(2)}\sim -\chi_{\rm init}rv^2\sin\theta_{\rm init}\cos\phi_{\rm init}.
\end{align}
\label{PN-curr-peak}%
\end{subequations}
Eqs.~(\ref{PN-curr-peak}) lead to a fitting formula
\begin{align}
S_{22}^{m2}=\chi_{\rm init}^2\sin^2\theta_{\rm init}[Q^{(1)}+Q^{(2)}\sin2(\phi_{\rm init}+\phi_S)], \label{s-pn}
\end{align}
where $Q^{(1)}$ and $Q^{(1)}$ correspond to $\mathcal{S}_{22}^{(1)}$ and $\mathcal{S}_{22}^{(2)}$, respectively. The fitted value of $Q^{(2)}$ and $Q^{(1)}$ are $9.43\times10^{-3}$ and $4.28\times10^{-2}$. The ratio, $Q^{(2)}/Q^{(1)}\sim 0.22$, is close to $v^2$ at the light ring (0.33), which is again consistent with the PN prediction $Q^{(2)}/Q^{(1)}\sim v^2$. Therefore, the peak of current quadrupole wave $\spp$ also inherits information from the PN regime.


\subsection{The phase difference $\Delta\Phi_{\rm IS}$}
\label{sec:multipolar-phi}

\begin{figure}[htb]
        \includegraphics[width=\columnwidth,height=6.4cm,clip=true]{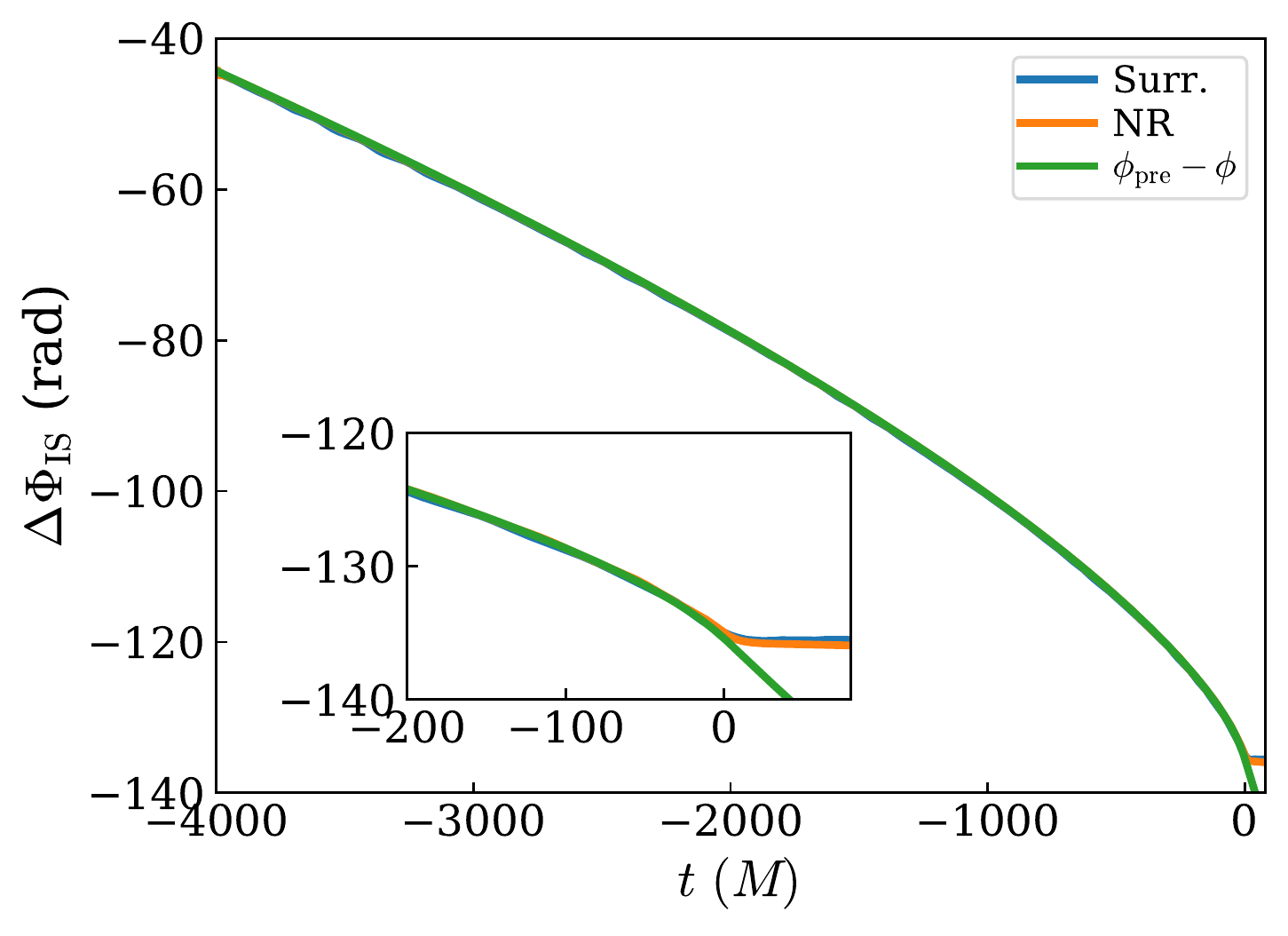}
  \caption{The time evolution of $\dphi$ for the SKd4--`03' system (orange curve). It is compared to NRSur7dq4 (blue curve) with the same initial condition. As expected, their results are close. Within the inspiral regime, PN theory predicts $\dphi=\phi_{\rm pre}-\phi$, which is shown as the green curve.}
 \label{fig:mass-curr-phase-diff}
\end{figure}

We finally study the phase difference between the mass and current quadrupole waves $\dphi$, which is the key factor that determines the final kick velocity. Fig.~\ref{fig:mass-curr-phase-diff} is $\dphi$ of the SKd4--`03' system (Table \ref{table:NR-pars}). During the inspiral stage, $\dphi$ accumulates monotonically over time. It then gradually settles down to a constant after the merger. In fact, one can use PN theory to understand the evolution of $\dphi$. Before the merger, we have [cf.~Eqs.~(\ref{pn-SI})]
\begin{align}
\dphi=\phi_{\rm pre}-\phi, \label{dphase-pn}
\end{align}
Here $\phi_{\rm pre}$ is the precession phase of the spins, and is  obtained by measuring the spins of  each individual BH; $\phi$ is the orbital phase. 
%
In Fig.~\ref{fig:mass-curr-phase-diff}, we compare Eq.~(\ref{dphase-pn}) to the NR result. Two results agree pretty well until $t\sim -50M$. Near the merger, $\phi_{\rm pre}$ is thought to be locked to $\phi$ \cite{Nichols:2011ih}, in order for the accumulation of $\dphi$ to be halted.  
An alternative way to think of this is based on the QNM decomposition. For the ringdown portion of $I_{22}$ and $S_{22}$, they must both be decomposed into $(2,2)$ QNMs. After higher overtones decay away $(t>20M)$, we are left with the fundamental mode [see Eq.~(\ref{IS-overtone}) for more details]
\begin{align}
&I_{22}\sim \mathscr{A}_0^{(I)}e^{i\varphi_0^{(I)}}e^{-i\omega_{220}t}, &S_{22}\sim \mathscr{A}_0^{(S)} e^{i\varphi_0^{(S)}}e^{-i\omega_{220}t},
\end{align}
which leads to $\dphi=\varphi_0^{(I)}-\varphi_0^{(S)}$, i.e., a constant. The fact that both $I_{22}$ and $S_{22}$ have the same QNM frequency is a consequence of the {\it isospectrality} feature of black holes.

\begin{figure}[htb]
        \includegraphics[width=\columnwidth,height=8.5cm,clip=true]{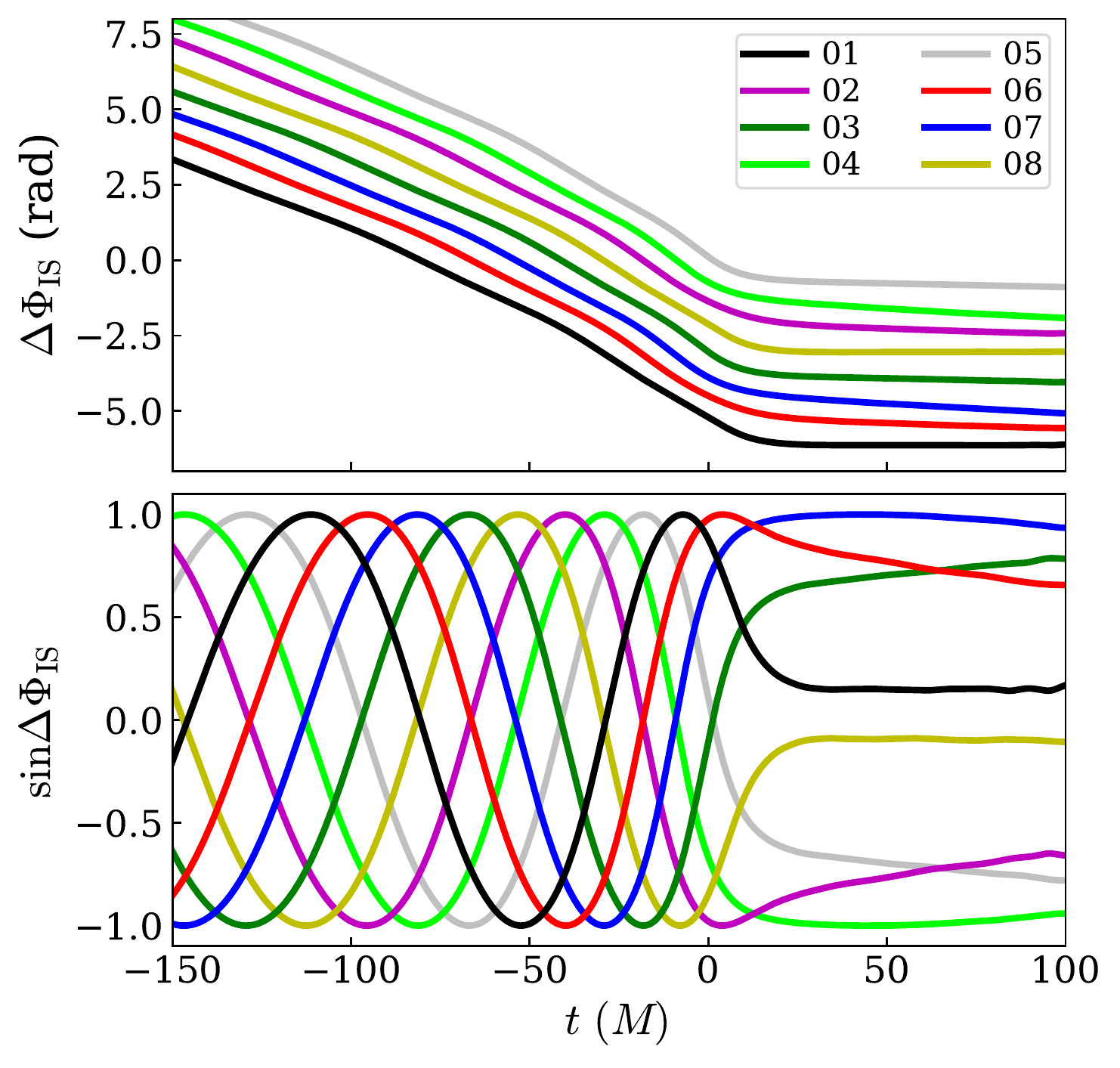}
  \caption{The evolution of $\dphi$ for SKd4 systems (Table \ref{table:NR-pars}). Eight runs start with different $\phi_{\rm init}$, and their $\dphi$ are finally locked to different values. The bottom panel is $\sin\dphi$. One can directly estimate the kick velocity from the final value of $\sin\dphi$, since the kick is roughly proportional to the integration of $\sin\dphi$ [Eq.~(\ref{vfinal})]. As for high-kick cases, their $\dphi$ change slowly during the late post-merger stage. This is due to the Doppler shift.}
 \label{fig:mass-curr-phase-diff-NRruns}
\end{figure}

Then we study how $\dphi$ depends on the progenitor's parameters. We first choose eight NR runs in Table \ref{table:NR-pars}, whose $\phi_{\rm init}$ are different. As shown in Fig.~\ref{fig:mass-curr-phase-diff-NRruns}, $\dphi$ with different $\phi_{\rm init}$ are finally locked to different values. The bottom panel is $\sin\dphi$. Recalling that the kick velocity can be roughly estimated by integrating $\sin\dphi$ [Eq.~(\ref{vfinal})], the final value of $\sin\dphi$ is a strong signature for the final kick velocity. For instance, one can directly read that SKd4--`07' leads to a positive largest kick, consistent with NR results (Table \ref{table:NR-pars}). Interestingly, $\dphi$ of several runs (e.g., `03') do not settle into a constant. Instead, there are slow changes over time. This is because the final BHs are boosted with relatively large kick velocities. As a result, there is a Doppler shift between the mode frequency of $h_{22}$ and of $h_{2,-2}$, recalling that $h_{22}$ is dominantly emitted upward, while $h_{2,-2}$ downward \cite{Gerosa:2016vip}. To test our statement, we pick four of SKu systems that are listed in Table \ref{table:NR-pars-all}. Here we choose SKu systems since they lead to larger kicks, thus the comparison is less impacted by numerical noises. The results are summarized in Table \ref{table:fit-overtone-kick}. We can see relative mass differences are close to the kick of final BHs. A slight difference in mass leads to a deviation between the mode frequency of $h_{22}$ and $h_{2,-2}$, i.e., [see Eq.~(\ref{qnm-overtone})]
\begin{subequations}
\begin{align}
    &\h_{22}\sim\mathcal{A}_{220}e^{i\psi_{220}} e^{-i\omega_{22 0}(1+\delta)t}, \\
    &\h_{2,-2}^*\sim\mathcal{A}_{2,-20}e^{-i\psi_{2,-20}} e^{-i\omega_{22 0}(1-\delta)t},
\end{align}
\label{h_late_time_overtone}%
\end{subequations}
where $\delta$ is a small parameter, and is proportional to the kick velocity. In the late time regime, Eq.~(\ref{h_late_time_overtone}) implies
\begin{align}
    \sin\dphi&=\sin[\varphi^{(I)}_0-\varphi^{(S)}_0]\notag\\
    &+\frac{4(\mathcal{A}_{220}^2-\mathcal{A}_{2,-20}^2)\mathcal{A}_{220}\mathcal{A}_{2,-20}}{|\mathcal{A}_{220}^2e^{-i(\psi_{220}+\psi_{2,-20})}-\mathcal{A}_{2,-20}^2e^{i(\psi_{220}+\psi_{2,-20})}|^2}\notag \\
    &\times\cos(\psi_{220}+\psi_{2,-20})\omega_{220}t\delta+\mathcal{O}(\delta^2).
\end{align}
The new term above gives rise to a linear change in time, and it is consistent with the feature which we observe in Fig.~\ref{fig:mass-curr-phase-diff-NRruns}.

We then use NRSur7dq4 to explore more parameter space of SKd systems, by varying three free parameters $\chi_{\rm init}$, $\theta_{\rm init}$ and $\phi_{\rm init}$, respectively. Results are shown in the bottom row of Fig.~\ref{fig:phasediff-surr}. We can see that $\chi_{\rm init}$ and $\theta_{\rm init}$ do not affect $\dphi$, even near and after the merger. Meanwhile, $\phi_{\rm init}$ gives rise to only a constant phase shift for $\dphi$, consistent with what we obtained in Sec.~\ref{sec:universality} [Eq.~(\ref{delta_phi_IS_universal})]. In fact, if we subtract $\phi_{\rm init}$ from $\dphi$, the rest of time dependence is still insensitive to $\phi_{\rm init}$, although not as good as the cases of $\chi_{\rm init}$ and $\theta_{\rm init}$. 

\begin{figure}[htb]
        \includegraphics[width=\columnwidth,height=7.1cm,clip=true]{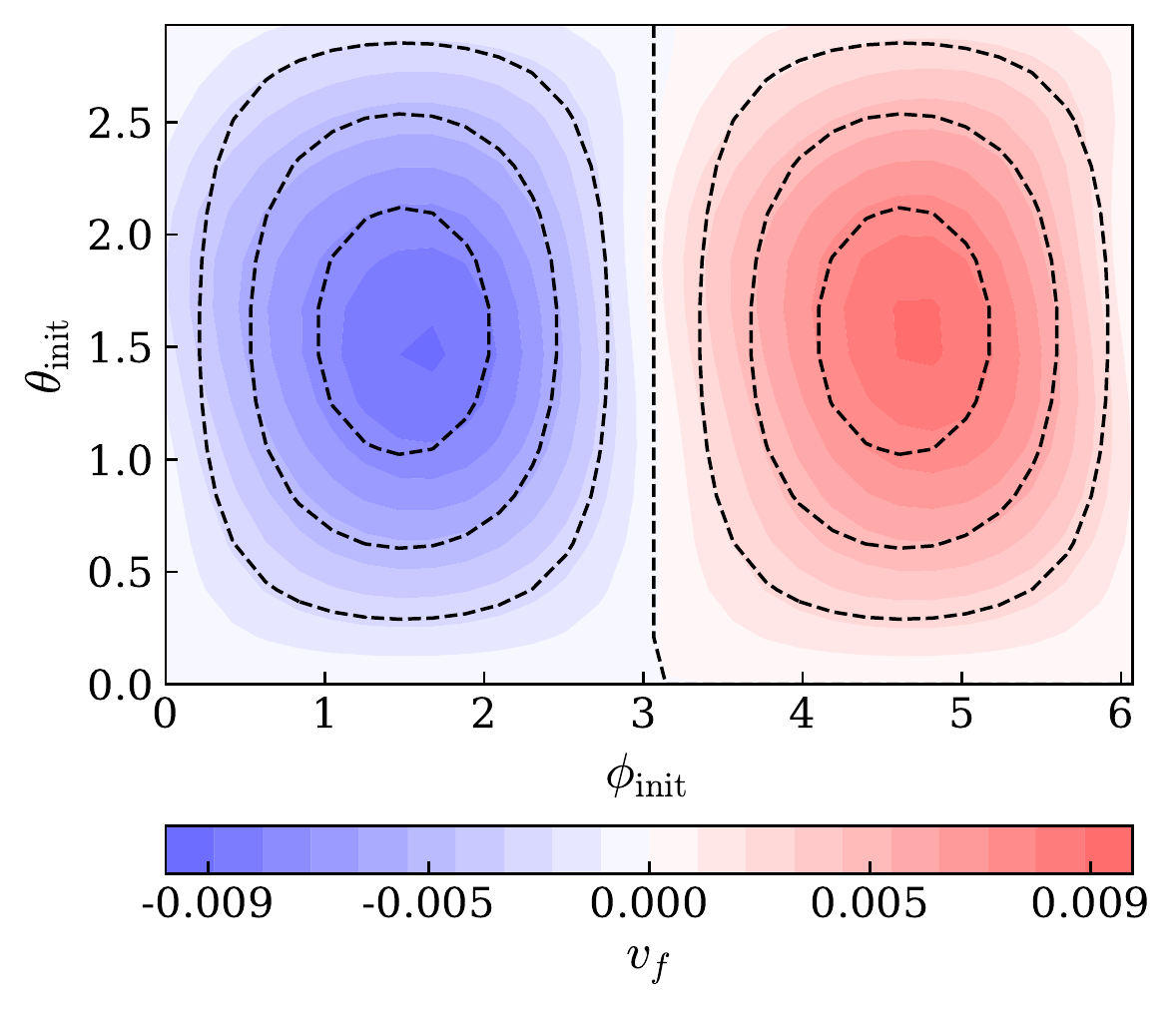}
  \caption{The final kick velocity as a function of $(\theta_{\rm init},\phi_{\rm init})$, predicted by NRSur7dq4Remnant. The component spin $\chi_{\rm init}$ is chosen to be 0.76. The contours with dashed lines are from Eq.~(\ref{v_IS_predict_mag})}
 \label{fig:vf_theta_phi}
\end{figure}

Recalling that the final kick velocity is given by [Eqs.~(\ref{vfinal}) and (\ref{universality})]
\begin{align}
    &v_f\sim {\rm Im}~\int \dot{I}_{22}\dot{S}_{22}^*dt\sim I_{22}^mS_{22}^{m}{\rm Im}~e^{i\phi_{\rm init}} \int \dot{T}_I(t)\dot{T}_S^*(t) dt \notag \\
    &\sim \chi_{\rm init}\sin\theta_{\rm init}\sin(\phi_{\rm init}-\phi_{\rm init}^{(0)}), 
    \label{v_IS_predict_mag}
\end{align}
where we have used the leading terms in Eq.~(\ref{I22-fit-leading}) and (\ref{s-pn}). This result is the same as Eq.~(\ref{map-kick}), as discussed in \cite{Gonzalez:2007hi,Campanelli:2007cga,Campanelli:2007ew}. To offer an illustration, we use NRSur7dq4Remnant to plot $v_f$ as a function of $(\theta_{\rm init},\phi_{\rm init})$ in Fig.~\ref{fig:vf_theta_phi}, with $\chi_{\rm init}=0.76$. Meanwhile, we use Eq.~(\ref{v_IS_predict_mag}) to fit the $v_f-(\chi_{\rm init},\theta_{\rm init},\phi_{\rm init})$ dependence, and the result is shown as dashed lines in Fig.~\ref{fig:vf_theta_phi}. We can see Eq.~(\ref{v_IS_predict_mag}) works properly.

\begin{table}
    \centering
    \caption{The mass of remnant BHs inferred from $h_{22}$ and $h_{2,-2}$, by fitting with QNMs (7 overtones, see Sec.~\ref{sec:overtone} for more details). Four runs below are in the SKu configuration (Table \ref{table:NR-pars-all}) with $\chi_{\rm init}=0.8$ and different $\phi_{\rm init}$. Among them, SKu8--`02' and SKu8--`04' are high-kick cases. The masses inferred from $h_{22}$ and $h_{2,-2}$ are quite different, and the relative difference are close to the final kick velocity.}
    \begin{tabular}{c c c c c} \hline\hline
Runs (SKu8) & `01' & `02' & `03' & `04' \\ \hline
Mass from NR & 0.941 & 0.939 & 0.941 & 0.939 \\ \hline
Mass from $h_{2,2}$ & 0.940 &  0.945 & 0.941 & 0.931 \\ \hline
Mass from $h_{2,-2}$ & 0.940 &  0.931 & 0.940 & 0.945 \\ \hline
Relative mass  & \multirow{2}{*}{$-7.8\times10^{-5}$} & \multirow{2}{*}{$-0.015$} & \multirow{2}{*}{$3.3\times10^{-4}$} & \multirow{2}{*}{$0.015$} \\ 
difference between $h_{2,\pm2}$ \\ \hline
Final kick & $-1.6\times10^{-3}$ & $-0.011$ & $1.3\times10^{-3}$& 0.011 \\ \hline\hline
     \end{tabular}
     \label{table:fit-overtone-kick}
\end{table}

\section{Backwards One-Body model}
\label{sec:BOB}
In this section, we shall focus on the time evolution of the mass and current quadrupole waves, $T_I(t)$ and $T_S(t)$, as defined in Eq.~(\ref{universality}). In particular, we use an analytic phenomenological model BOB, conceived by McWilliams \cite{McWilliams:2018ztb}, to model the ringdown evolution. We first give a brief introduction to BOB in Sec.~\ref{sec:BOB-intro}, and then compare it to NR results in Sec.~\ref{sec:BOB-compare}. 


\begin{figure*}[htb]
        \includegraphics[width=1.3\columnwidth,height=6.cm,clip=true]{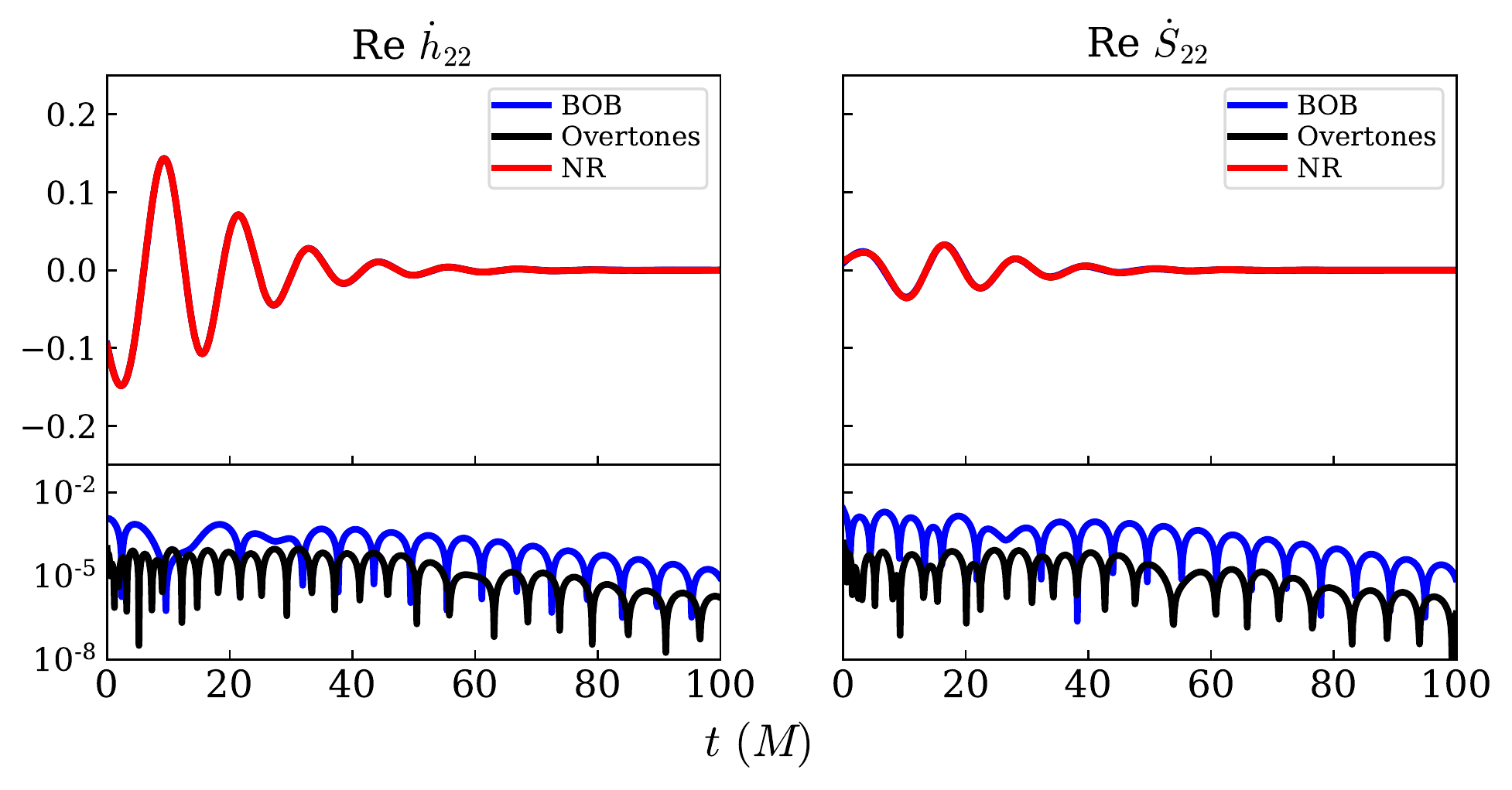}
        \includegraphics[width=0.7\columnwidth,height=5.cm,clip=true]{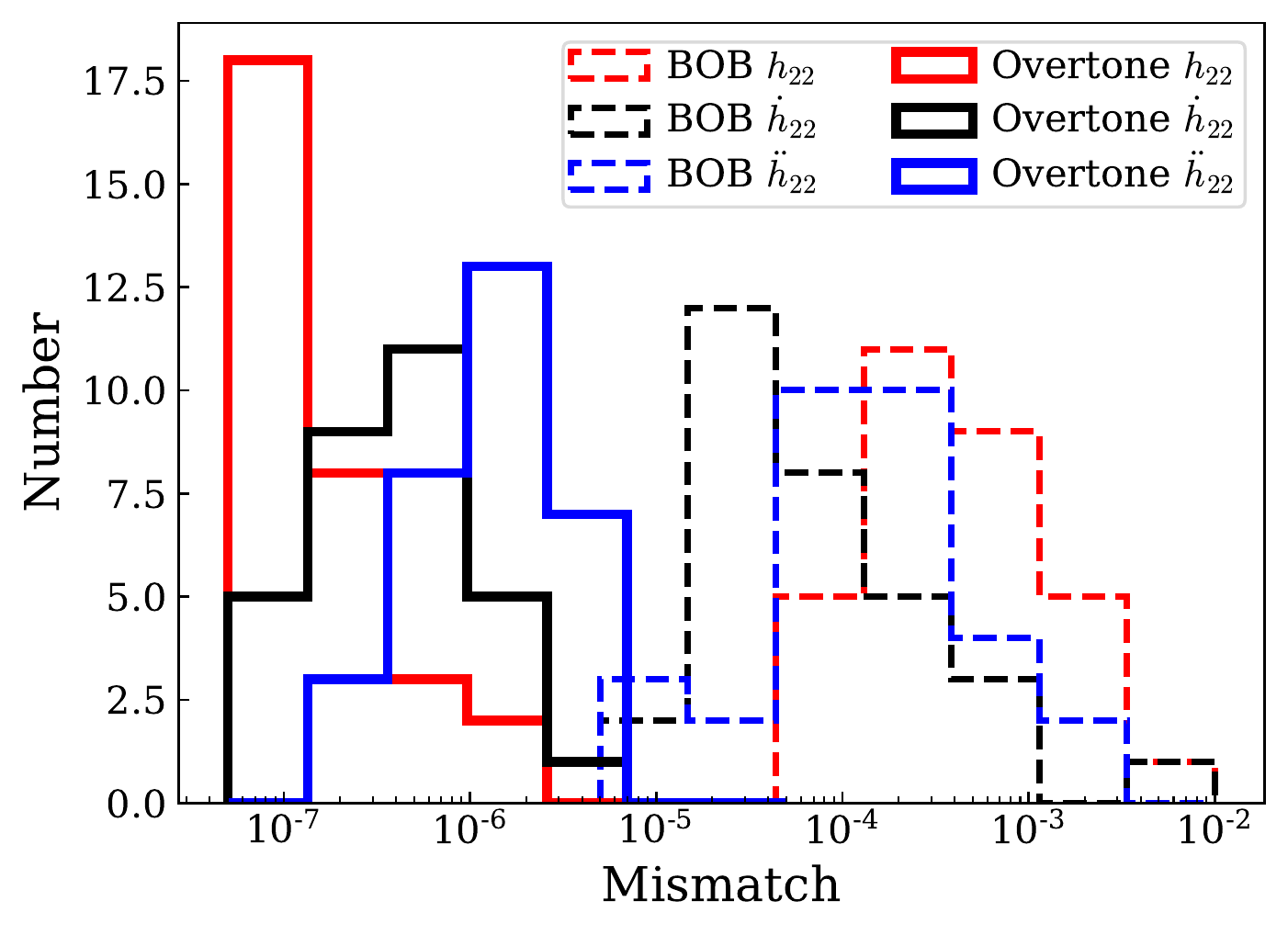}\\
        \includegraphics[width=1.3\columnwidth,height=6cm,clip=true]{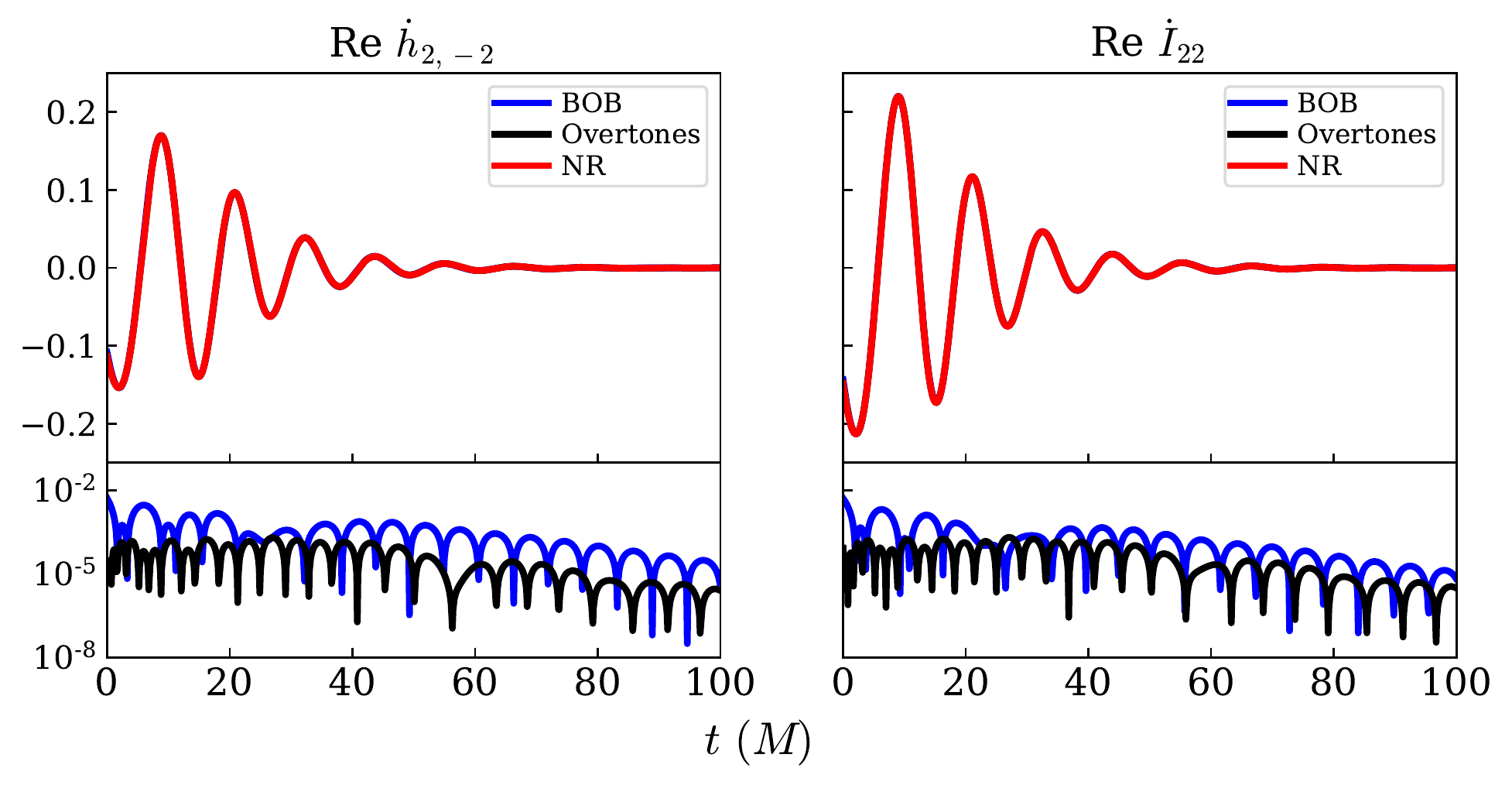} 
         \includegraphics[width=0.7\columnwidth,height=5.cm,clip=true]{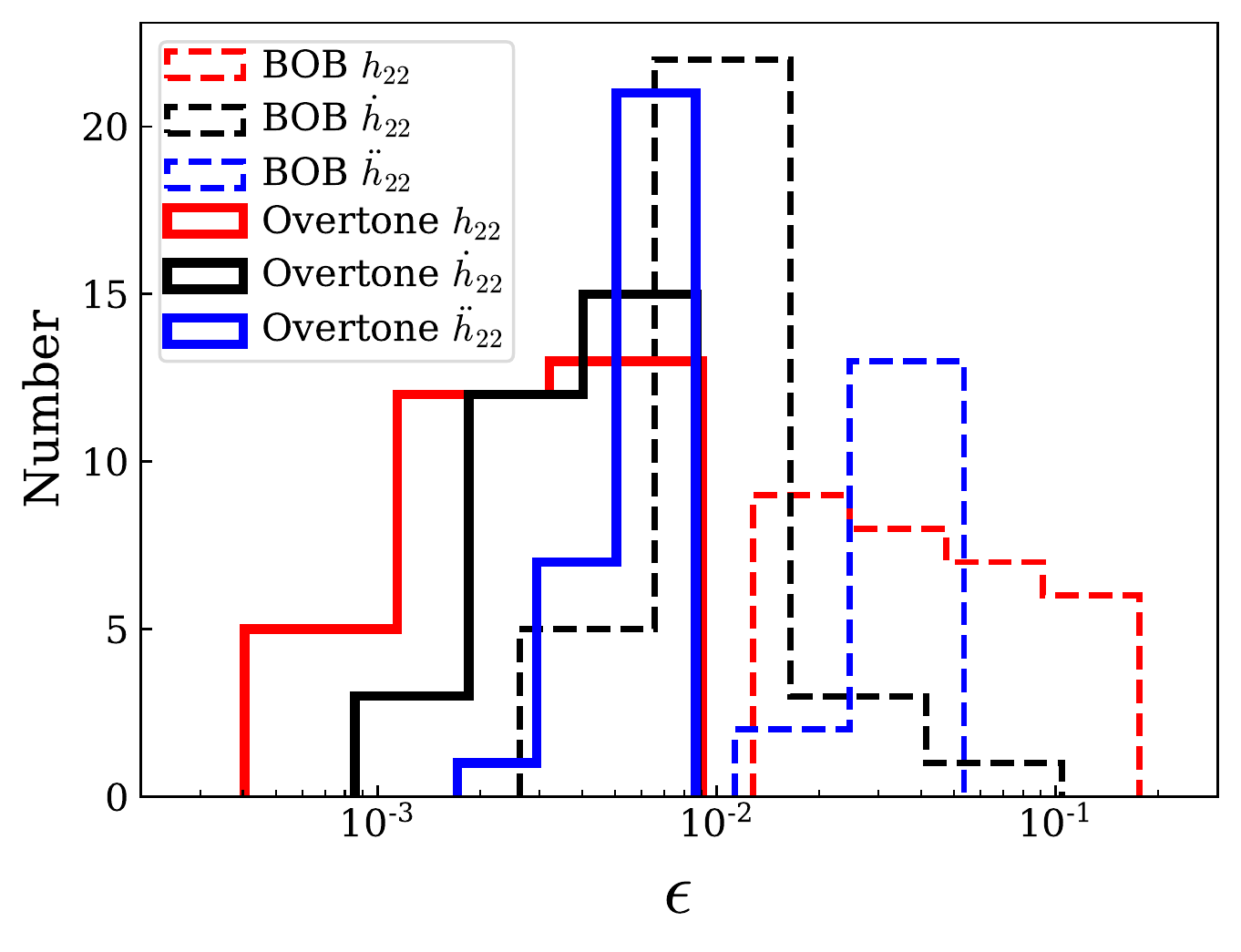}
  \caption{The BOB model for $\dot{h}_{22}$, $\dot{h}_{2,-2}$, $\dot{I}_{22}$ and $\dot{S}_{22}$ (the left and middle columns). They are compared to the ringdown portion of SKd4--`03'. We also fit data with QNMs. The residuals of BOB for four variables are all on order of $\sim10^{-3}$, an order of magnitude worse than the
fitting of QNMs. The right column corresponds to the distribution of mismatch [top panel, see Eq.~(\ref{mismatch})], and parameter deviation [bottom panel, see Eq.~(\ref{bob-par-dev})] for QNM decomposition and BOB, using our NR simulations listed in Table \ref{table:NR-pars} and \ref{table:NR-pars-all}. BOB is always worse than QNM fitting.}
 \label{fig:BOB-fit}
\end{figure*}

\subsection{A brief review of BOB}
\label{sec:BOB-intro}
The BOB model is an accurate, fully analytical GW waveform model for the late inspiral, merger and ringdown of BBH \cite{McWilliams:2018ztb}, which is able to match the waveform $\sim20M$ before the peak of strain. This feature enables people to avoid the extrapolation of inspiral models beyond their domain of validity. Here we restrict our attention to the ringdown portion.

As discussed in Refs.\ \cite{Baker:2008mj,McWilliams:2018ztb}, the amplitude of the News $|\dot{h}_{\ell m}|$ is related to its  frequency $\Omega_{\ell m}$ by
\begin{align}
|\dot{h}_{\ell m}|^2\propto \frac{d}{dt}\Omega_{\ell m}^2, \label{BOB-h-omega}
\end{align}
where the coefficient remains (approximately) constant throughout the merger and ringdown phase. It was found\footnote{In Ref.~\cite{McWilliams:2018ztb}, the author pointed out that this phenomenological formula works best for $|\ddot{h}_{\ell m}|$. For now, we try to make our statement general, and make comparisons later.} that either $|h_{2,\pm2}|$, $|\dot{h}_{2,\pm2}|$ or $|\ddot{h}_{2,\pm2}|$ can be modeled by 
\begin{align}
X\sech[\gamma (t-t_p)], \label{bob-intro}
\end{align}
with two free variables $X$ and $t_p$, where $\gamma=-{\rm Im} ~\omega_{220}$ is the decay rate of the fundamental mode, determined by the final mass $m_f$ and spin $\chi_f$. Applying Eq.~(\ref{bob-intro}) to $|\ddot{h}_{2,\pm2}|$, $|\dot{h}_{2,\pm2}|$, and $|h_{2,\pm2}|$ leads to three classes of BOB. Below we shall discuss the model for $\dot{h}_{2,\pm2}$, and refer the interested reader to Appendix \ref{app:BOB}  for $\ddot{h}_{2,\pm2}$ and $h_{2,\pm2}$.

We first write the News $\dot{h}_{22}$ as
\begin{align}
\dot{h}_{22}=X\sech[\gamma (t-t_p)] e^{-i\phi_{22}(t)}. \label{BOB-h22dot}
\end{align}
where $X$ is essentially the peak value of $|\dot{h}_{22}|$, and $t_p(>0M)$ is its peak time. Using the relation in Eq.~(\ref{BOB-h-omega}), we obtain
\begin{align}
\dot{\phi}_{22}(t)=\Omega_{22}(t)=\left\{\Omega_0^2+\frac{\omega_0^2-\Omega_0^2}{2}\left[\tanh\gamma(t-t_p)+1\right]\right\}^{1/2}, \label{omega-I}
\end{align}
where $\Omega_{0}$ is an integration constant and $\omega_0={\rm Re}~\omega_{220}$.
Eq.~(\ref{omega-I}) indicates that 
\begin{align}
\lim_{t\to\infty}\dot{\phi}_{22}(t)=\omega_0, \label{bob-hdot-fre-asy}
\end{align}
i.e., $\dot{h}_{22}$ oscillates at the fundamental QNM frequency during the late time of post-merger portion. Integrating Eq.\ (\ref{omega-I}) again gives
\begin{align}
\phi_{22}=\frac{1}{\gamma}\left(\omega_0\arctanh\frac{\dot{\phi}_{22}}{\omega_0}-\Omega_0\arctanh\frac{\Omega_0}{\dot{\phi}_{22}}\right)-\phi_0, \label{bob-phase-hdot}
\end{align}
where $\phi_0$ is another integration constant. We can see that $\dot{h}_{22}$ depends on 6 parameters 
\begin{align}
X, m_f, \chi_f, t_p, \Omega_0, \phi_0.
\end{align}
Similarly, Eqs.~(\ref{BOB-h22dot})--(\ref{bob-phase-hdot}) can also be applied to $\dot{h}^*_{2,-2}$, $\dot{I}_{22}$ and $\dot{S}_{22}$.

As $t\gg t_p$, we obtain an asymptotic expansion for $\dot{h}_{22}$
\begin{align}
\dot{h}_{22}=(2Xe^{\gamma t_p}) e^{i\psi_0}e^{-i\omega_{220}t}, \label{I22-aym}
\end{align}
where 
\begin{align}
\psi_0=\omega_0 t_p+\phi_0+\frac{\Omega_0}{\gamma}\arctanh\frac{\Omega_0}{\omega_0}-\frac{\omega_0}{\gamma}\frac{1}{2}\log\frac{4\omega_0^2}{\omega_0^2-\Omega_0^2}.
\end{align}
By comparing Eq.~(\ref{I22-aym}) with the overtone decomposition [e.g.~Eq.~(\ref{qnm-overtone})], $(2Xe^{\gamma t_p})$ is supposed to be equal to $|\omega_{220}\mathcal{A}_{220}|$.

\begin{table}
    \centering
    \caption{Fitting $\dot{h}_{22}$, $\dot{h}_{2,-2}$, $\dot{I}_{22}$, $\dot{S}_{22}$ to the BOB model, respectively. The original data is the ringdown portion of SKd4--`03'. The first four rows are the free parameters of BOB: peak magnitude $X$, peak time $t_p$, final spin $\chi_f$, and final mass $m_f$. Comparing with the NR prediction of final spin (0.685) and final mass (0.951), the BOB for $\dot{I}_{22}$ and $\dot{h}_{22}$ are more accurate to recover the final properties than the other two. The model for $\dot{S}_{22}$ is the worst. Using the BOB's asymptotic expansion in the late time limit [Eq.~(\ref{I22-aym})], $2Xe^{\gamma t_p}$ (the sixth row) is expected to be equal to $|\omega_{220}\mathcal{A}_{220}|$ (the seventh row). The agreement for $\dot{S}_{22}$ is the worst.  The last row is the mismatch between BOB and the original NR data. }
    \begin{tabular}{c c c c c} \hline\hline
& $\dot{h}_{22}$& $\dot{h}_{2,-2}$ & $\dot{I}_{22}$  &$\dot{S}_{22}$  \\ \hline
$X$ &0.153  &0.171  &0.227  &0.035  \\ \hline
$t_p/M$ &5.13  &7.31  &6.21  &12.53  \\ \hline
$\chi_f$ &0.684  &0.681  &0.686 & 0.559 \\ \hline
$m_f/M$& 0.954  &0.944  &0.951 &0.857  \\ \hline
$2Xe^{\gamma t_p}$& 0.474& 0.644 &0.771 &0.247  \\ \hline
$|\omega_{220}\mathcal{A}_{220}|$ &0.470 &0.622 &0.759 &0.175 \\ \hline
Mismatch $(\times10^{-5})$& 2.6 & 9.9 & 3.1 & 204.0  \\ \hline\hline
     \end{tabular}
     \label{table:BOB-fit}
\end{table}

\subsection{Numerical comparisons}
\label{sec:BOB-compare}
In this subsection, we use our NR simulations (Tables \ref{table:NR-pars} and \ref{table:NR-pars-all}) to study the accuracy of BOB. To begin with, we take the ringdown portion of SKd4--`03' (Tables \ref{table:NR-pars}), and fit $\dot{h}_{2,\pm2}$, $\dot{I}_{22}$, $\dot{S}_{22}$ to Eq.~(\ref{BOB-h22dot}), respectively. Similar to the previous QNM fitting algorithm (Sec.~\ref{sec:overtone}), we fit $X$ and $\phi_0$ with unweighted linear least squares, and fit $m_f$, $\chi_f$, $t_p$, $\Omega_0$ with nonlinear least squares. To give a comparison, we also fit the ringdown sector with QNMs. As shown in Fig.~\ref{fig:BOB-fit}, the BOB can capture the major feature of  $\dot{h}_{2,\pm2}$, $\dot{I}_{22}$, $\dot{S}_{22}$. Their residuals are all on the order of $\sim10^{-3}$, an order of magnitude worse than the fitting of QNMs. Note that there are fewer free parameters for the fitting of BOB than the QNM decomposition, a fairer comparison would be restricting to only 2 QNMs (so that there are 6 free parameters for both models) and studying the late ringdown portion\footnote{Based on TABLE I of Ref.~\cite{Giesler:2019uxc}, it corresponds to $\sim20M$ after the peak of strain.}. This is beyond the scope of this paper, and we leave the relevant discussions for future study. Table \ref{table:BOB-fit} is a summary for the fitting results, where the last row is the mismatch between BOB and NR, defined by
\begin{align}
{\rm Mismatch}=1-\frac{(h_{\rm B},h_{\rm NR})}{\sqrt{(h_{\rm NR},h_{\rm NR})(h_{\rm B},h_{\rm B})}}, \label{mismatch}
\end{align}
with 
\begin{align}
(h_{\rm B},h_{\rm NR})={\rm Re} \int_{0M}^{100 M}h_{\rm B}h_{\rm NR}^*dt,
\end{align}
where $h_{\rm B}$ and $h_{\rm NR}$ are the complex strains of BOB and NR in the time domain, respectively. The integration limit is taken to be the ringdown sector. We can see that the BOB for $\dot{I}_{22}$ and $\dot{h}_{22}$ lead to smaller mismatches than the other two. Meanwhile, the BOB behaves worst for $\dot{S}_{22}$: even though the mismatches can reach $2\times10^{-3}$, this is much higher than those achievable by $\dot{I}_{22}$ and $\dot{h}_{22}$; furthermore, the recovered estimations for spin and mass of the final black hole are substantially biased.

We then use Eq.~(\ref{I22-aym}) to make a connection between BOB and QNM decomposition, i.e., expanding BOB for the late-time ringdown $(t\gg t_p)$. The value of $(2Xe^{\gamma t_p})$ is expected to be close to $|\omega_{220}\mathcal{A}_{220}|$, so we make such a comparison in the sixth and seventh rows of Table \ref{table:BOB-fit}. We can see that $\dot{h}_{22}$ leads to the best agreement, while $\dot{S}_{22}$ the worst.

As we mentioned earlier, $X\sech[\gamma (t-t_p)]$ can also be used to describe the magnitude of $\ddot{h}_{2,\pm2}$ or $h_{22}$. Each of them leads to a class of BOB model (see Appendix \ref{app:BOB} for more details). We study their accuracy by fitting our simulations (cf.~Table \ref{table:NR-pars} and \ref{table:NR-pars-all}) to those three classes of BOB, and showing the distribution of mismatches (with NR waveforms) in the third column of Fig.~\ref{fig:BOB-fit}. Generally speaking, the mismatches of BOB are $10^{-5}-10^{-2}$, which are worse than those of QNM decomposition. Among the three classes, $\ddot{h}_{22}$ gives the smallest mismatch, while $h_{22}$ the largest.

Another way to quantify the accuracy of BOB is to compare the inferred spin and mass (from the fitting) to NR predictions. Similar to Ref.~\cite{Giesler:2019uxc}, we define a parameter deviation
\begin{align}
\epsilon=\sqrt{(\delta M_f/M)^2+(\delta \chi_f)^2}, \label{bob-par-dev}
\end{align}
and plot its distribution in Fig.~\ref{fig:BOB-fit}. We can see $\epsilon$ for BOB is also worse than the QNM decomposition. In addition, the distribution of $\epsilon$ show that the BOB works worst for $h_{22}$, and best for $\dot{h}_{22}$.

\section{Parameter-Estimation Contributions from Inspiral and Ringdown Stages}
\label{sec:fisher}


In this section, we demonstrate, with a few example sources, the impact of the ringdown portion of the waveforms --- as well as the correlation between the ringdown and the inspiral phases --- to parameter estimation errors. To do this, we will apply the  Fisher-matrix formalism to the NRSur7dq4 surrogate waveforms (for BBHs with $1<q<4$ and individual  dimensionless spin $\chi<0.8$) \cite{Varma:2018aht,Varma:2019csw},

%

In Sec.~\ref{sec:fisher:intro}, we will give a brief review for the Fisher-matrix formalism. In Sec.~\ref{sec:NRHybSur3dq8}, we discuss non-precessing BBH systems with varying total mass, illustrating how information contribution from the  ringdown stage gains more importance for more massive systems.   Finally, in Sec.~\ref{sec:NRSur7dq4}, we study parameter estimation errors of precessing systems, illustrating how estimations of individual spin components will benefit from information from the ringdown stage. 

\begin{figure}[htb]
         \includegraphics[width=\columnwidth,clip=true]{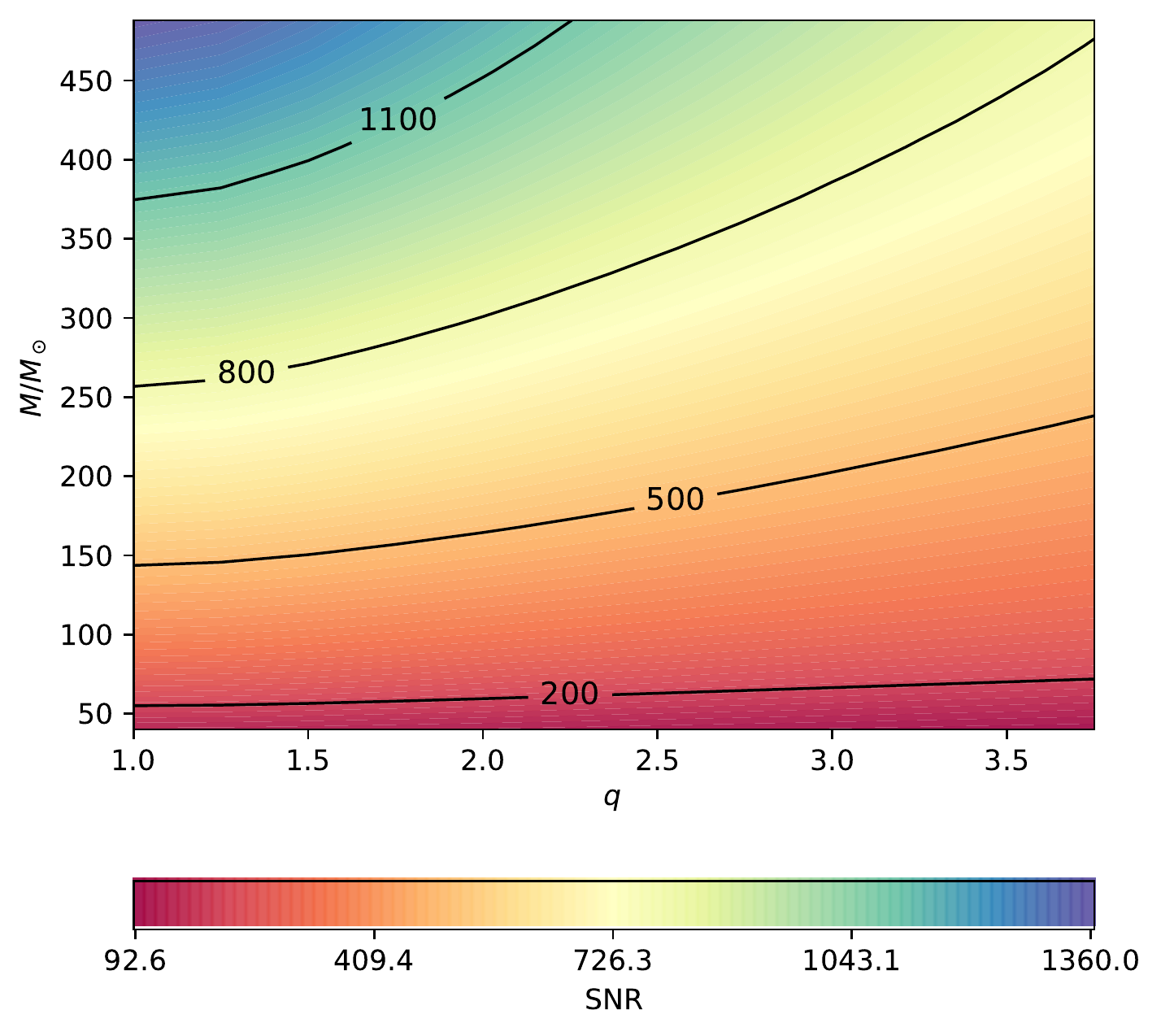}
  \caption{The SNR of an optimally oriented GW event with varying total (detector-frame) mass $M$ and mass ratio $q$, assuming the system is at redshift $z=1$ (6.7 Gpc) and using $S_n(f)$ of the CE.}
 \label{fig:snr}
\end{figure}

\begin{figure}[htb]
        \includegraphics[width=\columnwidth,height=10cm,clip=true]{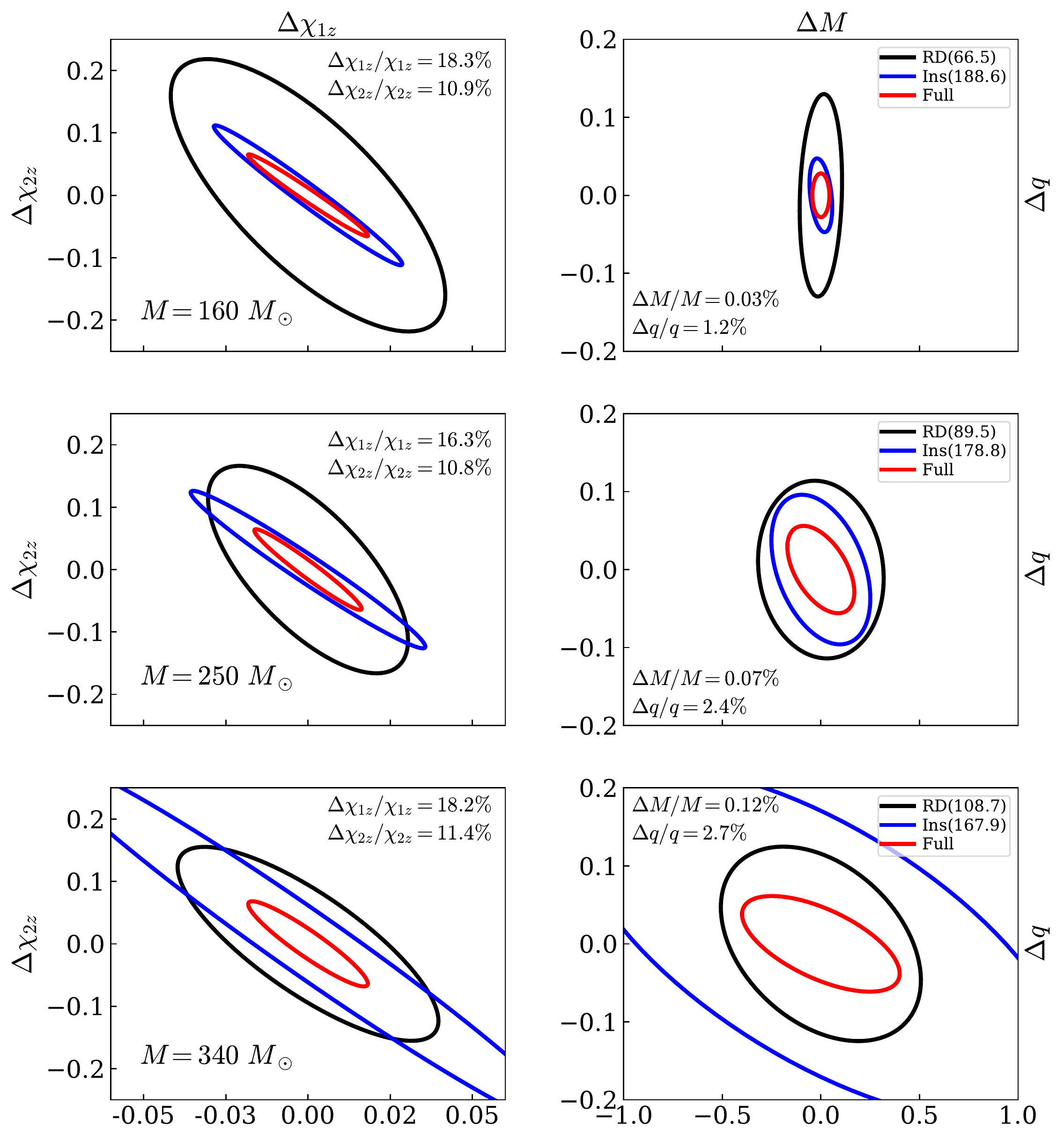}
  \caption{The error ellipses of $\chi_{1z}$ and $\chi_{2z}$ (the left column), as well as $M$ and $q$ (the right column), using the data from NRSur7dq4. Two individual spins are both aligned with the orbital angular momentum, and other parameters are $q=2.3,~\chi_{1z}=0.1,~\chi_{2z}=0.6,~ \iota=3\pi/10,~ \beta=\pi/2,~\chi_{1p}=\chi_{2p}=0$. The total mass is chosen to be $160M_\odot$ (the upper row), $250M_\odot$ (the middle row) and $340M_\odot$ (the bottom row). Three colors stand for the ringdown (black), inspiral (blue), and full sector (red), respectively. The numbers in parentheses are SNRs, where we normalize the total SNR of each event to 200 for comparison.}
 \label{fig:fim-hybrid-1}
\end{figure}

\subsection{The Fisher-Matrix Formalism and Waveform Models}
\label{sec:fisher:intro}
For a gravitational waveform $h(\theta^j)$ that depends on a list of parameters $\theta^j$, the Fisher matrix is given by
\begin{align}
\Gamma_{ij}=\left(\left.\frac{\partial h}{\partial \theta^i}\right|\frac{\partial h}{\partial \theta^j}\right). \label{fisher-def}
\end{align}
Here the inner product between two waveforms $(h|g)$ is defined as
\begin{align}
(h|g)=4{\rm Re} \int\frac{\tilde{h}^*(f)\tilde{g}(f)}{S_n(f)}df,
\end{align}
with the superscript $*$ standing for complex conjugation, and $S_n(f)$ the spectral density of the noise when detecting $h$.  In terms of this inner product, the signal-to-noise ratio (SNR) of a signal $h$ is given by $\sqrt{(h|h)}$. 

The covariance matrix for the estimated values of  $\theta^j$, in presence of noise, is given by the inverse of the Fisher matrix, 
\begin{align}
    {\rm Var}(\theta^i,\theta^j) = \left(\Gamma^{-1}\right)_{ij}\,.
\end{align}
From this, we obtain the individual estimation error for $\theta^j$, 
\begin{align}
\Delta\theta^i=\sqrt{(\Gamma^{-1})_{ii}}.
\end{align}
and the correlation coefficient between $\theta^i$ and $\theta^j$, 
\begin{align}
{\rm Corr}(\theta^i,\theta^j)=\frac{(\Gamma^{-1})_{ij}}{\sqrt{(\Gamma^{-1})_{ii}(\Gamma^{-1})_{jj}}}. \label{covariance}
\end{align}

Waveforms described by the NRSur7dq4 surrogate model are parametrized by 13 parameters: 
\begin{align}
\chi_{1z},\chi_{1p},\phi_1,\chi_{2z},\chi_{2p},\phi_2,M,q,\iota,\beta,t_c,\phi_c,D, \notag
\end{align}
Correspondingly, we have a 13-dimensional Fisher matrix. Here, 
the subscripts `1' and `2' stand for the two individual black holes in the binary system, $\chi_z$ is the spin component in the direction of orbital angular momentum, $M$ is the total mass in the detector frame, $q>1$ is the mass ratio, $D$ is the luminosity distance between the source and the detector, and $\iota$ and $\beta$ describe the wave emission direction in the frame of the source.
The spin component in the orbital plane is parameterized by the magnitude $\chi_p$ and the azimuthal angle $\phi$. Finally, $t_c$ and $\phi_c$ are the coalescence time and phase, respectively. 


\begin{figure}[htb]
         \includegraphics[width=\columnwidth,height=6.4cm,clip=true]{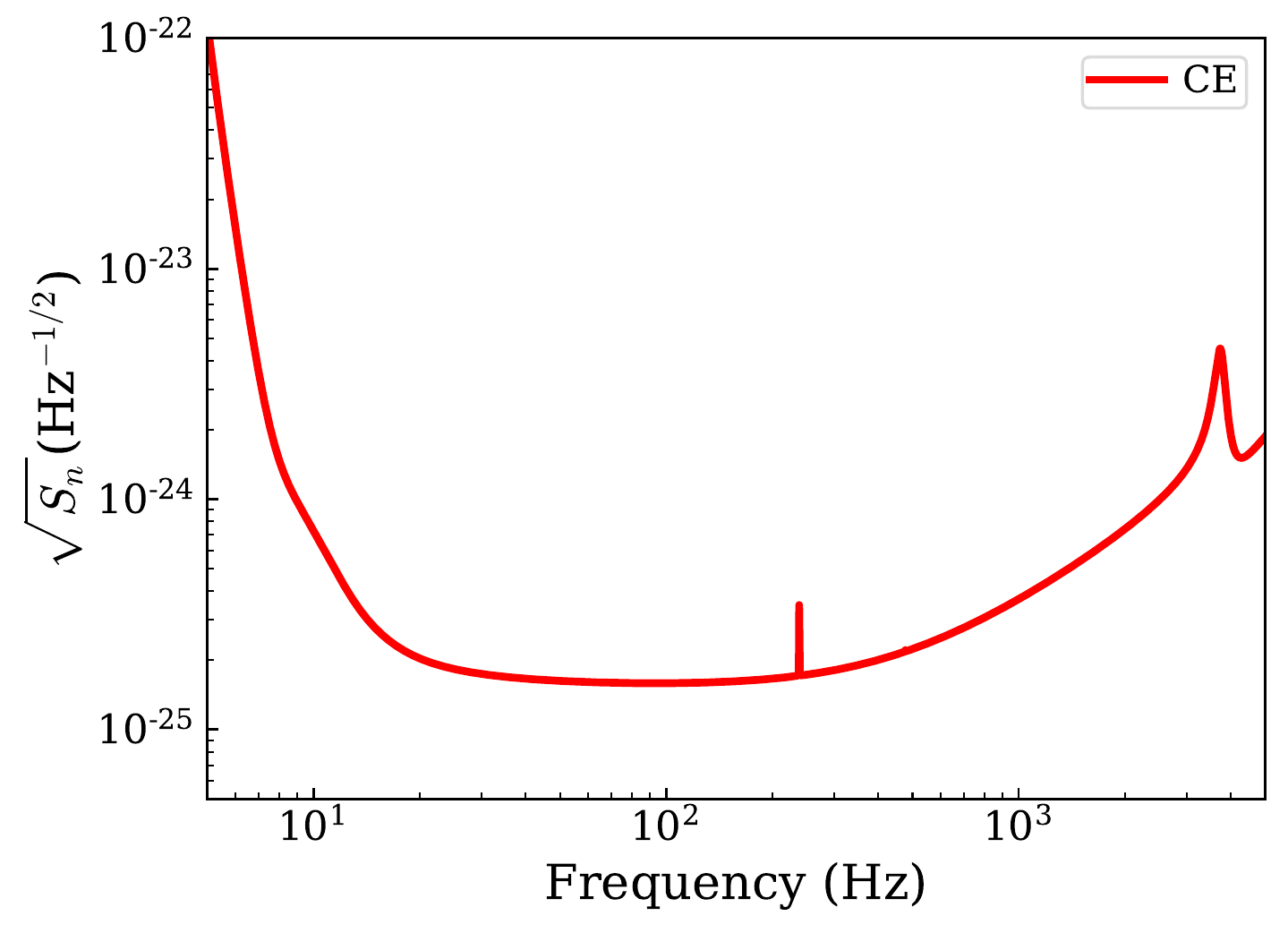}
  \caption{The noise spectral density of Cosmic Explorer.}
 \label{fig:psd}
\end{figure}

Throughout this paper, we adopt the Ansatz that the two gravitational-wave polarizations, $h_+$ and $h_\times$, can be individually measured, both with the noise spectrum $S_n$. This simplification allows us not to explicitly include sky location and orientation of the source; it can be justified in the situation of a three-detector network that can provide good source localization.  In this way, results given in this section should be more optimistic than the actual situation. 


\begin{figure*}[htb]
        \includegraphics[width=2\columnwidth,clip=true]{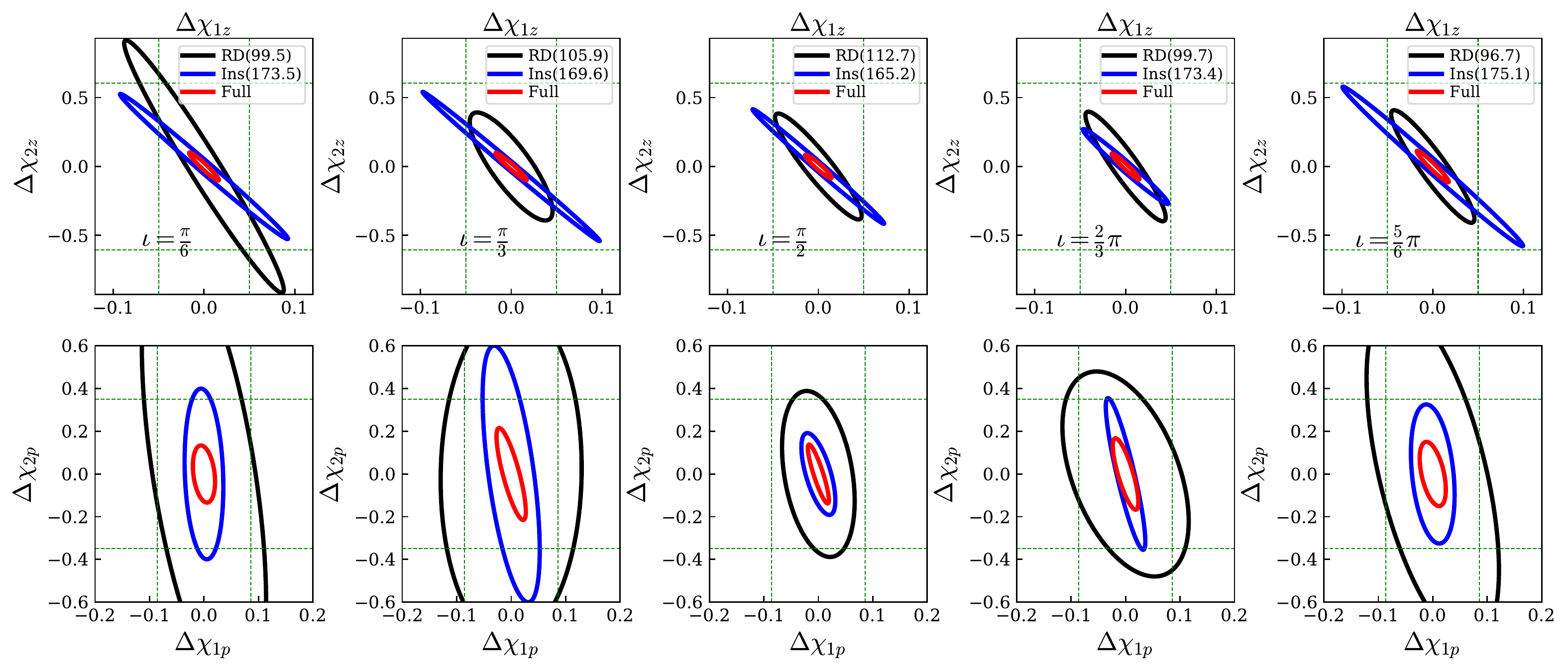}
  \caption{Similar to Fig.~\ref{fig:fim-hybrid-1}, the error ellipses of $\chi_{1z}$ and $\chi_{2z}$ (the first row), and $\chi_{1p}$ and $\chi_{2p}$ (the second row), with different $\iota$ (each column). The green dashed lines stand for the original value of each parameter. Thus we have a meaningful measurement (<100\%) on a parameter if the error ellipse is within the dashed lines. The BBH systems have parameters $M=300M_\odot$, $q=3.5$, $\chi_{1z}=0.05$, $\chi_{1p}=0.086$, $\chi_{2z}=0.606$, $\chi_{2p}=0.35$, $\phi_1=\pi/13$, $\phi_2=43\pi/52$, $\beta=\pi/2$. The error ellipses of ringdown and inspiral portions are not in the same direction, which implies different parameter correlations. After including the information of ringdown, the measurement accuracy of $\chi_z$ is improved by a factor of $\sim4-5$, whereas $\chi_p$ is improved by a factor of $\sim1.4$.}
 \label{fig:fim-NRSur7dq4-1}
\end{figure*}

\begin{figure*}[htb]
        \includegraphics[width=2\columnwidth,clip=true]{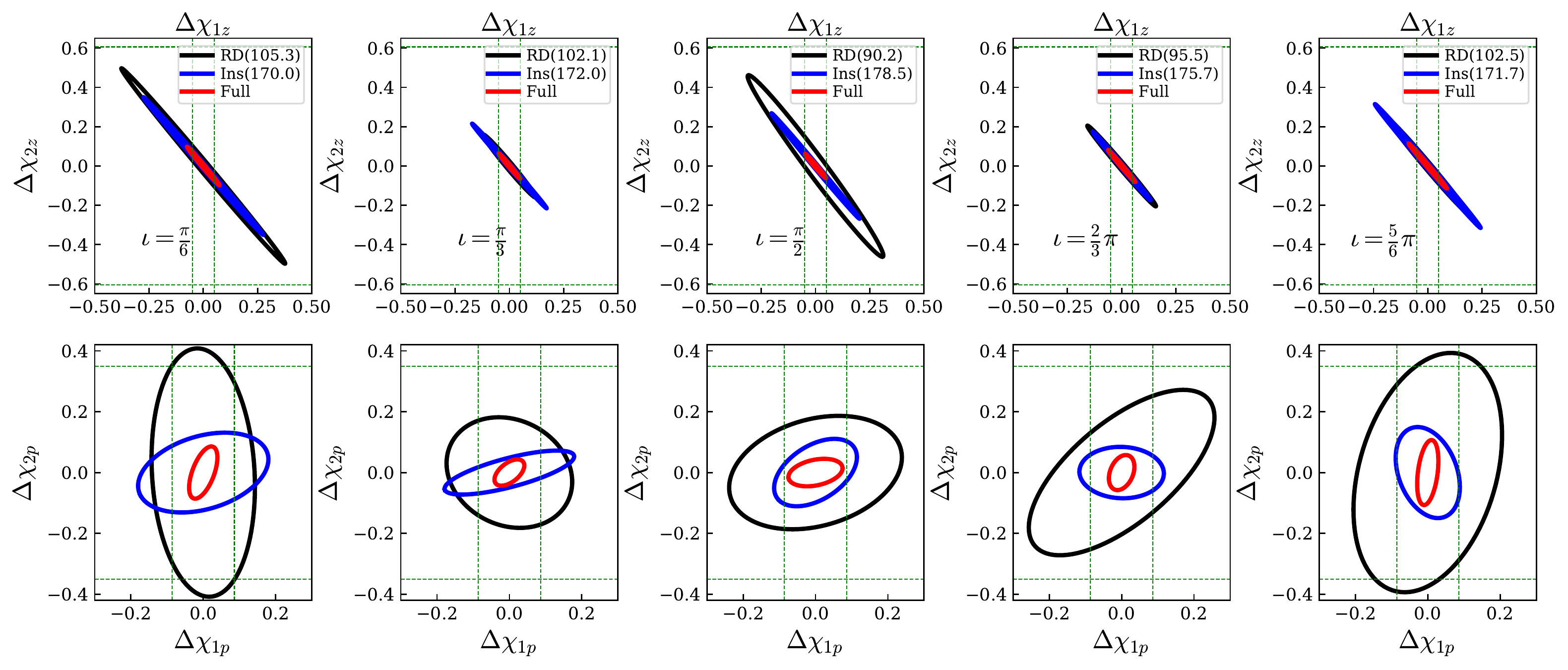}
  \caption{Same as Fig.~\ref{fig:fim-NRSur7dq4-1}, except $q=1.2$.}
 \label{fig:fim-NRSur7dq4-1-low-mass-ratio}
\end{figure*}
\subsection{Inspiral versus Ringdown: Non-precessing Binaries}
\label{sec:NRHybSur3dq8}

In this paper, we will focus mainly on the Cosmic Explorer (CE) \cite{LIGOScientific:2016wof}, whose $S_n(f)$ is shown in Fig.\ \ref{fig:psd}.  Using this sensitivity, in Fig.~\ref{fig:snr}, we show the SNR of an optimally oriented BBH with varying total (detector-frame) mass $M$ and mass ratio $q$, assuming the system is at redshift $z=1$ ($D_L=6.7$\,Gpc).  Note that the intrinsic total mass $M_{\bullet}$ is given by $M/(1+z)$. The high SNR shown in this figure indicates that in the 3G era, we will be most frequently be detecting binaries at cosmological distances of around $1\stackrel{<}{_\sim} z \stackrel{<}{_\sim} 3$. Correspondingly, we will be observing these binaries with higher detector-frame masses, with factor 2 to 4.  In this way, events like GW150914 can be redshifted to around $\sim 130$--$260\,M_\odot$, while heavy binaries like GW190521 can be shifted to $302$--$604\,M_\odot$. In the rest of this section, we shall study BBH systems with increasing total mass, in order to observe the increased importance of information contribution from the ringdown stage.

In order to study the ringdown and the inspiral portions individually, we separate two sectors (in the frequency domain) with the instantaneous $h_{22}$ frequency at $t=0M$ (where $\sqrt{\sum_{l,m}|h_{lm}|^2}$ is maximum). For non-precessing binaries, we will fix $q=2.3,~\chi_{1z}=0.1,~\chi_{2z}=0.6,~\chi_{1p}=\chi_{2p}=0,~ \iota=3\pi/10,~ \beta=\pi/2$, and consider $M/M_\odot = 160,\,250,\,340$. For comparison purposes, we normalize all waveforms so that the SNR of the entire waveform is 200. 
%
We consider joint parameter estimations errors of $(\chi_{1z},\chi_{2z})$ and $(M,q)$, with results  shown in Fig.~\ref{fig:fim-hybrid-1} (blue for inspiral alone, black for ringdown alone, and red for combined). As a reference, we also list the SNR of each sector in the figure (cf. numbers in parentheses).

Regarding the overall size of the error ellipses, for the BBH system with total mass $M = 160M_\odot$, constraints from the ringdown sector are worse than those from the inspiral portion. As $M$ increases to $250M_\odot$, constraints from the two sectors become comparable.  For more massive systems, the ringdown portion begins to dominate. It is remarkable that substantial parameter estimation can already be obtained from ringdown alone: this means not only the quasi-normal mode frequency, but also the excitation amplitudes, are providing the information \cite{Kamaretsos:2011um,Kamaretsos:2012bs,London:2014cma,Baibhav:2017jhs,Baibhav:2020tma}. We also note that detector-frame mass of $250\,M_\odot$ corresponds to intrinsic total mass of $\sim 125 M_\odot$ at $z=1$, which will not be a rare type of event in third-generation detectors. 

For spin measurements, both ringdown and inspiral sectors lead to somewhat degenerate measurement of $\chi_{1z}$ and $\chi_{2z}$. In particular,  the inspiral stage accurately measures the $\sim [q\chi_{1z}+(1-q)\chi_{2z}]$ direction (as can be argued from PN treatments~\cite{Racine:2008qv}), while the ringdown has a less degenerate measurement, although with a most accurately measured combination similar to that from the inspiral.  As for mass measurements,  the ringdown and inspiral sectors lead to $M-q$ error ellipses with different directions, but no substantial degeneracy breaking.

\subsection{Precessing BBH systems}
\label{sec:NRSur7dq4}
We now turn our attention to precessing systems. We set the total mass of the systems to $300M_\odot$, in order to make the contribution of ringdown sector comparable to the inspiral portion. Meanwhile, we choose $\chi_{1z}=0.05,~\chi_{2z}=0.606,~\chi_{1p}=0.086,~\chi_{2p}=0.35,~\phi_1=\pi/13,~\phi_2=43\pi/52$. The observation is made at $\beta=\pi/2$, with varying $\iota$.

We first study a BBH system with $q=3.5$. The results are shown in Figs.~\ref{fig:fim-NRSur7dq4-1}. We can see the relative size of ringdown and inspiral ellipses change with $\iota$. This is because the weights of different GW modes $\hlm$, i.e., $\sy$, are functions of $\iota$. At different observational locations, the contributions from different GW modes $\hlm$ are different. Secondly, the $\chi_{1z}-\chi_{2z}$ error ellipse computed from the ringdown portion is tilted relative to that of inspiral sector, which implies that the parameter correlation of these two sectors are different. After combining the information of ringdown and inspiral, the measurement accuracy of $\chi_{z}$ is around $30\%$, improved by a factor of $\sim4-5$ compared with using the inspiral signal only. This result agrees qualitatively with the discussion in Ref.~\cite{Biscoveanu:2021nvg}. We note that SNR is 1.14 times greater after incorporating the ringdown signal. Hence most of the improvement is contributed from the correlation between the inspiral and ringdown, which leads to reduction of parameter degeneracy. On the other hand, the measurement accuracy of $\chi_{p}$ is only improved by a factor of $\sim1.4$, not as good as the one of $\chi_{z}$. Nevertheless, the fractional error of $\chi_{p}$ is smaller than $100\%$, hence we can still put meaningful constraints on $\chi_{p}$.

We want to remark that values mentioned above depend heavily on properties of the BBH system in question. For instance, for a low-mass-ratio BBH system with $q=1.2$, as shown in Fig.~\ref{fig:fim-NRSur7dq4-1-low-mass-ratio}, the $\chi_{1z}-\chi_{2z}$ error ellipses computed from individual inspiral and ringdown sectors point along more similar directions.  This leads to much less degeneracy breaking between $\chi_{1z}$ and $\chi_{2z}$ than in the $q=3.5$ case above, in particular making the measurement error  $\Delta\chi_{1z}$ greater than the value of $\chi_{1z}$.  
%
%
Even so, incorporating ringdown, in addition to inspiral improves $\Delta\chi_{z}/\chi_z$  a factor of 2.8, substantially greater than the SNR improvement factor of around 1.16.   

\section{Conclusions}
\label{sec:conclusion}
In this paper, we studied the gravitational waveforms of SKd systems, using both NR simulations (\texttt{SpEC}) and surrogate models (NRSur7dq4, NRSur7dq4Remnant). We first decomposed the ringdown portion of GW signal into QNMs, and explored how mode amplitudes of overtones depend on the progenitor's parameters (for $I_{22}$, $S_{22}$, as well as $h_{2\pm2}$ contents). We then studied the features of the mass and current quadrupole waves, focusing on their time evolutions and peak values. This leads to a qualitative understanding of kick velocity. Next, we fitted the evolution of $I_{22}(t)$ and $S_{22}(t)$ to the Backward-One-Body (BOB) model. Finally, we used Fisher inforamtion matrix to study the role of the ringdown state in parameter correlation. Here we summarize our main results: 

(i) For SKd systems, the dependences of $I_{22}$ and $S_{22}$ on angular parameters $(\chi_{\rm init},\theta_{\rm init},\phi_{\rm init})$ can be separated from their  temporal dependences [Eq.~(\ref{universality})]. 

(ii) Similar to the case of EMRI \cite{Hughes:2019zmt}, the QNM amplitudes of SKd systems encode the information of progenitors' parameters. As an extension to  Ref.~\cite{Hughes:2019zmt}, we included more overtones to the QNM decomposition. We found that the spectra peak at the fourth overtone, and that the dependence of mode magnitudes on $\phi_{\rm init}$ is insensitive to the overtone index $n$ (up to a scaling factor).  We found that the dependence of mode amplitudes on progenitor parameters is more easily understood when decomposed into mass and current quadrupole waves, instead of $(2,2)$ and $(2,-2)$ modes. 

(iii) Peak values of  mass ($\lp$) and current ($\spp$) quadrupole waves encode the information of progenitors' spin. Enforced by the parity symmetry, the $(\lp,\spp)-(\chi_{\rm init},\theta_{\rm init},\phi_{\rm init})$ pattern is symmetric about $\theta_{\rm init}=\pi/2$ axis and has a period of $\pi$ in the direction of $\phi_{\rm init}-$axis. Quantitatively speaking, the $(\lp,\spp)-(\chi_{\rm init},\theta_{\rm init},\phi_{\rm init})$ dependence are consistent with the PN-inspired formulas. 

(iv) The phase difference between mass and current quadrupole waves $\dphi$ can lead to a qualitative understanding of kick velocity.  Its time evolution can be anticipated from PN and black-hole perturbation theories: in the inspiral regime, $\dphi$ is equal to the difference between the orbital and precession phases; near the merger, the spin precession rate is gradually locked to the orbital frequency --- until well into the ringdown regime, when $\dphi$ should become constant since both $I_{22}$ and $S_{22}$ oscillate at the fundamental QNM frequency.
%
%
%
However, we found that $\dphi$ does not always settle down to a constant value during the post-merger stage, especially for high-kick cases. Instead, there is a slow change over time. This is due to the Doppler shift caused by the kick. The QNM frequency of $h_{22}$ (emitted upwards) is slightly different from the one of $h_{2,-2}$ (emitted downwards), which leads to a slow time evolution. In fact, the relative frequency difference is on the same order as the kick velocity. 

(v) We verified that the BOB phenomenological model is accurate for the ringdown evolution of $\dot{h}_{2,\pm2}$, $\ddot{h}_{2,\pm2}$ and $I_{22}$, but much less so for $S_{22}$ and $h_{2,\pm2}$.  This calls for further, qualitative improvements of the current-quadrupole sector of the BOB model.

(vi) We found that in 3G detectors, the contribution of the ringdown part dominates over the inspiral part as the total detector-frame mass exceeds $\gtrsim 250-300 M_\odot$. We found that, as we combine both parts, the improvement in parameter estimation error is larger than the increase in SNR, indicating that the reduction of degeneracy due to the additional ringdown signal is the main reason for such improvement. 
As for $\chi_z$, in our examples,  incorporating the information from ringdown signal can lead to $\sim 4-5$ times improvement on the measurement accuracy, while the accuracy for $\chi_p$ is improved by a factor of $\sim1.4$. 

Our results indicate that the ringdown sector of a BBH event encodes plenty of information about the progenitor. It also plays a complementary role to PN theory in the study of BBH evolution. In our study, we primarily focused on the SKd configuration. Future work could include more generic BBH systems and other GW modes, which can lead to more comprehensive understandings of the ringdown signals. Another possible avenue for future work is to increase the precision of NR surrogate models for the ringdown sector, since our work has revealed that the current NR surrogate models are not accurate enough for BH spectroscopy. A more accurate ringdown surrogate model will be beneficial for both data analysis and theoretical studies.

Meanwhile, as revealed in Fig.~\ref{fig:overtone-h-scott}, as well as Eqs.~(\ref{IS-universal-overtone}) and (\ref{scott-explain}), it might also be interesting for future work to investigate the features of mass and current quadrupole waves of EMRIs, which may turn out to be simpler than features found in  Refs.~\cite{Hughes:2019zmt,Apte:2019txp,Lim:2019xrb}. Those further explorations could potentially provide us more physical understandings of EMRI ringdown spectra.
\begin{acknowledgments}
We want to thank Serguei Ossokine, Alvin Chua, Gregorio Carullo, and Sean McWilliams for useful discussions.
S.M.\ and Y.C.\ are supported by the Simons Foundation (Award Number 568762), the Brinson Foundation, and the National Science Foundation (Grants PHY--2011968, PHY--2011961 and PHY--1836809). 
V.V.\ is generously supported by a Klarman Fellowship at Cornell, the Sherman
Fairchild Foundation, and NSF grants PHY–170212 and PHY–1708213 at Caltech.
The computations presented here were conducted on the Caltech High Performance Cluster, partially supported by a grant from the Gordon and Betty Moore Foundation.
\end{acknowledgments}  

\appendix

\section{SpEC runs---SKu configuration }
\label{app:spec-run}
We summarize our NR simulations of SKu BBHs in Table \ref{table:NR-pars-all}. We remark that the SKu condition is not well preserved after the junk-radiation regime. Nevertheless, the maximum recoil velocity $v_f^z$ is 4050 km s$^{-1}$, and it is roughly proportional to $\chi_{\rm init}$.
\begin{table*}
    \centering
    \caption{A summary for SKu configurations. The convention is the same as the one used in Table \ref{table:NR-pars}, except that the fifth and sixth columns are the components of individual spin in the Cartesian coordinates, where the $z$-axis is in the direction of orbital angular momentum; the line of two BHs determines the $x$-axis; and the right-handed rule determines the $y$-axis. The dimensionless spin ranges from 0.6 to 0.95, specified at the orbital frequency $\Omega_{\rm orb}$.}
    \begin{tabular}{c c c c c c c c c c} \hline\hline
\multicolumn{3}{c}{Run label}  &$\Omega_{\rm orb}$ & \multirow{2}{*}{$\chi_1$}  & \multirow{2}{*}{$\chi_2$} &\multirow{2}{*}{$|\chi_1|=|\chi_2|$}  & \multirow{2}{*}{$m_f$} & $v_f^{z}$ & \multirow{2}{*}{$\chi_f$}   \\ 
 & This paper & SXS:BBH & $(\times 10^{-2})$ & & & & & $(\times 10^{-3})$ \\ \hline\hline
\multirow{9}{*}{SKu6}& `01' & 2428  & 1.63 & $(0.378,-0.378,0.273)$ & $(-0.413,0.389,0.200)$ &0.6 & 0.944 & $-1.46$ & $0.754$ \\ \cline{2-10}
&`02' & 2429  & 1.62 & $(0.390,0.359,0.281)$ & $(-0.402,-0.398,0.199)$ &0.6 & 0.942 & $-8.04$ & $0.749$ \\ \cline{2-10}
&`03' & 2430  & 1.63 & $(-0.374,0.383,0.271)$ & $(0.406,-0.395,0.199)$&0.6 & 0.944 & $-0.34$ & $0.754$ \\ \cline{2-10}
&`04' & 2431 &  1.62 & $(-0.386,-0.364,0.281)$ & $(0.397,0.402,0.201)$&0.6 & 0.942 & $8.03$ & $0.749$ \\ \cline{2-10}
&`05' & 2432  & 1.63 & $(0.254,0.465,0.282)$ & $(-0.254,-0.504,0.202)$&0.6 & 0.942 & $-7.54$ & $0.749$ \\ \cline{2-10}
&`06' & 2448 & 1.63 & $(0.533,-0.0207,0.275)$  & $(-0.568,2.00\times10^{-3},0.193)$&0.6 & 0.944 & $-6.80$ & 0.752 \\ \cline{2-10}
&`07' & 2449 & 1.63 & $(-4.75\times10^{-3},0.531,0.279)$ & $(0.0218,-0.564,0.203)$&0.6 & 0.942 & $-6.01$ & 0.750 \\ \cline{2-10}
&`08'& 2450  & 1.63 & $(0.0120,-0.531,0.280)$ & $(-0.0312,0.564,0.202)$&0.6 & 0.942 & $5.91$ & 0.750 \\ \hline
\multirow{4}{*}{SKu8} & `01' & 2433  & 1.63 & $(0.666,0.308,0.320)$ & $(-0.667,-0.314,0.311)$ & 0.8 & 0.941 & $-1.58$ & 0.773 \\ \cline{2-10}
& `02' & 2434 & 1.63 & $(-0.352,0.647,0.312)$  & $(0.360,-0.649,0.300)$ & 0.8 & 0.939 & $-11.0$ & 0.767 \\ \cline{2-10}
& `03' & 2435  & 1.63 & $(-0.669,-0.306,0.316)$  & $(0.667,0.309,0.316)$ & 0.8 &  0.941 & $1.31$ & 0.773 \\ \cline{2-10}
& `04' & 2436  & 1.63 & $(0.382,-0.629,0.315)$ & $(-0.390,0.630,0.301)$ & 0.8 & 0.939 & 11.0 & 0.766 \\ \hline
\multirow{14}{*}{SKu95} & `01' & 2437  & 1.63 & $(-0.793,-0.437,0.284)$ & $(0.792,0.437,0.290)$ & 0.95 & 0.942 & $2.43$ & 0.765 \\ \cline{2-10}
& `02' & 2438  & 1.62 & $(0.422,-0.803,0.280)$ & $(-0.423,0.804,0.279)$ & 0.95 & 0.938 & 13.5 & 0.752 \\ \cline{2-10}
& `03' & 2439  & 1.63 & $(0.800,0.423,0.288)$ & $(-0.800,-0.426,0.283)$ & 0.95 & 0.942 & $-4.29$ & 0.765  \\ \cline{2-10}
& `04' & 2440  & 1.63 & $(-0.428,0.802,0.277)$  & $(0.425,-0.801,0.283)$&0.95 & 0.938 & $-13.5$ & $0.753$ \\ \cline{2-10}
& `05' & 2441 & 1.63 & $(-0.826,-0.377,0.277)$ & $(0.824,0.376,0.284)$ & 0.95 & 0.941 & $10.6$ & $0.760$ \\ \cline{2-10}
& `06' & 2442  & 1.62 & $(0.390,-0.821,0.275)$ & $(-0.390,0.822,0.2733)$ & 0.95 & 0.938 & 13.5 & 0.750 \\ \cline{2-10}
& `07' & 2443  & 1.62& $(-0.358,0.837,0.272)$  & $(0.355,-0.836,0.278)$ & 0.95 & 0.938 & $-13.5$ & 0.751 \\ \cline{2-10}
& `08' & 2444 & 1.64 & $(0.293,-0.820,0.380)$ & $(-0.301,0.825,0.363)$ & 0.95 & 0.936 & $-2.80$ & 0.785 \\ \cline{2-10}
& `09' & 2445 & 1.65 & $(0.826,0.279,0.375)$ & $(-0.829,-0.281,0.368)$ & 0.95 & 0.933 & $13.0$ & 0.776 \\ \cline{2-10}
& `10' & 2446 & 1.64 & $(-0.229,0.842,0.376)$ & $(0.239,-0.850,0.351)$ & 0.95 & 0.936 & $3.35$ & 0.784  \\ \cline{2-10}
& `11' & 2447  & 1.65 & $(-0.836,-0.252,0.372)$ & $(0.837,0.251,0.372)$ & 0.95 & 0.933 & $-12.4$ & 0.776 \\ \hline\hline
     \end{tabular}
     \label{table:NR-pars-all}
\end{table*}

\begin{figure*}[htb]
        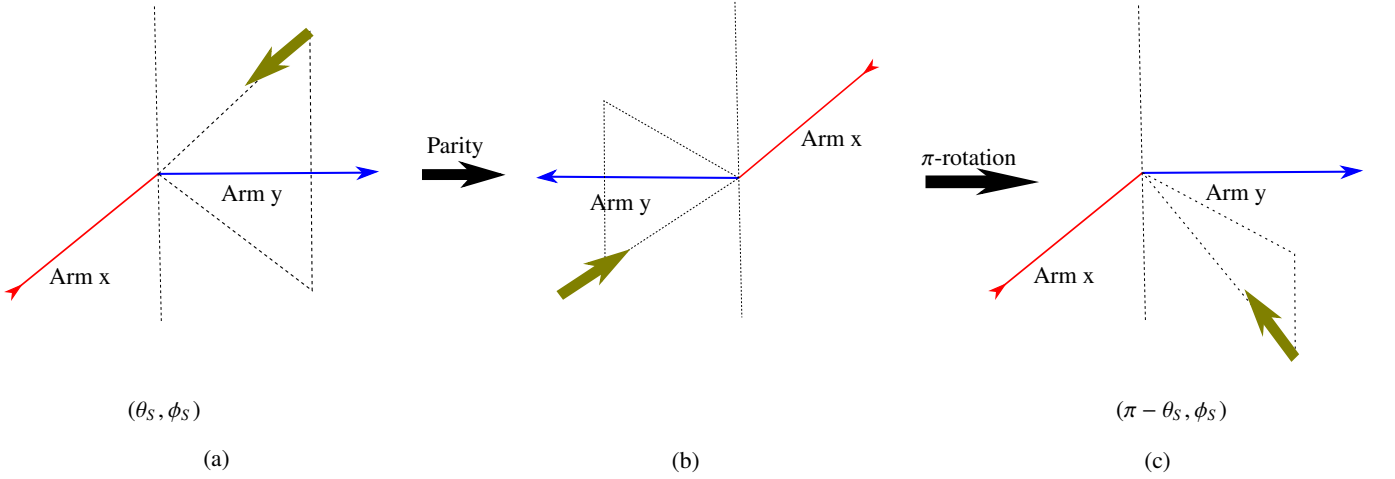
  \caption{Parity inversion of a SKd BBH binary system within the detector frame. The arrow stands for the direction of incoming GW.  The system undergoes a parity inversion from (a) to (b). We further rotate the whole system around the vertical dash line by $\pi$, which leads to (c). Comparing (a) and (c), the polar angle of sky location $\theta_S$ becomes supplementary under the transformation.}
 \label{fig:app-parity}
\end{figure*}

\section{BOB for $h_{22}$ and $\ddh_{22}$ }
\label{app:BOB}
In this section, we discuss the BOB model for $h_{22}$ and $\ddh_{22}$.
\subsection{ $\ddh_{22}$}
 Let us start from $\ddh_{22}$. As discussed in Ref.~\cite{McWilliams:2018ztb}
 \begin{align}
 \ddh_{22}=\frac{d}{dt}\dot{h}_{22}&=\frac{d}{dt}|\dot{h}_{22}|e^{-i\phi_{22}(t)}\sim-i\dot{\phi}_{22}|\dot{h}_{22}|e^{-i\phi_{22}(t)} \notag \\
 &=-i\Omega_{22}|\dot{h}_{22}|e^{-i\phi_{22}(t)},
 \label{bob-hddot}
 \end{align}
 where we have assumed that $|\dot{h}_{22}|$ changes much slower than $\phi_{22}$. The above equation implies that the frequency of $\ddh_{22}$ and $\dot{h}_{22}$ are roughly the same. Therefore, below we do not distinguish the frequency of $\ddh_{22}$ from that of $\dot{h}_{22}$, and use $\Omega_{22}$ to stand for both frequencies. Combining Eqs.~(\ref{BOB-h-omega}) with (\ref{bob-hddot}), we obtain
\begin{align}
|\dot{h}_{22}|^2\sim|\ddh_{22}|^2/\Omega_{22}^2\propto \frac{d}{dt}\Omega_{22}^2.
\end{align}
Then applying Eq.~(\ref{bob-intro}) to $|\ddh_{22}|$\footnote{We use the same notation as Eq.~(\ref{BOB-h22dot}) since this will not cause any confusion.}, i.e.,
\begin{align}
&|\ddh_{22}|=X\sech[\gamma (t-t_p)],
\end{align}
which leads to  
\begin{align}
\Omega_{22}=\left\{\Omega_0^{4}+\frac{\omega_0^4-\Omega_0^{4}}{2}\left[\tanh\gamma(t-t_p)+1\right]\right\}^{1/4}, \label{omega-hddot} 
\end{align}
The above equation implies 
\begin{align}
\lim_{t\to\infty}\Omega_{22}^\2=\omega_0,
\end{align}
which is the same as the case of $\dot{h}_{22}$ [Eq.~(\ref{bob-hdot-fre-asy})]. Integrating Eq.\ (\ref{omega-hddot}) again gives the time dependence of $\phi_{22}$, i.e., the phase of $\ddh_{22}$
\begin{align}
\phi_{22}&=\frac{1}{\gamma}\left\{\omega_0\left(\arctanh\frac{\Omega}{\omega_0}+\arctan\frac{\Omega}{\omega_0}\right)\right. \notag \\
&\left.-\Omega_0\left(\arccoth\frac{\Omega_0}{\Omega}+\arccot\frac{\Omega_0}{\Omega}\right)\right\}-\phi_0. \label{bob-phase-hddot}
\end{align}
This is the original form of BOB model [cf.~Eq.~(10) of Ref.~\citep{McWilliams:2018ztb}]. Clearly, Eq.~(\ref{bob-phase-hddot}) is different from Eq.~(\ref{bob-phase-hdot}).
\subsection{$h_{22}$}
Following the same line of reasoning, the frequency of $h_{22}$ is also approximately equal to $\Omega_{22}$. Therefore
\begin{align}
|\dot{h}_{22}|^2\sim|h_{22}|^2\Omega_{22}^2\propto \frac{d}{dt}\Omega_{22}^2.
\end{align}
Then using the assumption
\begin{align}
&|h_{22}|=X\sech[\gamma (t-t_p)],
\end{align}
we obtain
\begin{align}
\Omega_{22}=\Omega_{0}X\gamma^{-1/2}[\tanh\gamma(t-t_p)+1]^{1/2}.
\end{align}
Integrating the above equation again can lead to a tedious expression of $\phi_{22}$, we do not show it here.

The BOB model for $h_{22}$, $\ddh_{22}$, together with the one for $\dot{h}_{22}$ [Eq.~(\ref{BOB-h22dot})], are used to fit NR results, and are compared to QNMs in Fig.~\ref{fig:BOB-fit}. We can see the model works the worst for $h_{22}$.

\section{The parity transformation of a complex strain}
\label{app:parity}

In this section, we show that the complex strain $h=h_+-ih_\times$ is transformed to the complex conjugate if the whole system undergoes a parity transformation (including the BBH system and observer).

According to Fig.~\ref{fig:parity}, under the parity transformation two BHs exchange their locations, while have their individual spin fixed, since axial vectors are not changed by the parity transformation. Meanwhile, within the detector frame, the orientation of detector arms and the propagation direction are flipped simultaneously, as shown in Fig.~\ref{fig:app-parity} (a) and (b). We want to emphasize that the GW detector is a 2D plane (formed by two arms). Its parity transformation can be equivalently achieved by a $\pi$-rotation about the axis that is perpendicular to the detector plane. Therefore, we further rotate the whole system about the vertical dash line by $\pi$, as shown in Fig.~\ref{fig:app-parity} (c). We can see that the detector configuration changes back to the one of (a), while the sky location of GW source changes from $(\theta_S,\phi_S)$ (the northern hemisphere) to $(\pi-\theta_S,\phi_S)$ (the southern hemisphere). 

Since GR preserves the parity, the strain $h_{\rm obs}$ observed by a detector:
\begin{align}
h_{\rm obs}=h_+F_++h_\times F_\times,\notag 
\end{align}
is not affected by the abovementioned transformations. The antenna patterns $F_{+,\times}$ have forms \cite{Poisson+Will+14}
\begin{align}
F_+=\frac{1}{2}(1+\cos^2\theta_S)\cos2\phi_S, \quad
F_\times=\cos\theta_S\sin2\phi_S, \notag 
\end{align}
with $(\theta_S,\phi_S)$ the sky location of GW source relative to the detector. 
Under the transformation from Fig.~\ref{fig:app-parity} (a) to (c), i.e., $(\theta_S,\phi_S)\to(\pi-\theta_S,\phi_S)$ the antenna patterns $F_{+,\times}$ transform as
\begin{align}
F_+\to F_+, \quad F_\times \to -F_\times.\notag 
\end{align}
Recalling that the $h_{\rm obs}$ of Fig.~\ref{fig:app-parity} (a) and (c) are the same, we then have
\begin{align}
h_+\to h_+, \quad h_\times \to -h_\times.\notag 
\end{align}
As a result,
\begin{align}
h=h_+-ih_\times \to h^*=h_++ih_\times.
\end{align}


\bibliography{References}

\end{document}